\documentclass[11pt]{article}
\usepackage{geometry} 
\usepackage[parfill]{parskip}    % Activate to begin paragraphs with an empty line
\usepackage{graphicx}
\usepackage{amssymb}
\usepackage{epstopdf}
\usepackage{subfigure}
\usepackage{fancyhdr}
\usepackage{algorithms/algorithmic}
\usepackage{algorithms/algorithm}
\usepackage{array}
\usepackage{url}
\usepackage{comment}
\usepackage[cmex10]{amsmath}
\DeclareMathOperator*{\argmax}{argmax}
\DeclareMathOperator*{\argmin}{argmin}
\usepackage{multirow}
\usepackage{color}

\title{ADIS - A robust pursuit algorithm for probabilistic, constrained and non-square blind source separation with application to fMRI}

\author{Gautam Pendse\thanks{To whom correspondence should be addressed. e-mail: gpendse@mclean.harvard.edu}$^{\,\,\, 1}$  \and David Borsook$^{1}$ \and Lino Becerra$^{1}$ \\
\mbox{}\\ \\ \\ \\ 
$^1$ Imaging and Analysis Group (IMAG), Harvard Medical School}
\date{Feb 27, 2009}
\begin{document}

\maketitle

\newpage

\begin{abstract}
In this article, we develop an algorithm for probabilistic and constrained projection pursuit. Our algorithm called ADIS (automated decomposition into sources) accepts arbitrary non-linear contrast functions and constraints from the user and performs \textit{non-square} blind source separation (BSS). In the first stage, we estimate the latent dimensionality using a combination of bootstrap and cross validation techniques. In the second stage, we apply our state-of-the-art optimization algorithm to perform BSS. We validate the latent dimensionality estimation procedure via simulations on sources with different kurtosis excess properties. Our optimization algorithm is benchmarked via standard benchmarks from GAMS performance library. 
We develop two different algorithmic frameworks for improving the quality of local solution for BSS. Our algorithm also outputs extensive convergence diagnostics that validate the convergence to an optimal solution for each extracted component. The quality of extracted sources from ADIS is compared to other well known algorithms such as Fixed Point ICA (FPICA), efficient Fast ICA (EFICA), Joint Approximate Diagonalization (JADE) and others using the ICALAB toolbox for algorithm comparison. In several cases, ADIS outperforms these algorithms. Finally we apply our algorithm to a standard functional MRI data-set as a case study.
\end{abstract}

\section{Introduction}
The Generalized Linear Model (GLM) is a popular tool for analyzing functional MRI (fMRI) data. GLM analysis proceeds on a voxel by voxel basis using the same design matrix. One of the difficulties associated with GLM analysis is the construction of an appropriate design matrix. Unmodeled regressors that modulate the fMRI signal in addition to the EVs but are not a part of the design matrix will invalidate the analysis and the associated inferences. Further, these unmodeled regressors might be different in different brain regions and a voxel by voxel analysis with a common design matrix may not be appropriate. These considerations imply that model based analyses make very strong assumptions which are very likely violated in a real fMRI dataset.

Model free analysis on the other hand does not need any postulations as to the shape of the expected response. One such technique that has become popular in recent years, particularly for application to fMRI is Independent Component Analysis (ICA) \cite{Comon:1994}. See \cite{Hyvarinen:1999} for a survey on ICA. Popular software packages such as FSL (\cite{FSL:2004}) come with ICA software for doing non-square ICA via automatic latent dimensionality estimation also known as Probabilistic Independent Component Analysis (PICA) \cite{Beckmann:2004}. Essentially these techniques consists of a data reduction step using PCA followed by the application of standard ICA algorithms. In the signal processing community, many well established algorithms exist for doing ICA, the most popular ones being efficient FastICA \cite{EFICA:2006}, Fixed Point ICA (FPICA) \cite{FPICA:2000}, Joint Approximate Diagonalization of Cumulant Matrices (JADE) \cite{JADE:1996}, Extended Robust ICA (ERICA) \cite{ERICA:2002} and unbiased ICA (UNICA) \cite{UNICA:2000}. In this article, we will question the validity of solution produced by current ICA codes. Are these solutions really optimal? Can we afford to pay the price for using non-optimal solutions?

One of the challenges involved in applying ICA to real data is the confidence in the quality of estimated solution. This problem has been recognized before. For example the software package ICASSO \cite{ICASSO:2004} uses bootstrapping simulations to run an ICA algorithm multiple times and then clusters the estimated sources  to assess reliability. ICASSO takes into account the "algorithmic" variability and the variability in the original data induced due to sampling, but since it assumes square, noise free mixing it ignores the estimation errors induced due to noisy source mixing. The presence of non-square mixing in real data also introduces additional variability due to the unknown latent dimensionality of the sources.

While a bootstrapping strategy can always be used to test the "sensitivity" of estimated BSS solution from any algorithm, it is critical to have a reliable and verifiable optimization solver to solve the non-convex BSS problem in the first place.

Currently existing ICA codes perform optimization using techniques that have formulas for updating the unknown variables using a Newton step, gradient descent, natural gradient \cite{Bell:1997}, \cite{Amari:1996} \cite{Xavier:1998} or similar strategies. In addition they also potentially have a number of heuristic rules for updating the various "learning" parameters in these algorithms. No convergence diagnostics are used to check the optimality of estimated solution. To the best of our knowledge, such optimization codes are not benchmarked using standard optimization test cases. Such ad-hoc strategies along with non-verified optimality may severely affect the quality of solution produced by these algorithms and potentially impact the practical conclusions drawn from incorrect results. The popular FastICA algorithm \cite{EFICA:2006} uses an approximate Newton iteration where the approximate Hessian simplification reduces it to a gradient descent algorithm with a fixed step size. However there is no reason to use these approximations. One can use state of the art optimization software to compute near exact step sizes with locally varying Hessian approximations. In fact, it has been shown \cite{Zarzoso:2006} that these exact step size search significantly increases the estimation efficiency and robustness to initialization in comparison to the fixed update rules of FastICA. The optimization core in our algorithm ADIS efficiently handles local non-convexity as well as allows for infeasible steps, i.e., steps that violate the constraints leading to a more fuller exploration of parameter space and increasing the likelihood of converging to a global optimum.

Another issue is the modification of the default contrast function in ICA (e.g. Negentropy) to do other types of source extraction. It is also desirable to be able to add additional equality and/or inequality constraints to BSS estimation depending on the application at hand. These issues cannot be addressed using current ICA software. In this paper, we develop our algorithm ADIS and validate its various components. Finally, we compare ADIS to currently existing ICA codes on many standard benchmark datasets from ICALAB.

Our algorithm ADIS (section \ref{ppp}, \ref{ppproblems}):
\begin{enumerate}
\item Uses a state-of-the-art optimization algorithm at its core (inspired by LANCELOT software \cite{LANCELOT:1992}) (section \ref{optimalg}, \ref{optimalgdetails})
\item Uses a bootstrap simulation/cross-validation based approach for latent dimensionality estimation (in case of non-square BSS) (section \ref{latentdim})
\item Enables the user to use arbitrary contrast functions and nonlinear constraints for BSS
\item Produces "good quality" local solutions using a special multistage framework for BSS (section \ref{ppproblems})
\item Produces extensive convergence diagnostics for each extracted component to validate the "optimality" of the extracted source (section \ref{optimalg})
\end{enumerate}

We perform validation of each component of ADIS as follows:
\begin{enumerate}
\item Validation of the latent dimensionality estimation procedure using simulations on sources with different statistical properties (section \ref{latentdim})
\item Validation of our optimization core using standard benchmarks from the GAMS performance library (\url{http://www.gamsworld.org/performance}, \cite{COPS3}) (section \ref{supplemental})
\item We then use the "Negentropy" contrast function as a special case and compare the results of ADIS in terms of separation quality and robustness to other well known algorithms such as efficient FastICA, FPICA, JADE and others using the ICALAB toolbox \cite{ICALAB_BOOK:2003}, \cite{ICALAB} for BSS algorithm comparison. (section \ref{bssbench})
\item Finally we apply ADIS to real fMRI data as a case study (section \ref{realdata})
\end{enumerate}

\section{Probabilistic Projection Pursuit} \label{ppp}
Projection Pursuit is a standard statistical technique for data analysis \cite{Friedman:1974},\cite{Friedman:1987}, \cite{Huber:1985}. 
In this article we generalize projection pursuit in a probabilistic framework similar to the one proposed by Beckmann et. al. (\cite{Beckmann:2004}). We consider data generation at $n$ points via a noisy mixing process as follows:

\begin{equation}\label{1}
x = \mu +  A \, s + \eta, \,\, i = 1,2, \ldots, n
\end{equation}
where $x \in \mathbf{R}^p$, $s \in \mathbf{R}^q$ and $A \in \mathbf{R}^{p \times q}$,  $\eta \sim \mathbf{N}(0, \sigma^2 V)$. We assume that $p > q$ to achieve a compact representation of the observed data. 

The problem is to estimate automatically $q$, $A$, $s$ and $\mu$ given observations $x_i, i = 1 \ldots n$. This problem is also called a blind source separation (BSS) problem since $q$, $A$ and $s$ are all unknown. The inclusion of noise term $\eta$ makes the problem into a probabilistic one.

First consider the case when $V_i = I_p$ for all $i$. Section \ref{autocorr} shows how to handle the case $V_i \neq I_p, \forall i$.

If $1_p = [1,1,\ldots,1]^T \in \mathbf{R}^p$ be a vector of ones. Then

\begin{equation}
1_p^T x = 1_p^T \mu +  1_p^T A \, s + 1_p^T \eta, i = 1,2, \ldots, n
\end{equation}

Let
\begin{equation}
\bar{D} = D - 1_p 1_p^T D, \mbox{ where } D = x, A, \mu, \eta
\end{equation}

Then we can write
\begin{equation}\label{ppmodel}
\bar{x} = \bar{\mu} + \bar{A} s + \bar{\eta}
\end{equation}
where $\bar{\eta} \sim \mathbf{N}\left(0, \sigma^2 (I_p - 1_p 1_p^T/p)\right)$
Since the scaling of $A$ and $s$ is arbitrary we assume without loss of generality that 

\begin{equation}\label{2}
\mbox{E} (s) = 0 \mbox{ and } \mbox{Cov} (s) = \mathbf{I}_q
\end{equation}

No other assumptions are made about the joint source density $p(s)$ other than the ones in \ref{2}. From equations \ref{ppmodel} and \ref{2}:
\begin{equation}\label{4}
\bar{\mu} = E(\bar{x}) 
\end{equation}

\begin{equation}\label{5}
E[(\bar{x} - \bar{\mu})(\bar{x} - \bar{\mu})^T] = \bar{A} \bar{A}^T + \sigma^2 (I_p - 1_p 1_p^T/p) 
\end{equation}

Using the law of large numbers (LLN) we can approximate the covariance matrix as:
\begin{equation} \label{covapprox}
E[(\bar{x} - \bar{\mu})(\bar{x} - \bar{\mu})^T] \approx \frac{1}{n} \sum_{i = 1}^n (\bar{x}_i - \bar{\mu})(\bar{x}_i - \bar{\mu})^T = U \Sigma U^T
\end{equation}
In \ref{covapprox},  $\Sigma = diag(\lambda_k)$ with $\lambda_k, k = 1,\ldots,p$ the singular values and $U$ a matrix containing the corresponding singular vectors of the covariance matrix. The estimate of $\mu$ is easily obtained from \ref{4}.

\begin{equation}
\hat{\bar{\mu}} = \frac{1}{n} \sum_{i = 1}^n \bar{x}_i
\end{equation}

Without knowing the true source densities $p(s)$ it is not possible to estimate the maximum likelihood estimates of $\bar{A}$ and $\sigma^2$. However one can find estimates that try to satisfy the second order condition
\ref{5} as closely as possible by minimizing the Frobenius norm:

\begin{eqnarray}\label{frobnorm}
\hat{ \bar{A} }, \hat{ \sigma^2 } = \argmin_{\bar{A}, \sigma^2}  || \frac{1}{n} \sum_{i = 1}^n (\bar{x}_i - \hat{\bar{\mu}})(\bar{x}_i - \hat{\bar{\mu}})^T -   \bar{A} \bar{A}^T - \sigma^2 (I_p - 1_p 1_p^T/p)  ||_F^2 \nonumber
\end{eqnarray}

It is easily shown that if $U_q$ is a $p \times q$ submatrix of $U$ containing the first $q$ singular vectors corresponding to the $q$ largest singular values and $\Sigma_q$ is the $q \times q$ submatrix of $\Sigma$ then the solution to \ref{frobnorm} is given by:
\begin{equation}
\hat{\bar{A}} = U_q (\Sigma_q - \hat{\sigma^2} I_q)^{1/2} Q^T
\end{equation}

and

\begin{equation}
\hat{\sigma^2} = \frac{1}{p - q - 1} \sum_{i = q + 1}^{p-1} \lambda_i
\end{equation}

where $Q$ is an arbitrary $q \times q$ orthogonal matrix ($Q^T Q = I_q$). Given $\hat{\bar{A}}$ and $\hat{\sigma^2}$, the least squares estimate of $\hat{s_i} \in \mathbf{R}^q$ are given by:

\begin{eqnarray}
\hat{s_i} = (\hat{\bar{A}}^T \hat{\bar{A}})^{-1} \hat{\bar{A}}^T (\bar{x}_i - \hat{\bar{\mu}})  = Q (\Sigma_q - \hat{\sigma^2} I_q)^{-1/2} U_q^T (\bar{x}_i - \hat{\bar{\mu}})  = Q \tilde{x}_i
\end{eqnarray}

where 
\begin{equation}
\tilde{x}_i =  (\Sigma_q - \hat{\sigma^2} I_q)^{-1/2} U_q^T (\bar{x}_i - \hat{\bar{\mu}}) 
\end{equation}

\section{Problems solved in Projection Pursuit}\label{ppproblems}
In Projection Pursuit (PP), we parameterize the orthogonal matrix $Q$ as 
\begin{equation}
Q = [w_1, w_2 \ldots ,w_q]^T
\end{equation}
where $w_i \in R^q$.
The $k$th PP projection is defined as for each point $i$:

\begin{equation}
\hat{s_{ki}} = w_k^T \tilde{x}_i, \, i = 1 \ldots n \mbox{ and } k = 1 \ldots q
\end{equation}

In vector form we can write the $k$th projection as:
\begin{equation}
\hat{s}^{k} = w_k^T \tilde{x}
\end{equation}
where $\hat{s}^{k} = [\hat{s}_{k1},\ldots,\hat{s}_{kn}]$ is a $1 \times n$ vector and $\tilde{x} = [\tilde{x}_1,\tilde{x}_2,\ldots,\tilde{x}_n]$ is a $q \times n$ matrix.
We define an objective function $f(s^{1},s^{2},\ldots,s^{q} )$ and optimize for the projection vectors. Mathematically we solve the optimization problem:
\begin{eqnarray}\label{optimization_problem}
[w^*_1, \ldots, w^*_q] =  \argmax_{w_1,\ldots,w_q} f(w_1^T \tilde{x},\ldots, w_q^T \tilde{x} ) + b(w_1^T \tilde{x},\ldots, w_q^T \tilde{x}; \Theta) \nonumber
\end{eqnarray}
Here $b$ is function accounting for supplementary information that we want to include in the optimization problem. $\Theta$ is all supplementary information of interest to the optimization problem. For example, in spatial problems $\Theta$ could be a spatial location of points and $g$ could
be a function accounting for spatial smoothness (such as a Markov random field). Many such problem specific functions can be proposed based on user objectives. There could also be additional user defined equality and inequality constraints, for example:
\begin{equation}\label{ppequality}
c_i(w_k^T \tilde{x}) = 0, \,\, i = 1 \ldots m, \,\,, k = 1 \ldots  q
\end{equation}

\begin{equation}\label{ppinequality}
g_i(w_k^T \tilde{x}) \ge 0, \,\, i = 1 \ldots L, \,\,, k = 1\ldots q
\end{equation}

Thus in general, we get a constrained projection pursuit problem. Constraints \ref{ppequality} and \ref{ppinequality} can be arbitrary non-linear constraints not necessarily parameterized by $w_k^T \tilde{x}$.

\subsection{Separability}
The optimization problem in \ref{optimization_problem} must involve a joint optimization of the vectors $w_1,\ldots,w_q$ in general. In many important practical cases (such as for example when $f$ is the joint negentropy index) the objective function $f$ has a separable structure \cite{FPICA:2000} such that
\begin{equation}\label{separable}
f(w_1^T \tilde{x}, w_2^T \tilde{x} ,\ldots, w_q^T \tilde{x}) = \sum_{k = 1}^q h(w_k^T \tilde{x})
\end{equation}
for some function $h$, then the optimization can proceed sequentially where at each stage we solve
\begin{equation}\label{separable_single}
w^*_k = \mbox{ arg min }_{w_k} h(w_k^T \tilde{x})
\end{equation}

At each stage $w^*_k$ is a unit vector orthogonal to the previously calculated vectors i.e

\begin{equation}\label{constraints}
w^{*T}_k w^*_l = 
\left\{
\begin{array}{ccc}
0 & \mbox{ when } & l  < k \\
1 & \mbox{ if } & l = k \\
\end{array}
\right.
\end{equation}

\subsection{Multistage Optimization stragegy}
In this section we describe a 2 or 3 stage strategy which we have found via experience to converge to good "local" solutions. It is well known that if an optimization problem that has multiple optima then the "local" solution found by an algorithm is strongly dependent on how the algorithm is initialized. For applications such as BSS, even though an algorithm finds a "local" solution as indicated by convergence diagnostics, it may not be a "global" solution.  In order to increase our chances of finding a solution that is global, we propose a random sampling strategy for initialilzation of an optimization algorithm first.

\subsubsection{Stage 0: Search for good seed points}
For concreteness, suppose $w^*_1,w^*_2,\ldots,w^*_{k-1}$ are the optimal solutions found previously and suppose we are tying to find $w^*_k$. Let $\tilde{W}$ be a $q \times (q - k + 1)$ matrix that is the orthogonal complement of $[w^*_1,w^*_2,\ldots,w^*_{k-1}]$ as determined by say Gram-Schmidt orthogonalization. Then
\begin{equation}
\tilde{W}^T w^*_l = 0, \,\, l = 1,2,\ldots, k - 1
\end{equation}
\begin{enumerate}
\item Generate $n_s$ vectors in $\mathbf{R}^q$ whose elements are drawn from a uniform distribution on $(-1,1)$.
\item Standardize each vector to have unit norm to get the set of vectors $Z = [z_1, z_2, \ldots, z_{n_s}]$ where each $z_i \in \mathbf{R}^q$.
\end{enumerate}

For each $z_i$ we define seed points as
\begin{equation}
u_i = \tilde{W} z_i
\end{equation}
It is easy to see that $u_i^T w^*_l = 0, l = 1,2,\ldots, k -1$ and $u_i^T u_i = 1$. Thus $u_i$ satisfies the constraints in \ref{constraints}. We then compute the objective function $h(w_k^T \tilde{x})$ at each of these points $u_i, i = 1,2,\ldots,n_s$ and choose $R$ points $t_1,\ldots,t_R$ that give the highest objective function values as candidate seed points for the next step. 

\subsubsection{Stage 1: Computing local optimum at R best points from Stage 1}
In this stage, we compute the local solutions $w^{*1}_k, w^{*2}_k, \ldots, w^{*R}_k$ of the optimization problem \ref{separable_single} starting from $t_1,\ldots,t_R$ and choose the best local solution that has the highest function value.
\begin{equation}
w^*_k = \argmax h_k(w_k^{*i}), i = 1,2, \ldots, R
\end{equation}
ADIS uses $n_s = 1000$ and $R = 2$ as the defaults.

\subsubsection{Stage 2: Joint optimization with initialization from Stage 2}
After Stage 1 has been applied from $k = 1, \ldots, q$ we have an estimate of the solution vectors $w_k^{*}, k = 1,2,\ldots, q$. In this stage we solve the joint optimization problem \ref{separable} subject to the single joint constraint:
\begin{equation}\label{single_con}
\sum_{i = 1}^q \sum_{j = i}^q (w_i^T w_j - \delta_{ij})^2 = 0
\end{equation}
where $\delta_{ij} = 1$ if $i = j$ and $0$ otherwise. We initialize the algorithm with the solution $w_k^*, k = 1,\ldots, q$ from Stage 1. The algorithm converges to a local joint solution only in a few iterations.

\section{Optimization Algorithm}\label{optimalg}
For flexible and powerful BSS algorithms, a primary requirement is a fast and robust optimization algorithm that can handle non-linear user defined constraints as well as handle non-convexity in the objective function or the constraints. Furthermore, extensive convergence diagnostics should be a standard output of the optimization process to ensure convergence to a local solution. The optimization core in ADIS (coded in MATLAB, \verb+www.mathworks.com+) uses a modified augmented lagrangian algorithm (inspired by the implementation in LANCELOT package \cite{Conn:1991}, \cite{LANCELOT:1992}) to solve equality constrained problems generated in constrained non-convex BSS problems. Inequality constraints are handled by first transforming them to equality constraints via slack variables and solving the resulting bound constrained optimization problem. Some features of interest are as follows:
\begin{enumerate}
\item A Trust region based approach \cite{More:1983} is used to generate search directions at each step (for both equality constrained and inequality constrained problems).
\item For equality constraints only, the subproblems above are solved using a conjugate gradient approach (Newton-CG -Steihaug) \cite{Steihaug:1983} that is fast and accurate even for large problems and can handle both positive definite and indefinite hessian approximations. If both equality and inequality constraints are present then we solve the trust region problem with a non-linear gradient projection technique \cite{byrd95limited} followed by subspace optimization using Newton-CG-Steihaug. Our algorithm allows for infeasible iterates i.e., those that do not satisfy the problem constraints during optimization. This allows for a fuller exploration of parameter space and increases the likelihood of converging to a global optimum.
\item A symmetric rank 1 (SR1) quasi-Newton approximation to the hessian \cite{Conn:SR1} is used which is known to generate good hessian approximations for both convex and non-convex problems. As suggested in \cite{Nocedal:book} we do the update also on the rejected steps to gather curvature information about the function. We provide options for BFGS \cite{Broyden:1970} especially for convex problems and an option for preconditioning the CG iterations. We also implement limited memory variants of SR1 and BFGS for large problems.
\item Our algorithm accepts vectorized constraints so that multiple constraints can be programmed simultaneously. Only gradient information is required. Hessian information is optional but not required. Optionally, it is easy to interface our code with INTLAB software package \cite{rump95intlab} for automatic differentiation in which case the user only codes the function  and constraints and the gradients/hessians are generated automatically.
\end{enumerate}

We tested the performance of our algorithm using standard optimization benchmarks from the GAMS performance benchmark problems (\url{http://www.gamsworld.org/performance}, \cite{COPS3}). The appendix shows some sample benchmarks as well as provides more technical details of the algorithm.

\subsection{Convergence Diagnostics}
It is \textit{critical} to verify that the optimization algorithm has found a local solution by checking convergence diagnostics. These diagnostics help us determine if the Karush-Kuhn-Tucker (KKT) \cite{Nocedal:book} necessary conditions for optimality have been satisfied or not. Profile plots are plots of a diagnostic measure versus iteration number. We propose the following checks for all BSS algorithms:
\begin{enumerate}
\item Convergence to a point satisfying necessary conditions for optimality can be accessed by looking at profile plots for 
\begin{itemize}
\item Objective function
\item Optimality error (such as "gradient of the lagrangian" for equality constraints or "KKT optimality checks" for general constraints)
\item Feasibility error (checking constraint satisfaction)
\item Lagrange multipliers
\item Other parameters in the algorithm (such as a barrier or penalty parameter)
\end{itemize}

The user should at the very least check these diagnostic plots to make sure convergence is attained. If possible the second order sufficient conditions for optimality should also be verified at the solution point using Hessian information.
\item The algorithm used for optimization should \textbf{flag an error and stop running} in case convergence is not attained at any intermediate stage. This prevents the user from getting access to incorrect results.
\end{enumerate}
These convergence diagnostics are a standard feature of ADIS. Any solution returned by ADIS is guaranteed to be optimal.

\section{Statistics on estimated sources}\label{sourcestats}

In this section we develop equations that enable us to apply ADIS to a real data-set and make inferences from extracted sources.

Once $\hat{s_i}$ are estimated we can compute their variance using the GLM estimate
\begin{equation}
\hat{\mbox{Cov}}(\hat{s}_i) = (\hat{\bar{A}}^T \hat{\bar{A}})^{-1} \hat{\sigma_i^2}
\end{equation}

where $\hat{\sigma_i^2}$ is the estimated variance at point $i$ 
\begin{equation}
\hat{\sigma_i^2} = \frac{ (\bar{x}_i - \hat{ \bar{ \mu } }- \hat{\bar{A}} \hat{s_i})^T (\bar{x}_i - \hat{ \bar{ \mu } } - \hat{\bar{A}} \hat{s_i} ) } { p - q}
\end{equation}

We can create maps of contrasts of interest using the above equations. We also estimate the relative variance (RV) contribution at point $i$ using the component $k$ as follows:

If $\hat{\bar{A}} = [a_1, a_2, \ldots, a_q]$ then

\begin{equation}
RV(k,i) = \frac{ \mbox{Var}(a_k) \hat{s_{ki}}^2 } { \sum_{k = 1}^q \mbox{Var}(a_k) \hat{s_{ki}}^2 }
\end{equation}

Inspection of these voxelwise variance explained maps is very useful in searching through the estimated sources for application based relevance.

\subsection{Correcting for Autocorrelation}\label{autocorr}
Extending to the case of autocorrelated noise is straightforward. When $V_i \neq I$ then we proceed in an iterative fashion as follows:

\begin{enumerate}
\item Set $V_i = I$ and estimate $\hat{ \bar{A} }$, $\hat{s}$ and compute the pointwise residual
\begin{equation}
\bar{r}_i = \bar{x}_i - \hat{ \bar{ \mu } }- \hat{\bar{A}} \hat{s_i}
\end{equation}
\item Compute autocorrelation in $\bar{r}_i$ and prewhiten the data $\bar{x}_i$ using the estimated correlation matrix to get $\bar{x}^{pw}_i$. Run the PP algorithm on $\bar{x}^{pw}_i$ until $\bar{x}^{pw}_i$ does not change from one iteration to the next in an average sense. Various prewhitening schemes such as $AR(p)$ models or non-parametric approaches can be used.  ADIS uses the non-parametric approach proposed in \cite{Woolrich:2001} to do prewhitening. By default ADIS will not do iterative prewhitening unless explicitly specified by the user.
\end{enumerate}

\section{Latent dimensionality estimation}\label{latentdim}

Sophisticated bayesian strategies exist for estimating the latent dimensionality in the case of Gaussian sources \cite{MINKA}. In this section we develop a latent dimensionality estimation procedure that works very well both with Gaussian \textit{and} non-Gaussian sources using a bootstrap/cross-validation procedure.

The latent dimensionality $q$ is estimated in two steps. We first estimate a lower bound on the latent dimensionality (\ref{latentstage1}) followed by a cross validation analysis to refine the lower bound (\ref{latentstage2}). In subsection \ref{latentvalidation} we validate this approach via extensive numerical simulations.

\subsection{Stage 1 - Estimating the lower bound} \label{latentstage1}
Let 
\begin{equation}
X = [\bar{x}_1 - \hat{\bar{\mu}},\bar{x}_2 - \hat{\bar{\mu}},\ldots,\bar{x}_n - \hat{\bar{\mu}}]
\end{equation}
and suppose $\lambda_1 \ge \lambda_2 \ge \ldots \lambda_p$ are the $p$ eigenvalues of $X X^T/n$.

\begin{enumerate}
\item Randomly permute each column of the $p \times n$ matrix $X$ to get the matrix $X^{b}$. 
\item Compute the $p$ eigenvalues $\lambda^b_i$ of $X^b {X^b}^T/n$ such that $\lambda^b_1 \ge \lambda^b_2 \ge \ldots \lambda^b_p$
\item Choose $q_l$ to be the largest value of $i$ such that $\lambda_i > \lambda^b_i$.
\end{enumerate}
First we destroy the systematic correlations between the columns of $X$ via random permutations of each column. Then we estimate the singular values of the permuted covariance matrix and compare these with the singular values of the unpermuted covariance matrix. Only those singular values that exceed the ones from random permutation are deemed significant and the lower bound on latent dimensionality $q_l$ is estimated to be the cardinality of those singular values.

If $P_i \in \mathbf{R}^{p \times p}$ are permutation matrices then we can write:

\begin{equation}
X^{b} = \frac{1}{n} \sum_{i = 1}^n P_i (\bar{x}_i - \hat{\bar{\mu}}) (\bar{x}_i - \hat{\bar{\mu}})^T P_i^T
\end{equation}

Suppose
\begin{equation}
\bar{A} = U_a \Sigma_a V_a^T 
\end{equation}
is the singular value decomposition of A with $\Sigma_a = \mbox{diag}(\sigma_{a_i})$

Let $q$ be the true latent dimensionality. Then for large $n$ it can be shown (and verified by simulation) that: %(see appendix) 
\begin{equation}\label{eig1}
\lambda_i = 
\left\{
\begin{array}{ccc}
 \sigma_{a_i}^2 + \sigma^2  &   \mbox{ if } & i \le q  \\
 \sigma^2  &   \mbox{ if } & q + 1\le i \le (p - 1)  \\
  0 & \mbox{ if }  &   i = p
\end{array}
\right.
\end{equation}

and
\begin{equation}
\lambda^b_i = 
\left\{
\begin{array}{ccc}
 \sigma^2 + \frac{1}{p - 1} \sum_{i = 1}^q \sigma_{a_i}^2  &   \mbox{ if } & i \le (p - 1)  \\
  0 & \mbox{ if }  &   i = p
\end{array}
\right.
\end{equation}

Thus the non-zero eigenvalues of $X^b$ satisfy $\lambda^b_i > \sigma^2$, the noise variance. Hence the largest index $i$ such that $\lambda_i > \lambda^b_i$ is a lower bound for the latent dimensionality $q$.

\subsection{Stage 2 - Cross Validation} \label{latentstage2}
Suppose the true latent dimensionality is assumed to be $q$. Then for large $n$ the eigenvalues $\lambda_i$ will follow equation (\ref{eig1}) i.e, the eigenvalues \{$\lambda_i, i = (q + 1) \ldots (p-1)$\} should be well approximated by a constant $\sigma^2$. We estimate the quality of this model using leave one out cross validation \cite{Hastie:2001}. Let $M_q^{-k}$ be the mean of the eigenvalues $\lambda_i$ from $q + 1$ to $p-1$ excluding the index $k$.
\begin{equation}
M_q^{-k} = \frac{1}{p - 2 - q}\sum_{j = q + 1, j \neq k}^{p - 1} \lambda_i
\end{equation}
The leave one out cross validation error assuming the true latent dimensionality to be $q$ at point $k$ is given by:

\begin{equation}
E(q, k) = \left(\lambda_k -M_q^{-k} \right)^2, k = q + 1, \ldots, p - 1
\end{equation}

The mean cross validation error and its variance can be estimated from these pointwise values as follows:

\begin{equation}
\bar{E}(q) = \frac{1}{p - 1 - q} \sum_{ k = q + 1}^{p- 1} E(q,k)
\end{equation}

\begin{equation}
\mbox{Var}(\bar{E}(q)) = \frac{1}{p - 1 - q} \mbox{Var} \{E(q,k), k = q + 1, \ldots, p - 1\}
\end{equation}

When $q$ is smaller than $q_{true}$ then both $\bar{E}(q)$ and $\mbox{Var}(\bar{E}(q))$ will be large and when $q$ is greater than $q_{true}$ then both $\bar{E}(q)$ and $\mbox{Var}(\bar{E}(q))$ will be small. Since the eigenvalues are expected to remain constant beyond $q_{true}$ the change in $\bar{E}(q)$ will be small beyond $q_{true}$. We define the following index for a given value of $q$ quantifying the change in cross validation error from $q$ to $q + 1$.

\begin{equation}
\Delta(q) = \frac{ \bar{E}(q) - \bar{E}(q + 1)}{ \sqrt{ \mbox{Var}(\bar{E}(q))  + \mbox{Var}(\bar{E}(q + 1)) } }
\end{equation}

If $q_{true}$ is the true latent dimensionality then $\Delta(q)$ will tend to have a maximum at $q_{true} - 1$ since this is the point which will show the largest change in cross validation error in going from $q_{true} - 1$ to $q_{true}$. We thus propose the estimate

\begin{equation}
q_{true} = 1 + \argmax_{q\,} \Delta(q), q = q_l, \ldots ,p - 4
\end{equation}

where $q_l$ is a lower bound on $q$ calculated from stage 1.

The estimation of $\bar{E}(q)$ uses a smaller number of points when $q$ gets very close to $p-1$. Thus the estimate $\Delta(q)$ becomes unstable when $q$ is within a few time points of $p - 1$. In order to robustify our estimate against this instability we define the cumulative maximum index function which calculates the index of maximum of $\Delta(q)$ from $q = q_l, \ldots, r$.

\begin{equation}
f(r) = \argmax_{q \,} \Delta(q), q = q_l,\ldots, r
\end{equation}

Then we count the number of times that a maximum is detected at $y$
\begin{equation}
g(y) = \mbox{Card} \{ r : f(r) = y \}
\end{equation}
and define the estimate of dimensionality to be
\begin{equation}
q_{true} = 1 + \argmax_{y\,} g(y)
\end{equation}

\subsection{Validation of the approach}\label{latentvalidation}

To test this latent dimensionality algorithm we generate data as per equation (\ref{1}). Details of the simulation are as follows:
\begin{enumerate}
\item The true mixing matrix $A$ was chosen to be a $p \times q$ matrix where the elements were drawn from a uniform distribution in (0,1). This random matrix was then scaled so that its minimum singular value $\sigma_{min}(A) = 1$.

\item The sources $s$ in (\ref{1}) were generated from three different types of distributions based on their kurtosis "excess" values 
$\gamma$ . The chosen distributions were Gaussian ($\gamma = 0$), Uniform ($\gamma < 0$) and Gamma ($\gamma > 0$).

\item The ratio of noise standard deviation in (\ref{1}), $\sigma$ to the minimum singular value of $A$, $\frac{\sigma_{min}(A)}{\sigma}$ was varied between $0.75,1,\ldots,2$.

\item The ratio of the true latent dimensionality to the dimensionality of the observed data $\frac{q_{true}}{p}$ was varied between 0.1,0.2,\ldots,0.5.

\item This process was repeated 20 times for each combination of source type, $\frac{\sigma_{min}(A)}{\sigma}$ ratio and $\frac{q_{true}}{p}$ ratio.

\item For each individual simulation we estimated the latent dimensionality $\hat{q}$ based on the 2 stage estimation strategy described above.
\end{enumerate}

\begin{figure*}[htbp]
\centering
%\raggedleft
\begin{tabular}{cc}

\subfigure[Gaussian sources (kurtosis excess $= 0$)]
{
\hspace{-1in}
\label{latent1}
\includegraphics[width = 90mm, angle = -90]{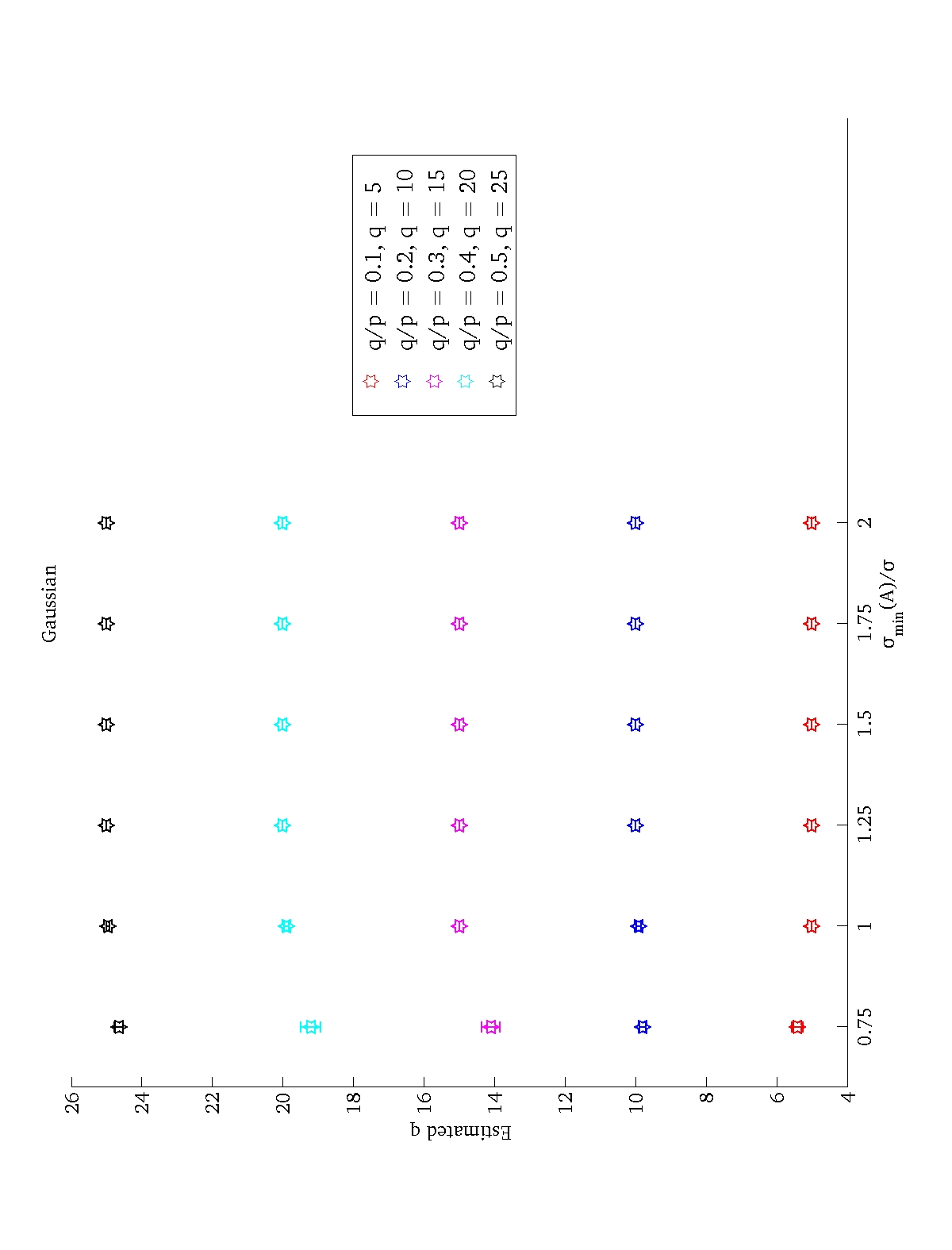}
} 

&
\subfigure[Uniform sources (kurtosis excess $< 0$)]
{
\hspace{-0.8in}
\label{latent3}
\includegraphics[width = 90mm, angle = -90]{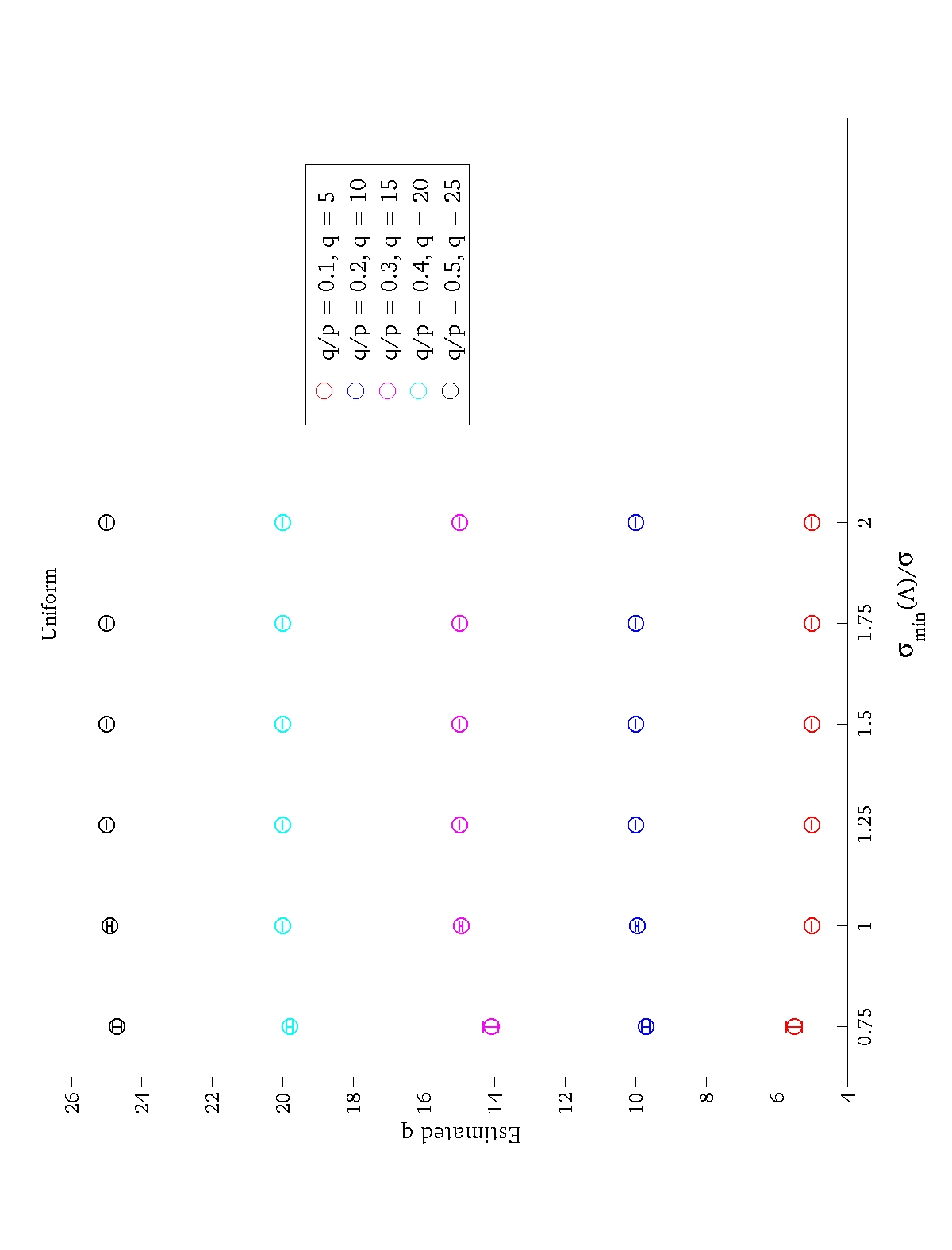}

}

\\
\subfigure[Gamma sources (kurtosis excess $> 0$) ]
{
\hspace{-1in}
\label{latent4}
\includegraphics[width = 90mm, angle = -90]{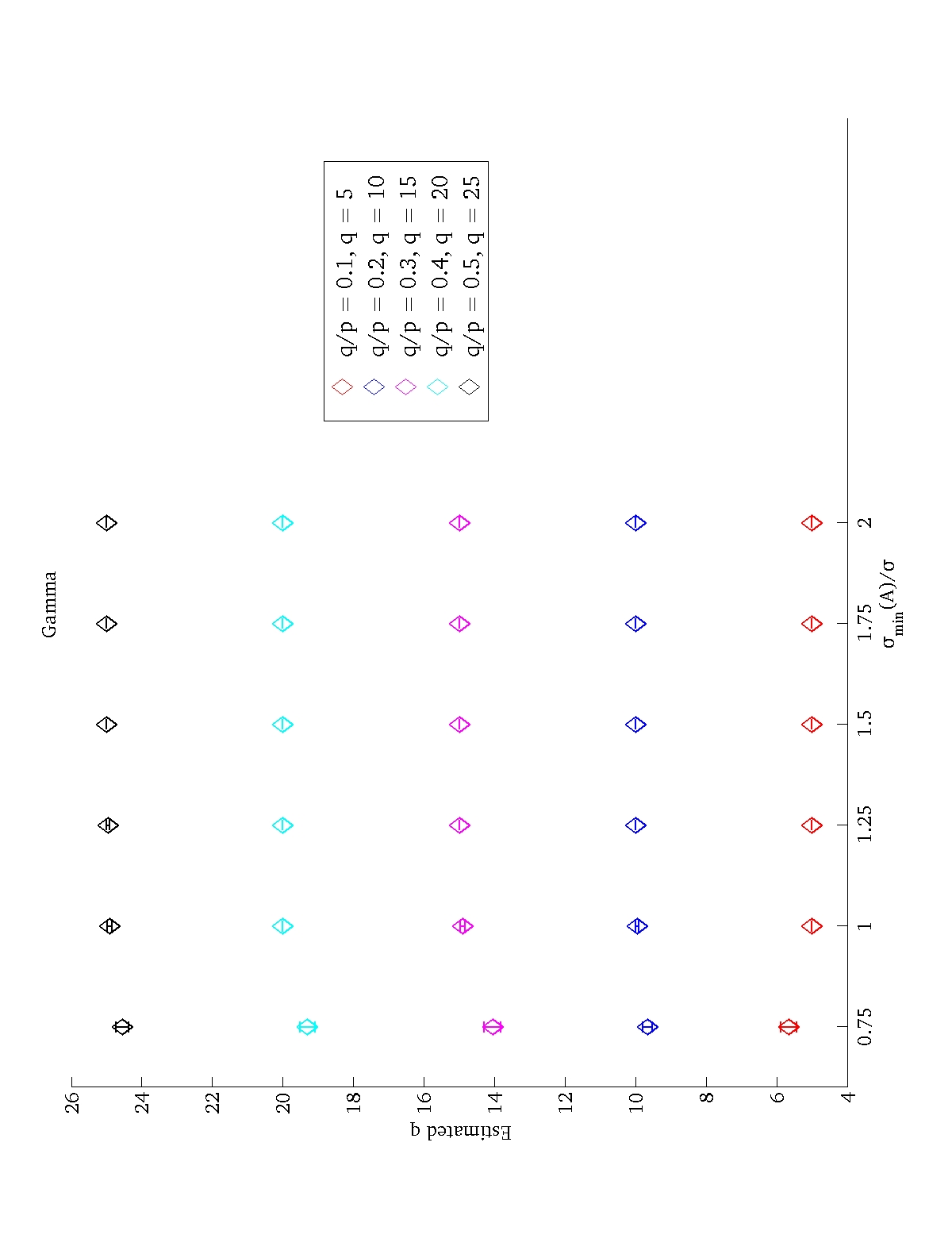}
}

&
\subfigure[Illustration of latent dimensionality estimation]
{
\hspace{-0.8in}
\label{latent2}
\includegraphics[width = 90mm, angle = -90]{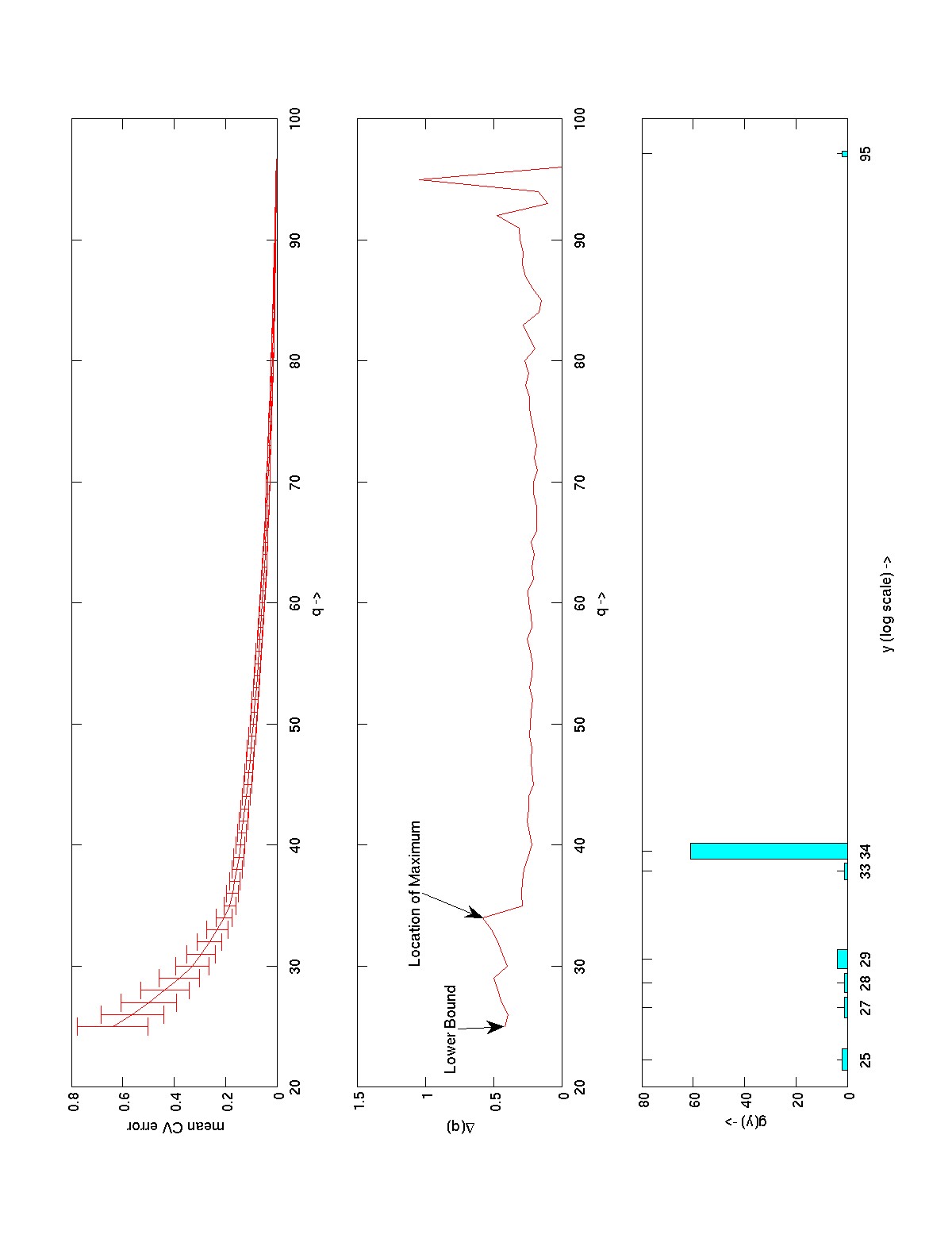}
}
\end{tabular}
\caption{ (a), (b) and (c) depict simulations showing the performance of latent dimensionality estimation procedure on various source types at different $\frac{\sigma_{min}(A)}{\sigma}$ ratios parameterized by various $q/p$ ratios. (d) Latent dimensionality estimation for a Gaussian sources with $q_{true} = 35$, $p = 100$, $n = 1000$ and $\frac{\sigma_{min}(A)}{\sigma} = 0.75$. $\Delta(q)$ attains a maximum for $q = 34$ and so $\hat{q} = 1 + 34 = 35$}
\label{figlatent}
\end{figure*}

We chose $p = 50$ and $n = 1000$ as fixed parameters of the simulation.
Simulations show that this 2 stage estimation is almost unbiased for both Gaussian and non-Gaussian embedded sources for all values of the ratios $\frac{\sigma_{min}(A)}{\sigma}$ and $\frac{q_{true}}{p}$.
Results are shown in figure \ref{latent1} - \ref{latent4}.

%--------
%----

\section{Benchmarking: Comparison with other BSS algorithms}\label{bssbench}

Choosing the negative entropy index as our objective function and without imposing any additional constraints, our algorithm attempts to estimate sources that are independent. To test and compare our algorithm with others, we used an approximation to negative entropy as proposed in \cite{FPICA:2000} (see appendix for details).

ICALAB \cite{ICALAB_BOOK:2003}, \cite{ICALAB} (available from \url{http://www.bsp.brain.riken.go.jp/ICALAB/}) is a Matlab package for comparing algorithms for BSS. We used ICALAB to compare the performance of our algorithm with other standard BSS algorithms such as FJADE \cite{JADE:1996}, FPICA \cite{FPICA:2000}, \cite{FPICA:1999}, EFICA \cite{EFICA:2006}, \cite{EFICA2:2006} , ERICA \cite{ERICA:2002} and UNICA \cite{UNICA:2000} which use higher order statistics to separate sources. 

The quality of source extraction is measured using the Source to Interferences Ratio (SIR) \cite{SIR:2006} (of the estimated mixing matrix) which measures the ratio of the energy of the estimated source projected onto the true source to the energy of the estimated source projected onto the other sources. Higher values of (SIR) indicate better performance. Please see the appendix for details.
ICALAB also comes with standard benchmarking datasets (\url{http://www.bsp.brain.riken.jp/ICALAB/ICALABSignalProc/}). 

A Monte Carlo analysis was performed using ICALAB by generating uniformly distributed random matrices $A_i$ and creating a mixed source data-set $X_i$ for a given set of sources $S$. 
\begin{equation}
X_i = A_i \, S, \,\, i = 1,2,\ldots, n_b
\end{equation}

To get a baseline measure of performance for each algorithm, we use square mixing without additional noise for the simulation. Each algorithm was then run on this mixed data set. This process was repeated $n_b = 100$ times for each dataset using a \textbf{\emph{new}} mixing matrix every time. In ICALAB, most of the algorithms are given default parameters that are tuned optimum values for typical data. As suggested in ICALAB, we use these default algorithmic parameters for benchmarking purposes.

ADIS is able to perform non-square BSS in the presence of noise. However, since the other algorithms in our benchmarking test have not been designed to do this, we think its unfair to compare non-square ability of ADIS with other algorithms.

The 13 benchmarking datasets and their short descriptions are given in the appendix. In order to evaluate the effect of different types of mixing matrices, we ran additional Monte Carlo simulations when $A$ was chosen to be one of the following (a) Random sparse (b) Random bipolar (c) Symmetric random (d) Ill conditioned random (e) Hilbert (f) Toeplitz (g) Hankel (h) Orthogonal (i) Nonnegative symmetric (j) Bipolar symmetric (k) Skew symmetric.

\begin{figure*}[htbp]
\centering
\begin{tabular}{cc}

\subfigure[nband5]
{
\hspace{-1in}
\label{bench1a}
\includegraphics[width = 85mm, angle = -90]{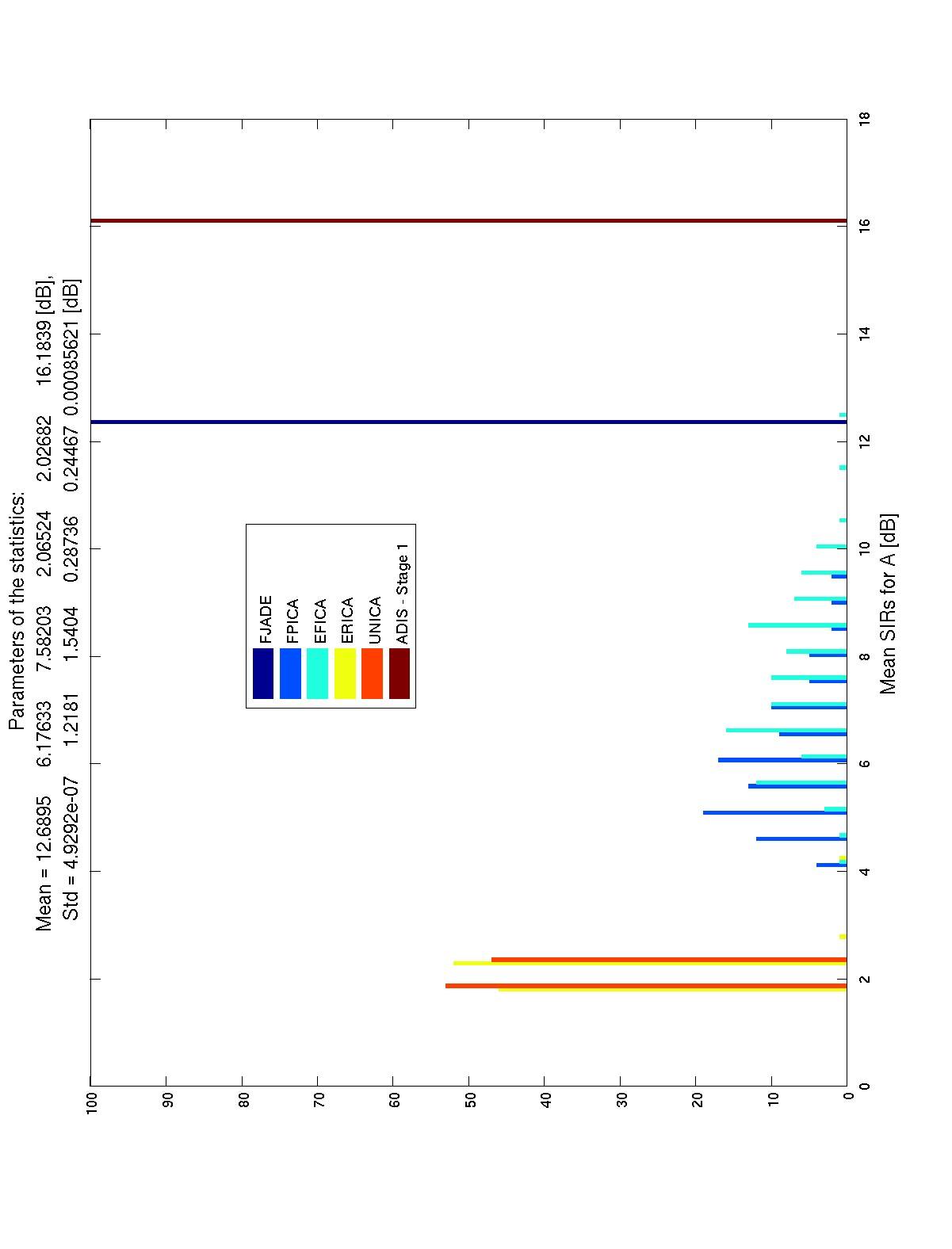}
}

& 

\subfigure[10halo]
{
\hspace{-0.5in}
\label{bench1b}
\includegraphics[width = 85mm, angle = -90]{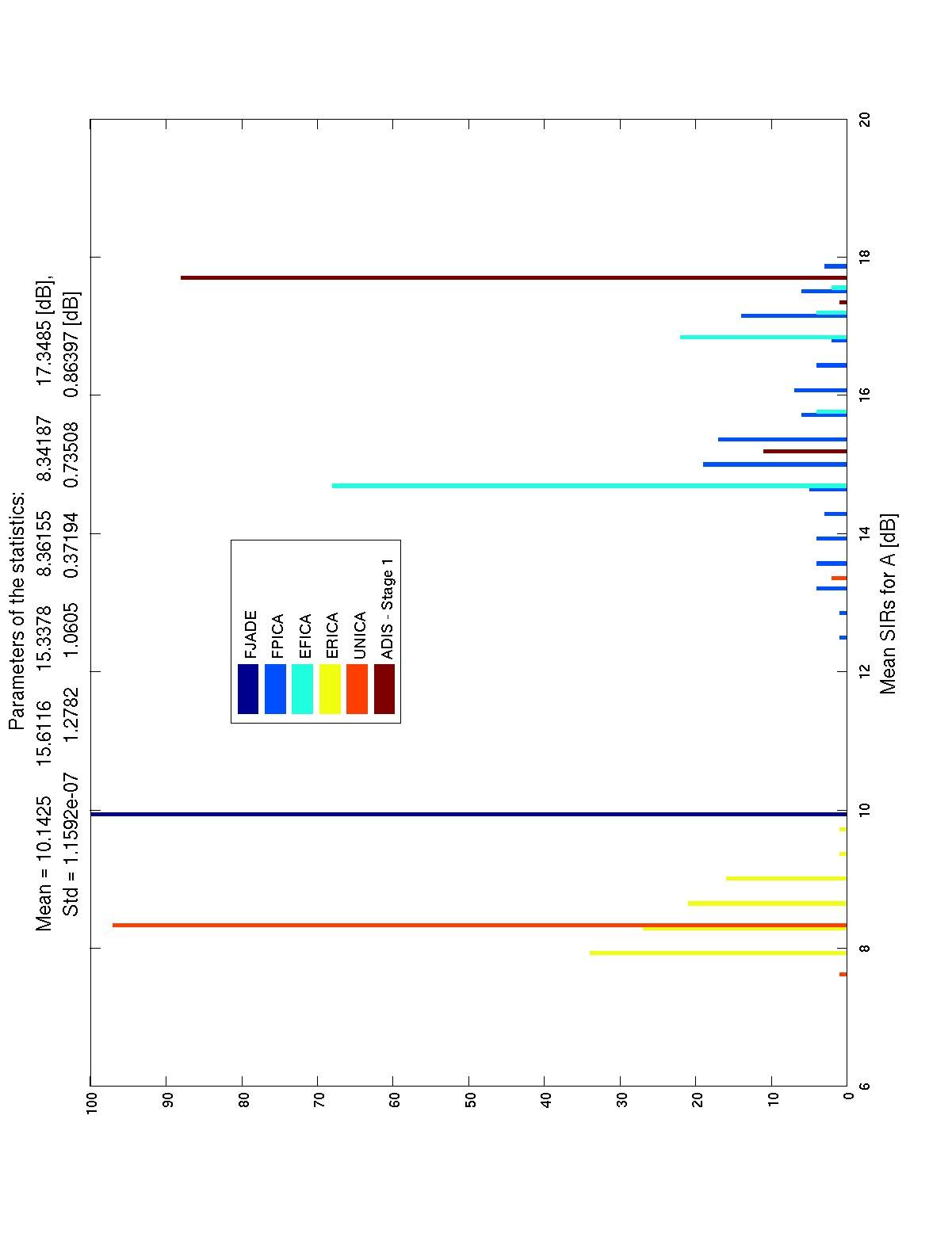}
}

\\

\subfigure[GnBand]
{
\hspace{-1in}
\label{bench1c}
\includegraphics[width = 85mm, angle = -90]{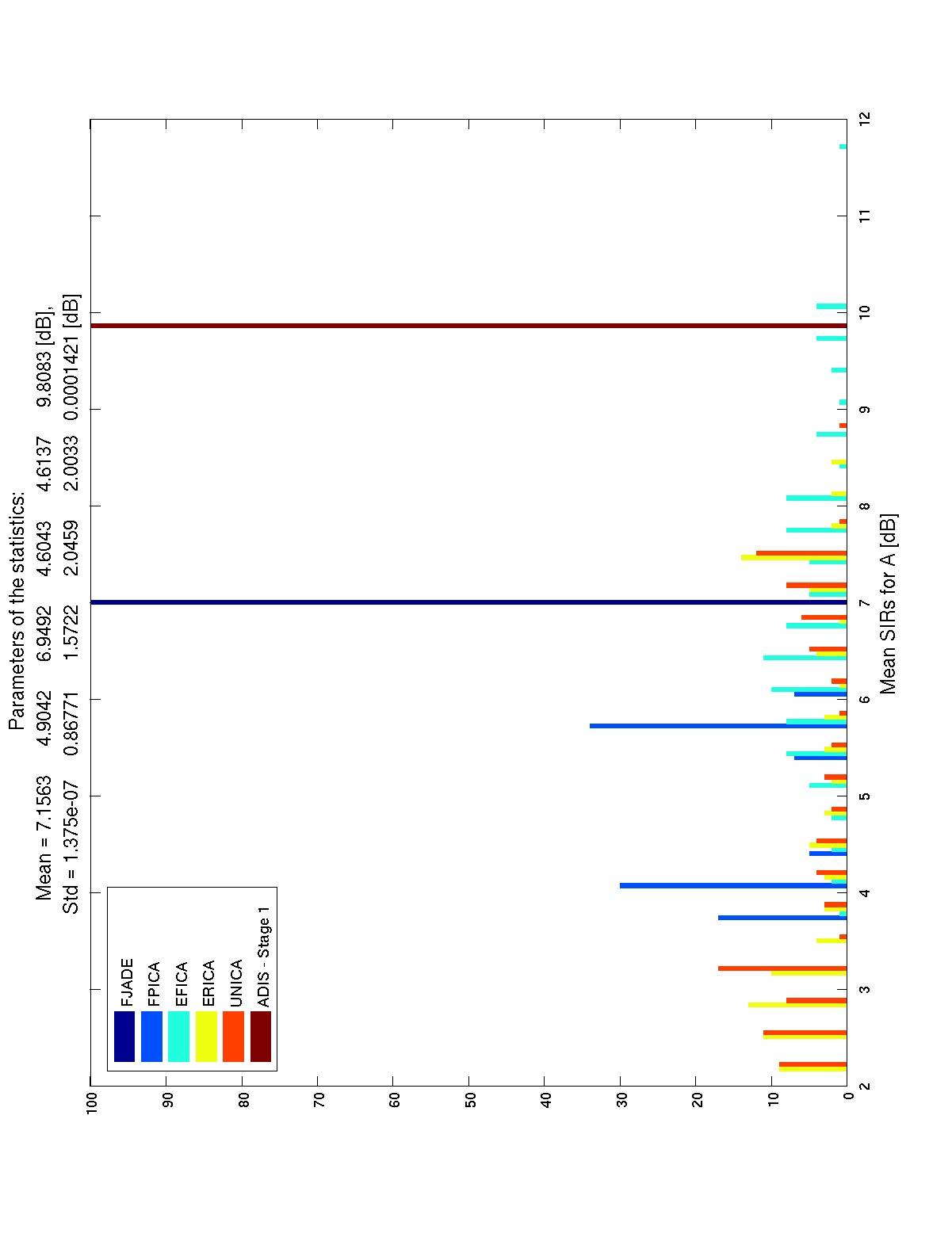}
}

& 

\subfigure[acspeech16]
{
\hspace{-0.5in}
\label{bench1d}
\includegraphics[width = 85mm, angle = -90]{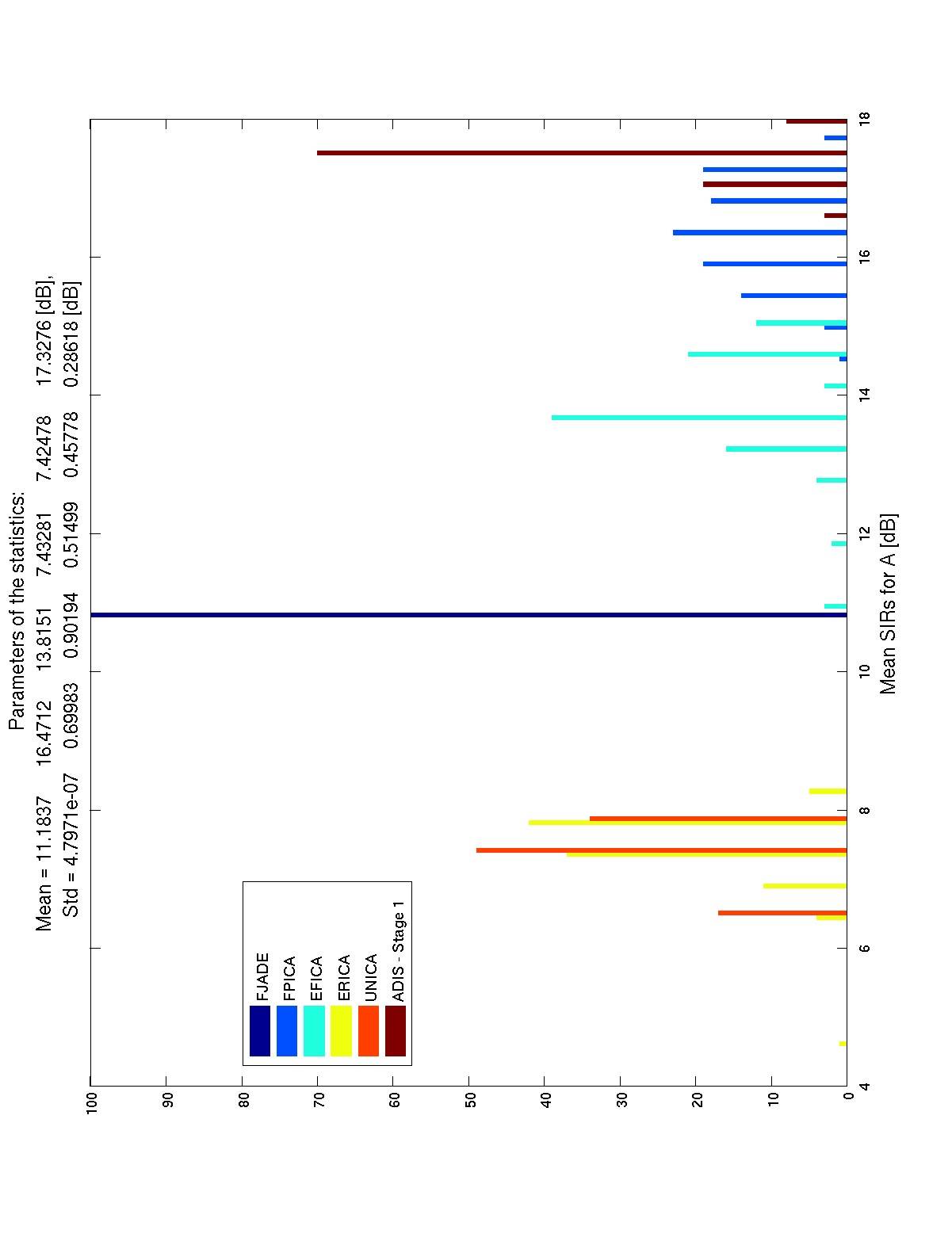}
}

\end{tabular}
\label{bench1}
\caption{Histograms of mean SIR for each algorithm over 100 Monte Carlo simulations using randomly generated mixing matrices for various benchmarking datasets.}
\end{figure*}
%---

\begin{figure*}[htbp]
\centering
\begin{tabular}{cc}

\subfigure[ABio5]
{
\hspace{-1in}
\label{bench2a}
\includegraphics[width = 85mm, angle = -90]{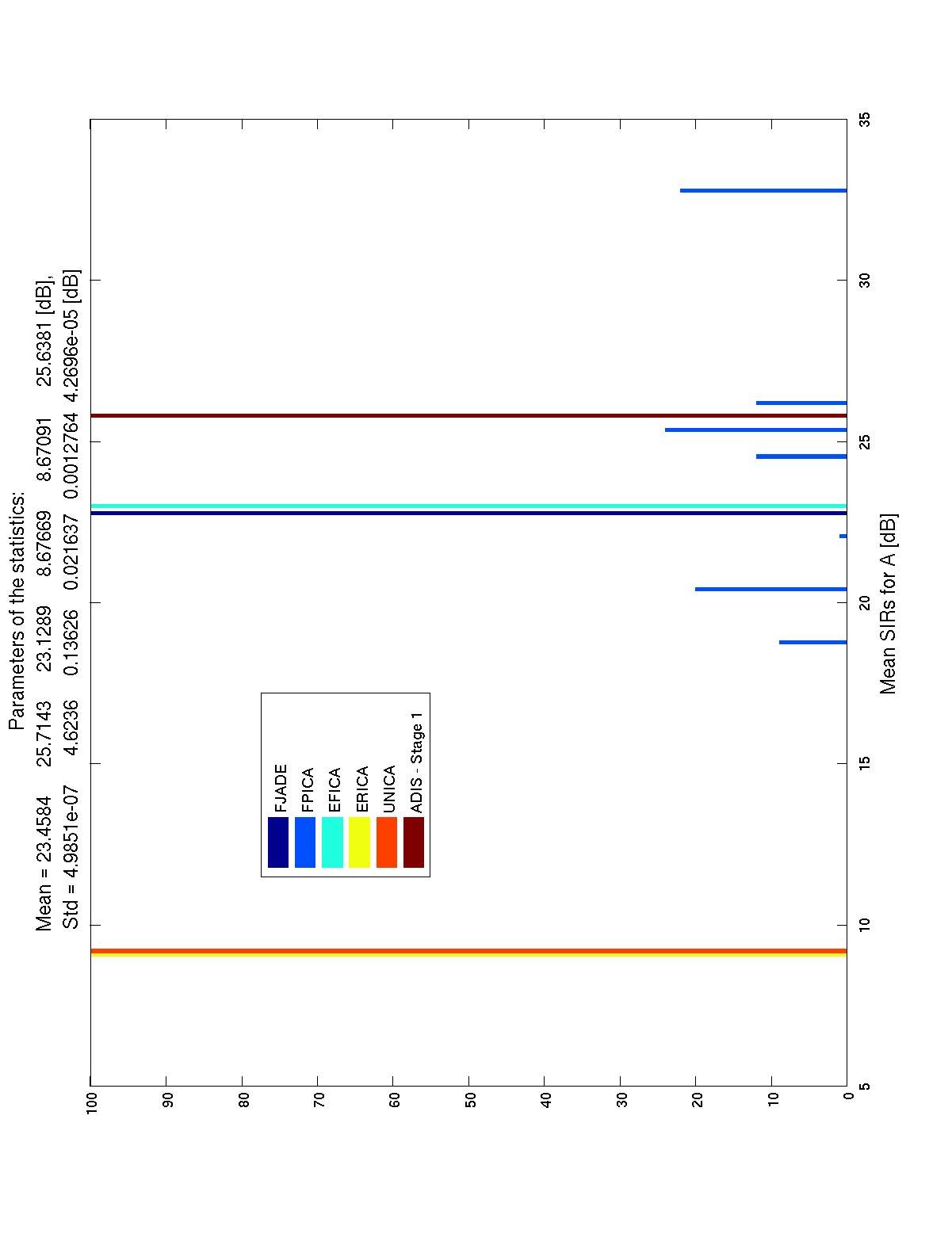}
}

& 

\subfigure[ACsparse10]
{
\hspace{-0.5in}
\label{bench2b}
\includegraphics[width = 85mm, angle = -90]{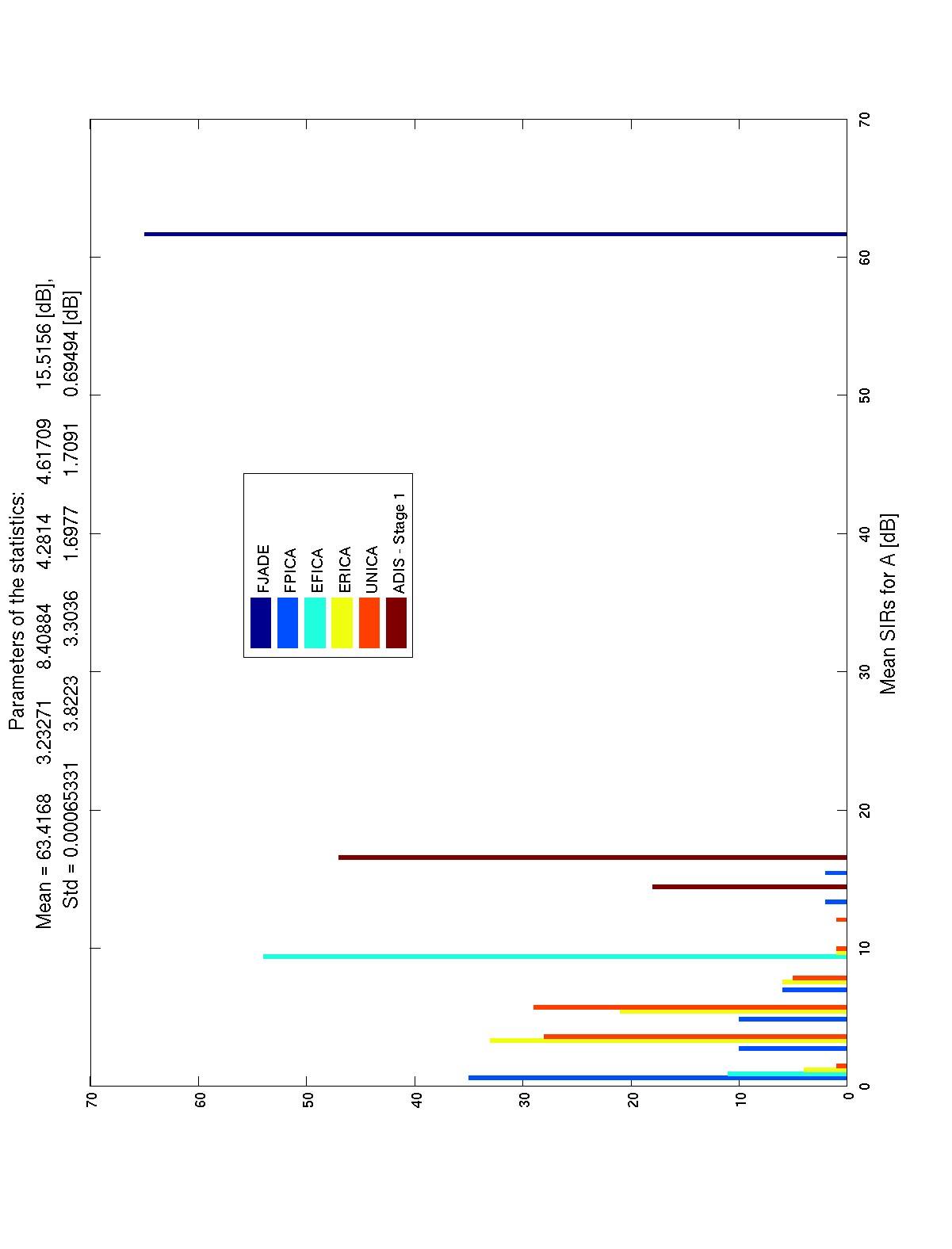}
}

\\

\subfigure[25SpeakersHALO]
{
\hspace{-1in}
\label{bench2c}
\includegraphics[width = 85mm, angle = -90]{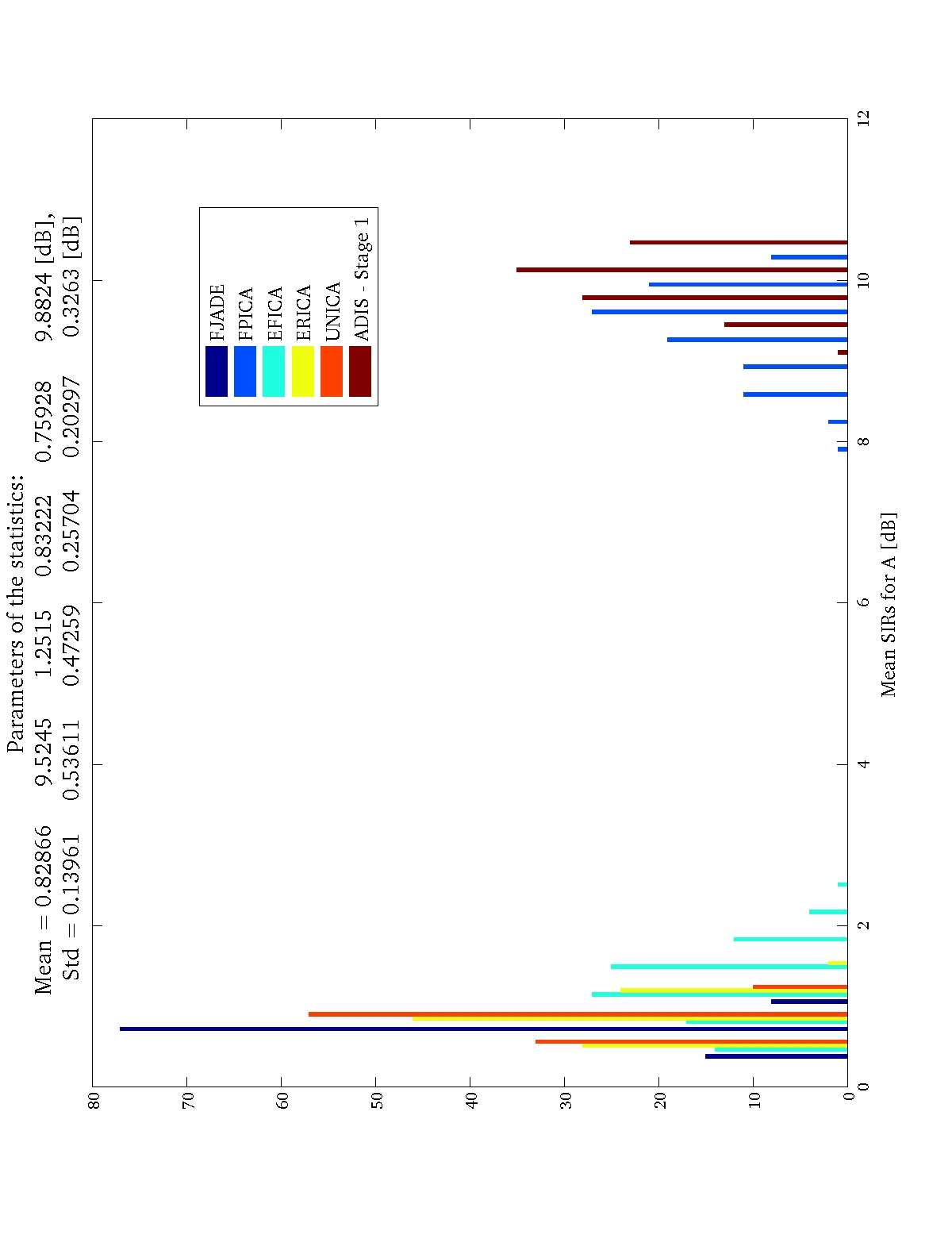}
}

& 

\subfigure[Vsparserand10]
{
\hspace{-0.5in}
\label{bench2d}
\includegraphics[width = 85mm, angle = -90]{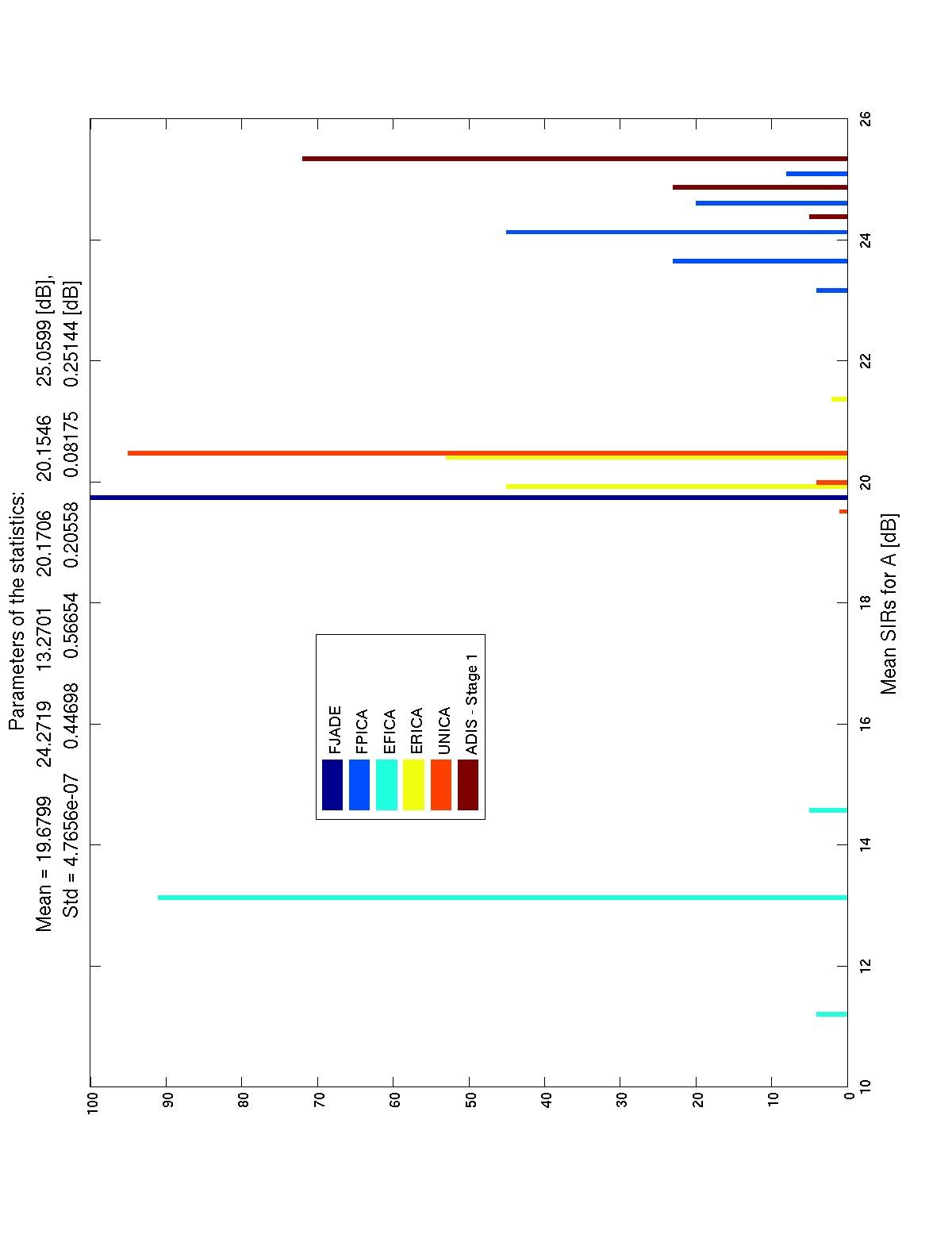}
}

\end{tabular}
\label{bench2}
\caption{Histograms of mean SIR for each algorithm over 100 Monte Carlo simulations using randomly generated mixing matrices for various benchmarking datasets.}
\end{figure*}
%-----

\begin{figure*}[htbp]
\centering
\begin{tabular}{cc}

\subfigure[ACsincpos10]
{
\hspace{-1in}
\label{bench3a}
\includegraphics[width = 85mm, angle = -90]{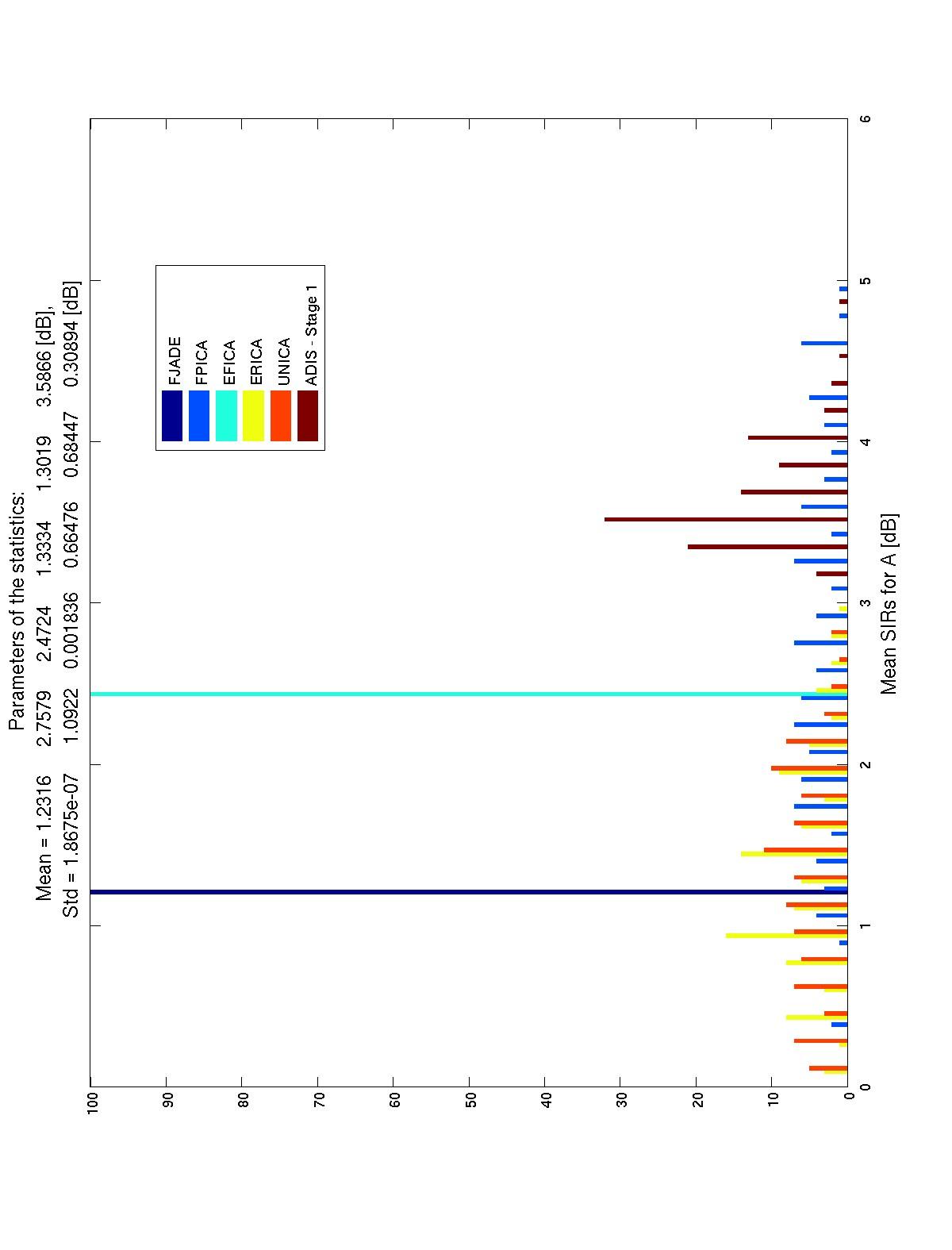}
}

& 

\subfigure[X5smooth]
{
\hspace{-0.5in}
\label{bench3b}
\includegraphics[width = 85mm, angle = -90]{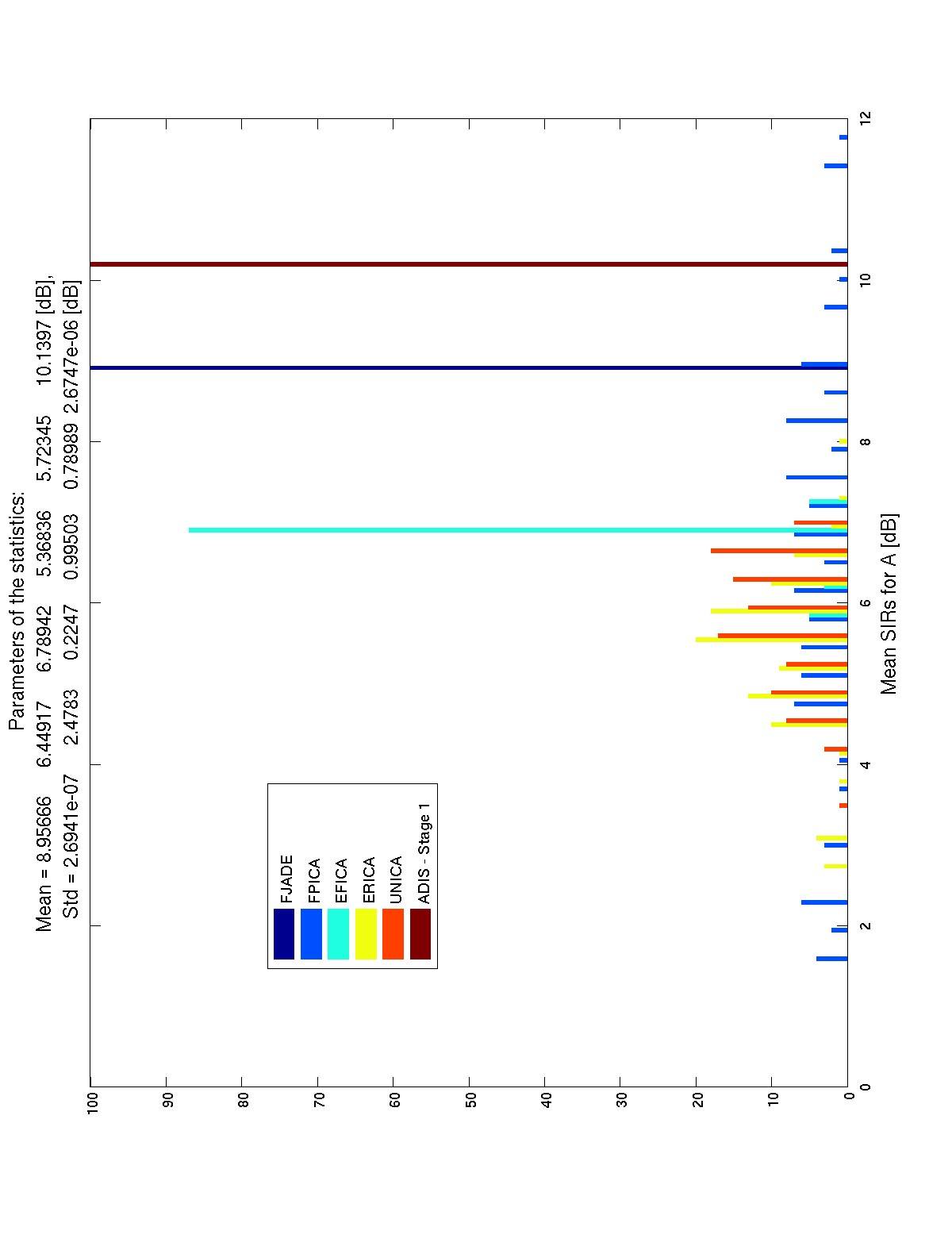}
}

\\

\subfigure[speech20]
{
\hspace{-1in}
\label{bench3c}
\includegraphics[width = 85mm, angle = -90]{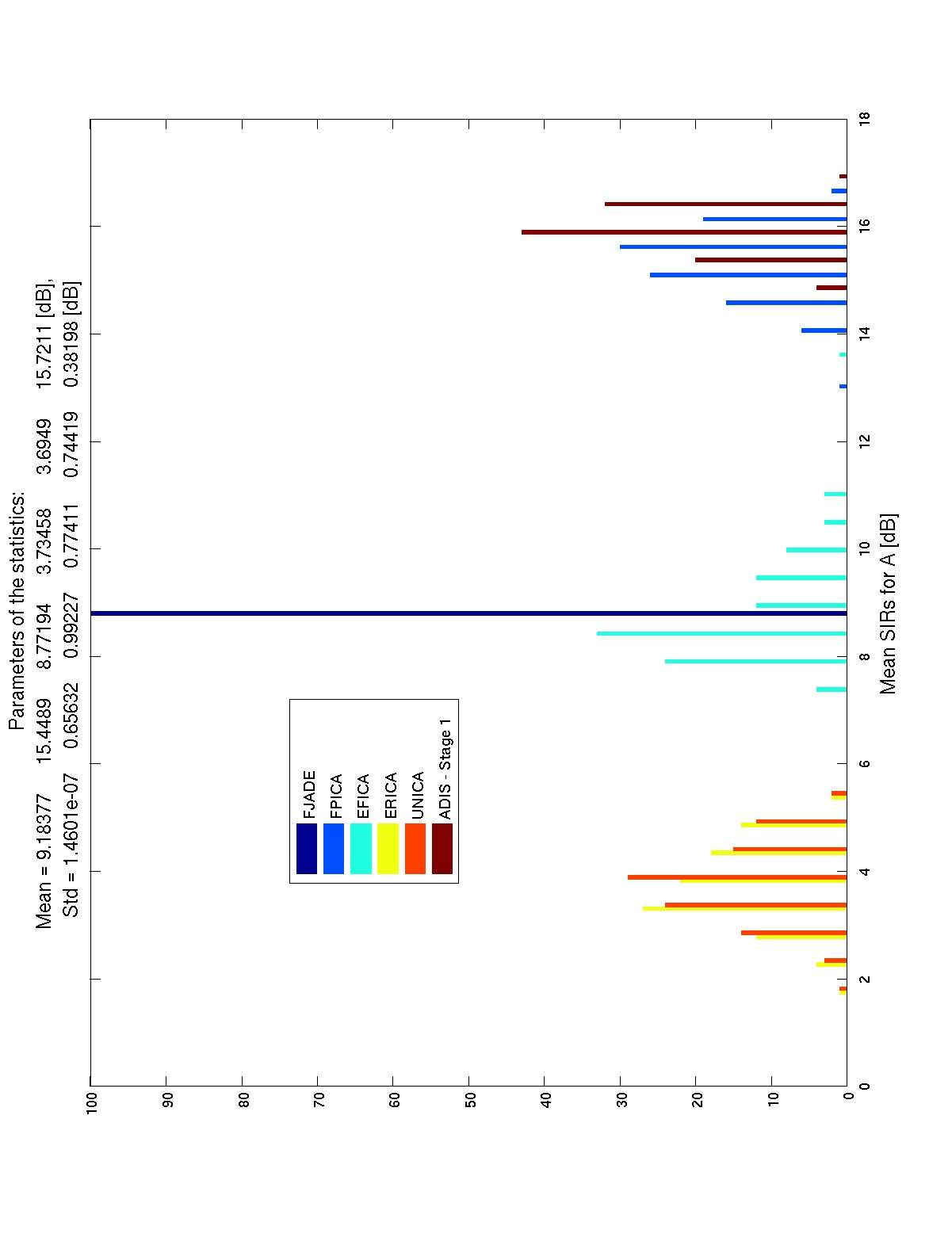}
}

& 

\subfigure[X10randsparse]
{
\hspace{-0.5in}
\label{bench3d}
\includegraphics[width = 85mm, angle = -90]{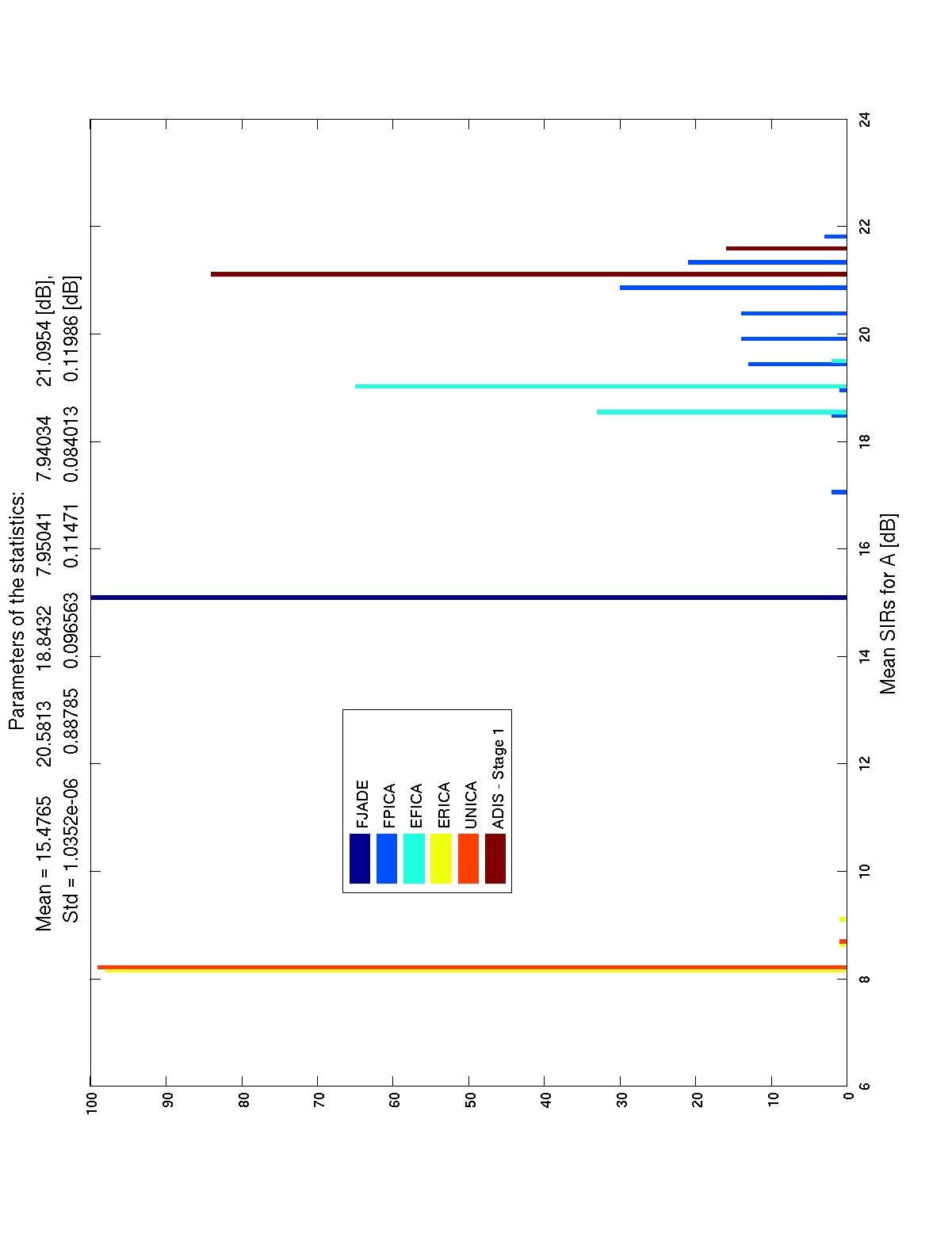}
}

\end{tabular}
\label{bench3}
\caption{Histograms of mean SIR for each algorithm over 100 Monte Carlo simulations using randomly generated mixing matrices for various benchmarking datasets.}
\end{figure*}

\begin{table*}[htdp]
\begin{center}
\begin{tabular}{|c|c|c|c|c|c|c|c|c|}
\hline
               & \multicolumn{7}{|c|}{ Algorithm, $M = \bar{SIR}$ [dB] and $S = std(\bar{SIR})$ [dB]} \\
\hline
\multicolumn{2}{|c|} {Dataset} & FJADE & FPICA & EFICA & ERICA & UNICA & ADIS - Stage 1 \\
\hline
\multirow{2}{*}{nband5} & M & \textcolor{blue}{$12.6895$} & $6.17633$ & $7.58203$ & $2.06524$ & $2.02682$ & \textcolor{red}{ $16.1839$ } \\
\cline{2-8}
& S & $4.9292e{-07}$ & $1.2181$ & $1.5404$ & $0.28736$ & $0.24467$ & $8.5621e{-04}$ \\
\hline
\multirow{2}{*}{10halo} & M & $10.1425$ & \textcolor{blue}{$15.6116$} & \textcolor{blue}{$15.3378$} & $8.36155$ & $8.34187$ & \textcolor{red}{ $17.3485$ }\\
\cline{2-8}
& S & $1.1592e{-07}$ & $1.2782$ & $1.0605$ & $0.37194$ & $0.73508$ & $0.86397$ \\
\hline
\multirow{2}{*}{GnBand} & M & \textcolor{blue}{$7.1563$} & $4.9042$ & $6.9492$ & $4.6043$ & $4.6137$ & \textcolor{red}{ $9.8083$ } \\
\cline{2-8}
& S & $1.375e{-07}$ & $0.86771$ & $1.5722$ & $2.0459$ & $2.0033$ & $1.421e{-04}$ \\
\hline
\multirow{2}{*}{acspeech16} & M & $11.1837$ & \textcolor{blue}{$16.4712$} & $13.8151$ & $7.43281$ & $7.42478$ & \textcolor{red}{ $17.3276$ }\\
\cline{2-8}
& S & $4.7971e{-07}$ & $0.69983$ & $0.90194$ & $0.51499$ & $0.45778$ & $0.28618$ \\
\hline
\multirow{2}{*}{ABio5} & M & $23.4584$ & \textcolor{blue}{$25.7143$} & $23.1289$ & $8.67669$ & $8.67091$ & \textcolor{red}{ $25.6381$ }\\
\cline{2-8}
& S & $4.9851e{-07}$ & $4.6236$ & $0.13626$ & $0.021637$ & $1.2764e{-03}$ & $4.2696e{-05}$ \\
\hline
\multirow{2}{*}{ACsparse10} & M & \textcolor{red}{ $63.4168$ } & $3.23271$ & $8.40884$ & $4.2814$ & $4.61709$ & \textcolor{blue}{$15.5156$} \\
\cline{2-8}
& S & $6.5331e{-04}$ & $3.8223$ & $3.3036$ & $1.6977$ & $1.7091$ & $0.69494$ \\
\hline

\multirow{2}{*}{25SpeakersHALO} & M & $0.82866$ & \textcolor{blue}{$9.5245$} & $1.2515$ & $0.83222$ & $0.75928$ & \textcolor{red}{$9.8824$} \\
\cline{2-8}
& S & $ 0.13961$ & $0.53611$ & $0.47259$ & $0.25704$ & $0.20297$ & $0.3263$ \\
\hline

\multirow{2}{*}{VSparserand10} & M & $19.6799$ & \textcolor{blue}{$24.2719$} & $13.2701$ & $20.1706$ & $20.1546$ & \textcolor{red}{$25.0599$} \\
\cline{2-8}
& S & $4.7656e{-07}$ & $0.44698$ & $0.56654$ & $0.20558$ & $0.08175$ & $0.25144$ \\
\hline

\multirow{2}{*}{sincpos10} & M & $1.2316$ & \textcolor{blue}{$2.7579$} & \textcolor{blue}{$2.4724$} & $1.3334$ & $1.3019$ & \textcolor{red}{$3.5866$} \\
\cline{2-8}
& S & $1.8675e{-07}$ & $1.0922$ & $0.01836$ & $0.66476$ & $0.68447$ & $0.30894$ \\
\hline

\multirow{2}{*}{X5smooth} & M & \textcolor{blue}{$8.95666$} & $6.44917$ & $6.78942$ & $5.36836$ & $5.72345$ & \textcolor{red}{$10.1397$} \\
\cline{2-8}
& S & $2.6941e{-07}$ & $2.4783$ & $0.2247$ & $0.99503$ & $0.78989$ & $2.6747e{-06}$ \\
\hline

\multirow{2}{*}{Speech20} & M & $9.18377$ & \textcolor{blue}{$15.4489$} & $8.77194$ & $3.73458$ & $3.6949$ & \textcolor{red}{$15.7211$} \\
\cline{2-8}
& S & $1.4601e{-07}$ & $0.65632$ & $0.99227$ & $0.77411$ & $0.74419$ & $0.38198$ \\
\hline

\multirow{2}{*}{X10randsparse} & M & $15.4765$ & \textcolor{blue}{$20.5813$} & $18.8432$ & $7.95041$ & $7.94034$ & \textcolor{red}{$21.0954$} \\
\cline{2-8}
& S & $1.0352e{-06}$ & $0.88785$ & $0.096563$ & $0.11471$ & $0.084013$ & $0.11986$ \\
\hline

\multirow{2}{*}{64soundstd} & M & x & \textcolor{blue}{$5.9408$} & $3.4533$ & $0.41501$ & $0.43314$ & \textcolor{red}{$6.3234$} \\
\cline{2-8}
& S & x & $0.20937$ & $0.13021$ & $0.10444$ & $0.11409$ & $0.1699$ \\
\hline

\end{tabular}
\end{center}
\caption{Mean SIR ($M$) and its standard deviation ($S$) for various benchmark datasets over 100 Monte Carlo simulations for different algorithms. The benchmark datasets are a part of ICALAB \cite{ICALAB_BOOK:2003}, \cite{ICALAB}. An 'x' means that the algorithm failed to converge. The algorithms were ranked based on not only their mean SIR ($M$) and standard deviation ($S$) but also the entire SIR histogram. The color \textcolor{red}{red} is used to mark the best performing algorithm and the color \textcolor{blue}{blue} is used to mark the 2nd best. In cases where more than one algorithm is marked with the same color, both algorithms were judged to perform equally well.}
\label{sirtable}
\end{table*}%

\begin{figure*}[htbp]
\centering
\begin{tabular}{c}

\subfigure[64sounds mean SIR]
{
\label{fig64sounds}
\includegraphics[width = 100mm, angle = -90]{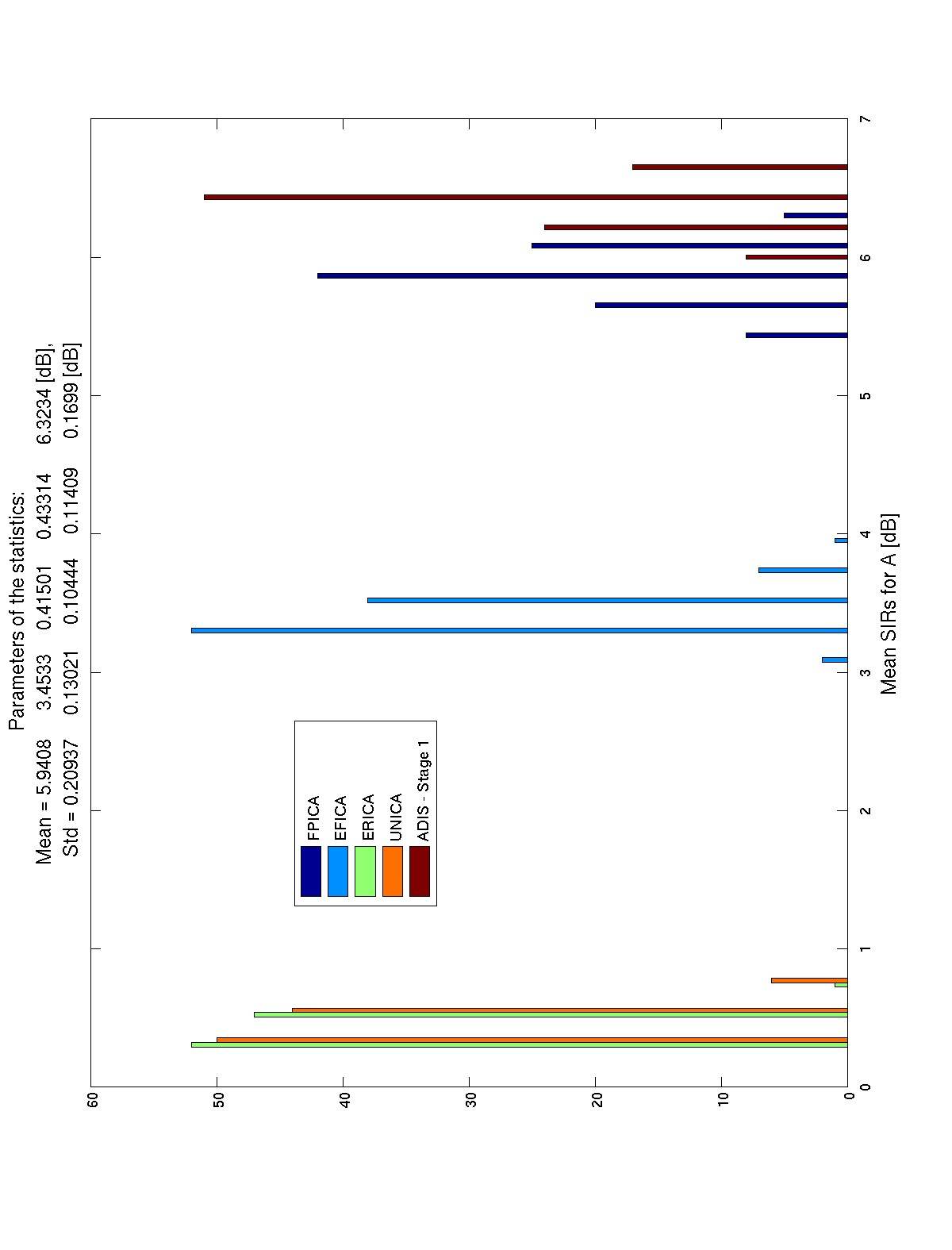}
}

\\

\subfigure[ACsparse10 ADIS Stage 1+2]
{
\label{ADIS2stage}
\includegraphics[width = 100mm, angle = -90]{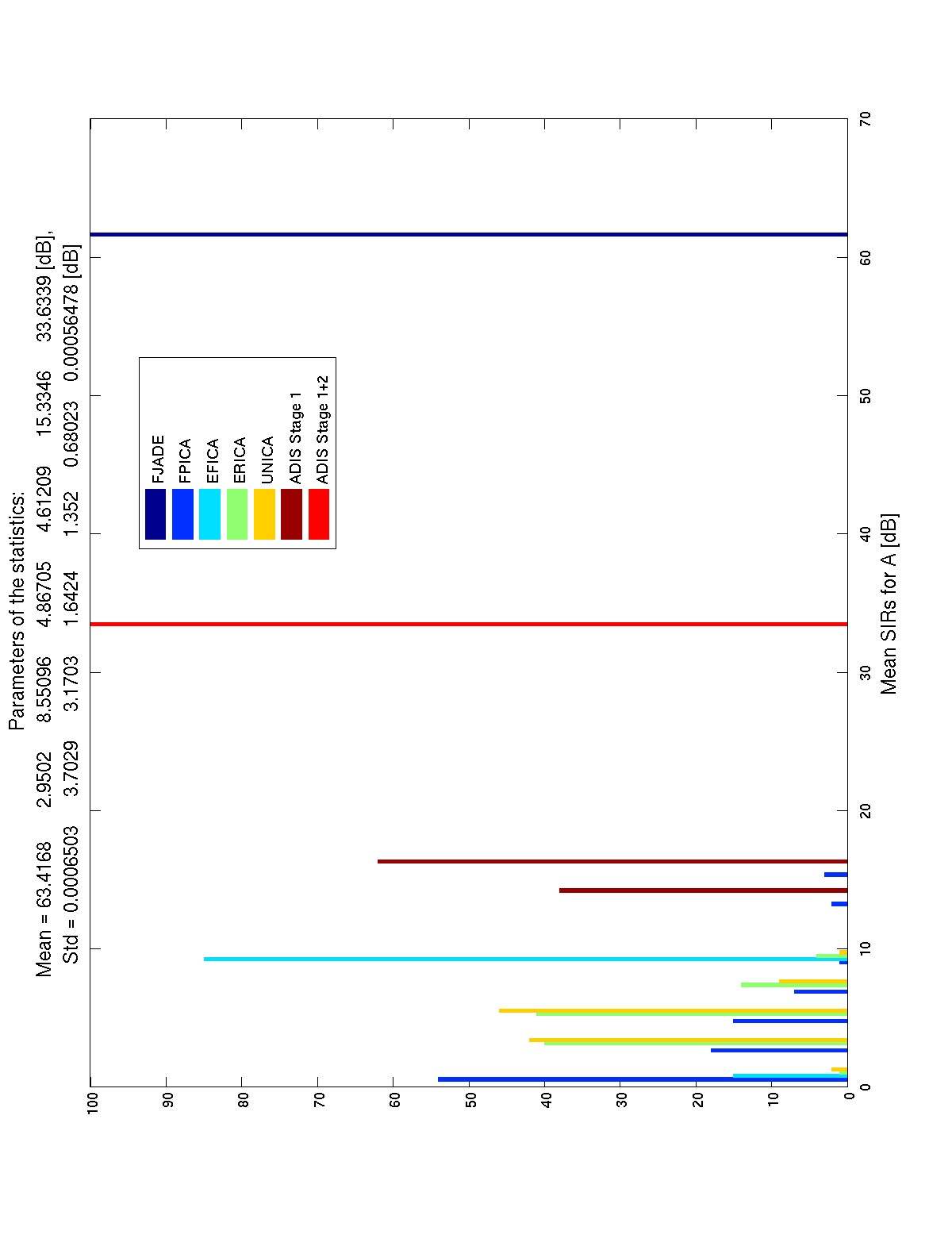}
}

\end{tabular}
\label{fig64soundsandADIS2stage}
\caption{Histograms of mean SIR for each algorithm over 100 Monte Carlo simulations using randomly generated mixing matrices on (a) the 64sounds benchmark dataset. This is a relatively large dataset with 64 sources. FJADE fails on this test case. (b) the ACsparse10 benchmarking dataset. This figure illustrates the improvement in SIR using the joint optimization of ADIS Stage 1+2 over ADIS Stage 1}
\end{figure*}

\begin{figure*}[htbp]
\centering
\begin{tabular}{c}

\subfigure[convergence diagnostics]
{
\label{samplediagnostic}
\includegraphics[width = 95mm, angle = -90]{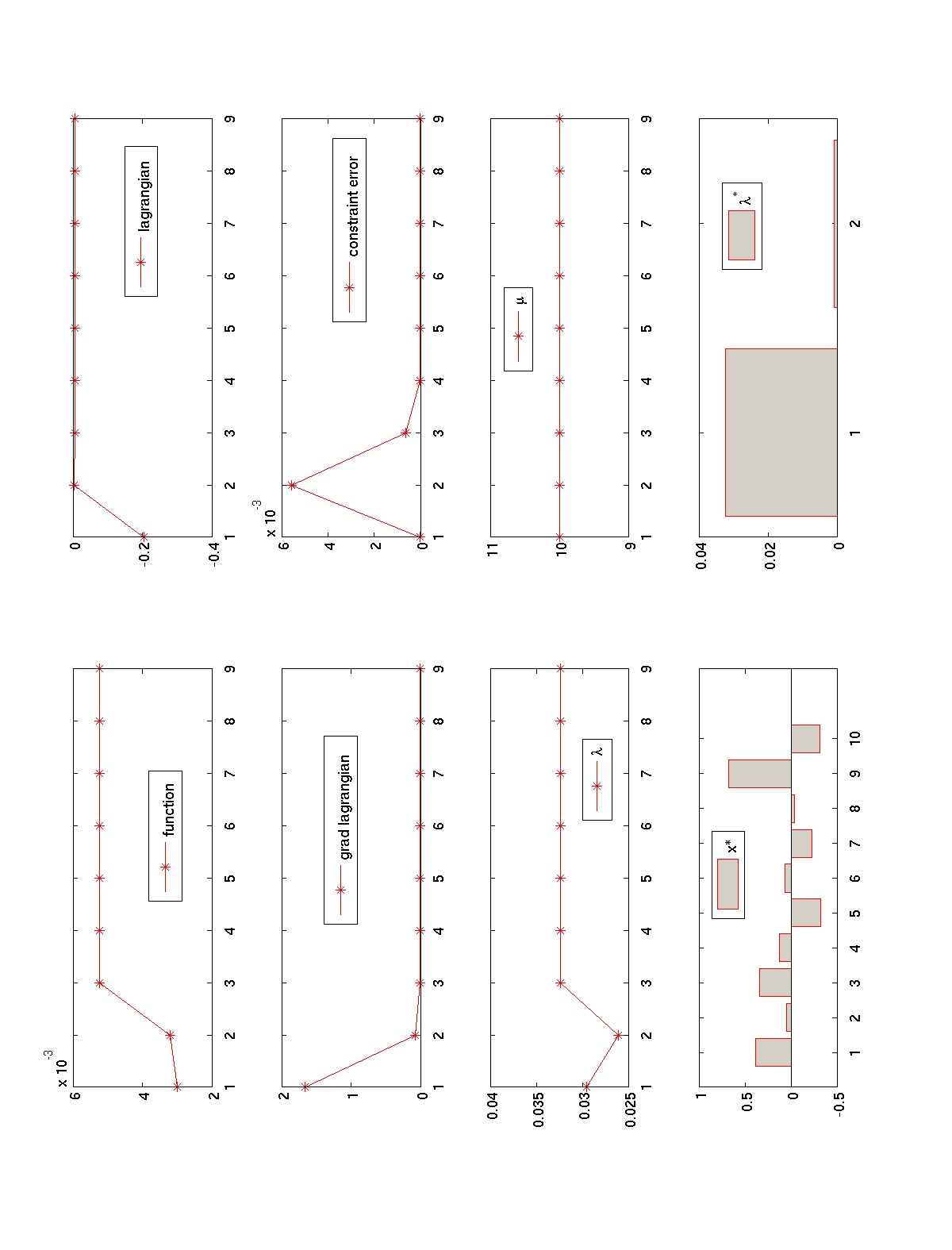}
}

\\

\subfigure[2 stage convergence diagnostics]
{
\label{ADIS2stagediagnostic}
\includegraphics[width = 95mm, angle = -90]{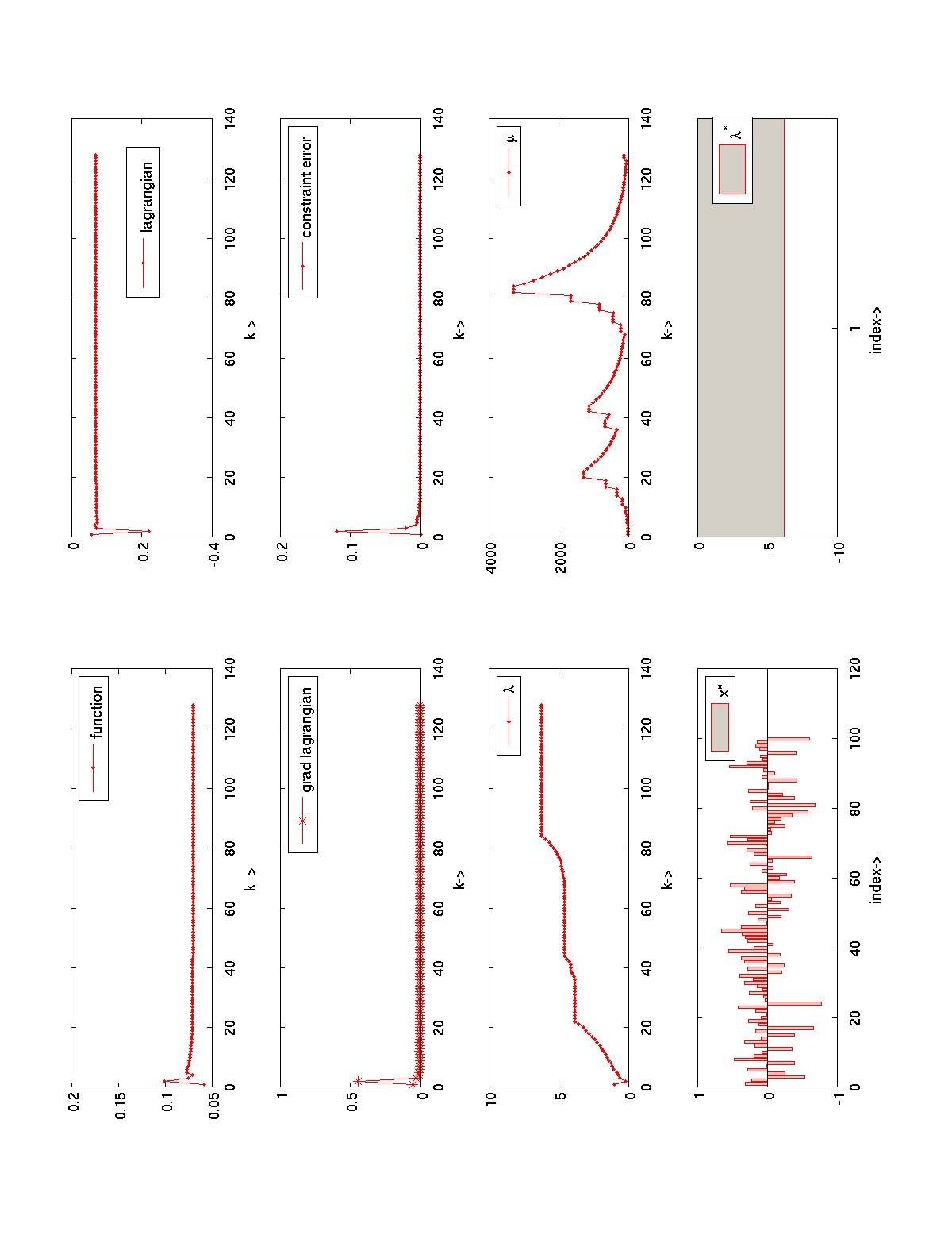}
}

\end{tabular}
\label{figsamplediagnosticandADIS2stagediagnostic}
\caption{(a) Sample Convergence Diagnostic plot for the 10halo benchmark in estimating the 2nd component. Figure shows the evolution of objective function, the Lagrangian, norm of the gradient of the Lagrangian, norm of the constraint satisfaction error, norm of the Lagrange multipliers and penalty parameter over algorithm iterations. ADIS \textit{guarantees} the optimality of the estimated sources. (b) Convergence diagnostics for ADIS Stage 1+2 for ACsparse10 dataset. The joint optimization was initialized using the optimal solution from ADIS Stage 1.}
\end{figure*}

\begin{figure*}[htbp]
\centering
\begin{tabular}{cc}

\subfigure[nband5]
{
\hspace{-1in}
\label{bench4a}
\includegraphics[width = 85mm, angle = -90]{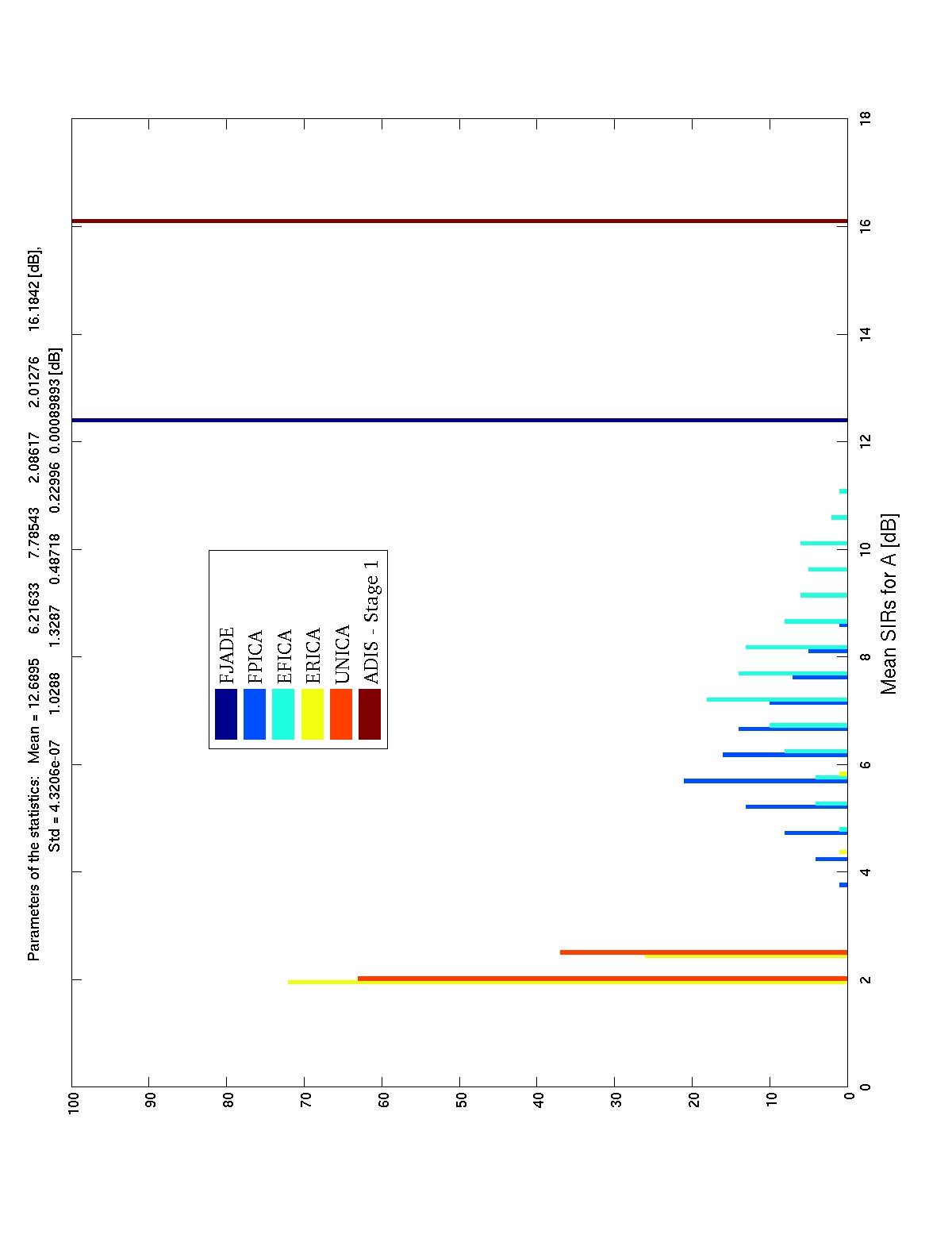}
}

& 

\subfigure[10halo]
{
\hspace{-0.5in}
\label{bench4b}
\includegraphics[width = 85mm, angle = -90]{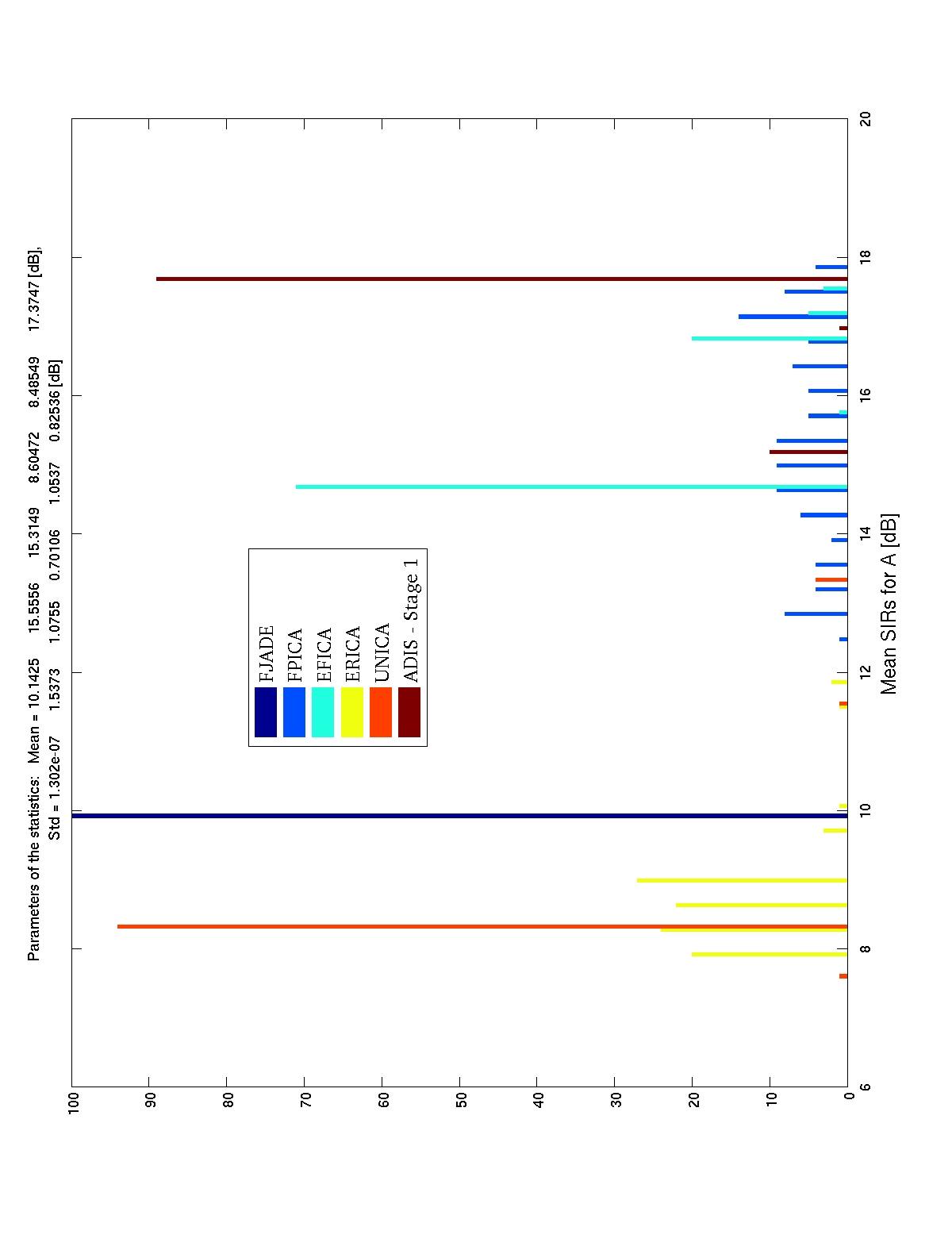}
}

\\

\subfigure[GnBand]
{
\hspace{-1in}
\label{bench4c}
\includegraphics[width = 85mm, angle = -90]{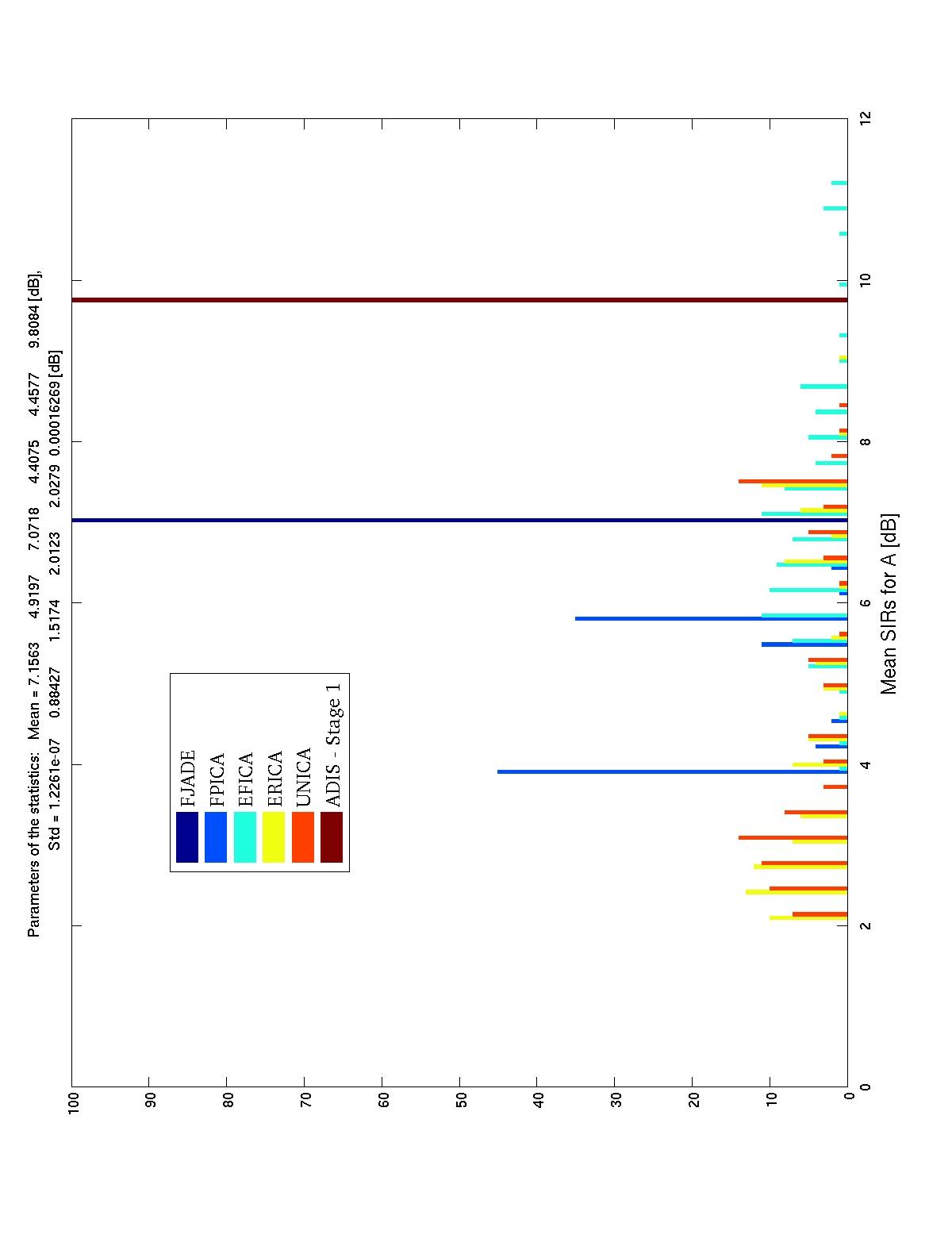}
}

& 

\subfigure[acspeech16]
{
\hspace{-0.5in}
\label{bench4d}
\includegraphics[width = 85mm, angle = -90]{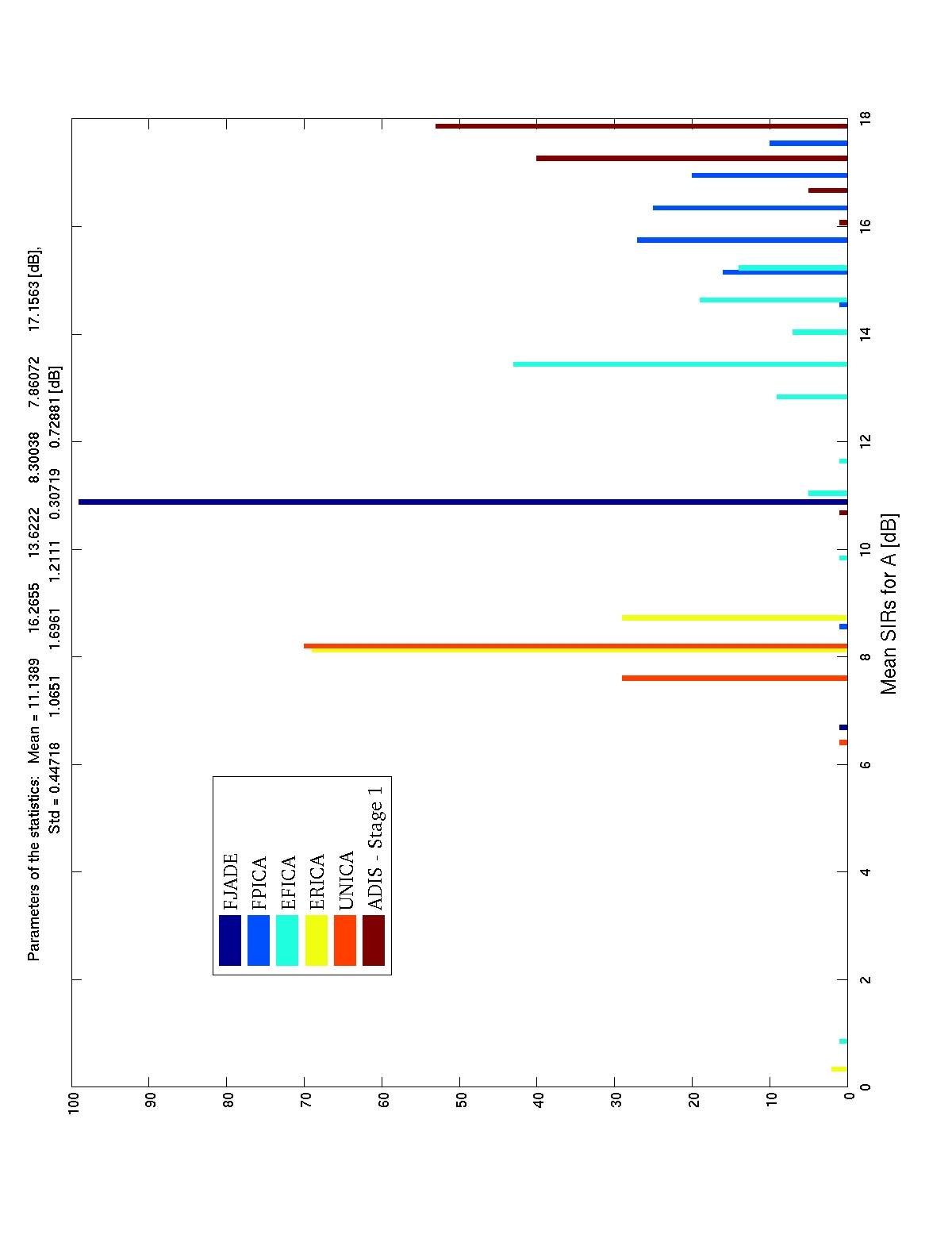}
}

\end{tabular}
\label{bench4}
\caption{Histograms of mean SIR for each algorithm over 100 Monte Carlo simulations using randomly generated mixing matrices for various benchmarking datasets. The properties of mixing matrices $A$ used for the simulations are as follows:
(a) Random sparse (b) Random sparse (c) Random bipolar (d) Symmetric random}
\end{figure*}

\begin{figure*}[htbp]
\centering
\begin{tabular}{cc}

\subfigure[ABio5]
{
\hspace{-1in}
\label{bench5a}
\includegraphics[width = 85mm, angle = -90]{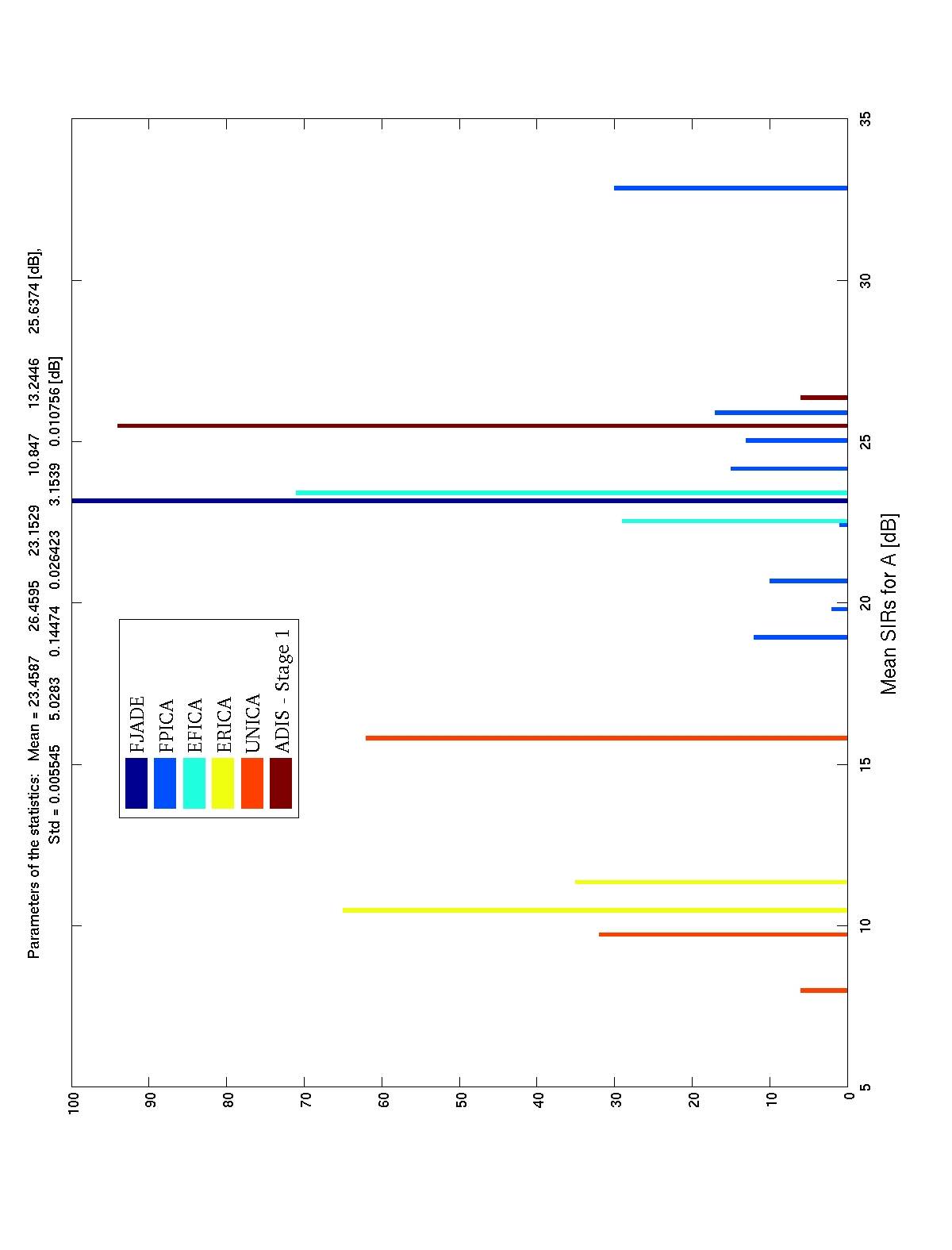}
}

& 

\subfigure[ACsparse10]
{
\hspace{-0.5in}
\label{bench5b}
\includegraphics[width = 85mm, angle = -90]{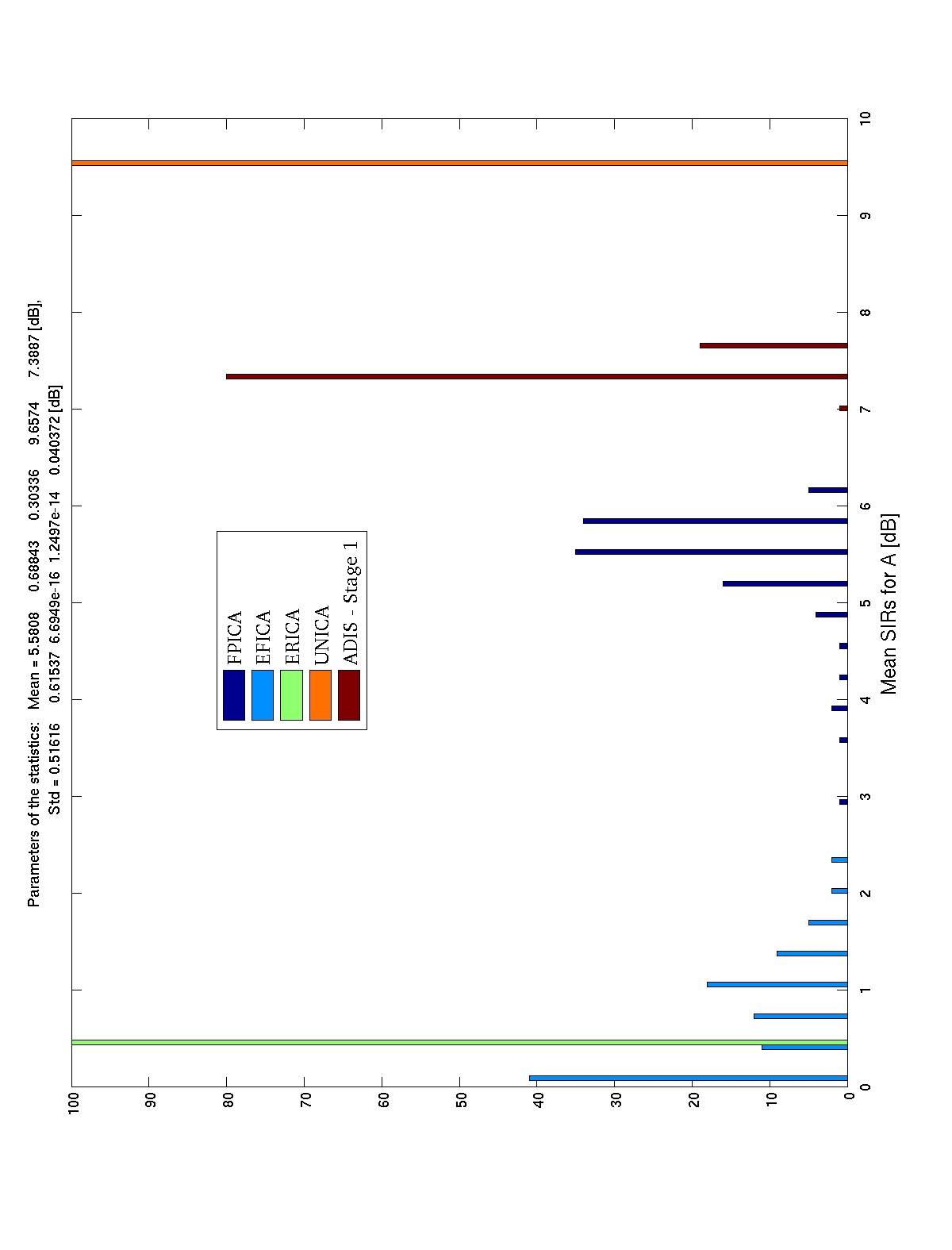}
}

\\

\subfigure[25SpeakersHALO]
{
\hspace{-1in}
\label{bench5c}
\includegraphics[width = 85mm, angle = -90]{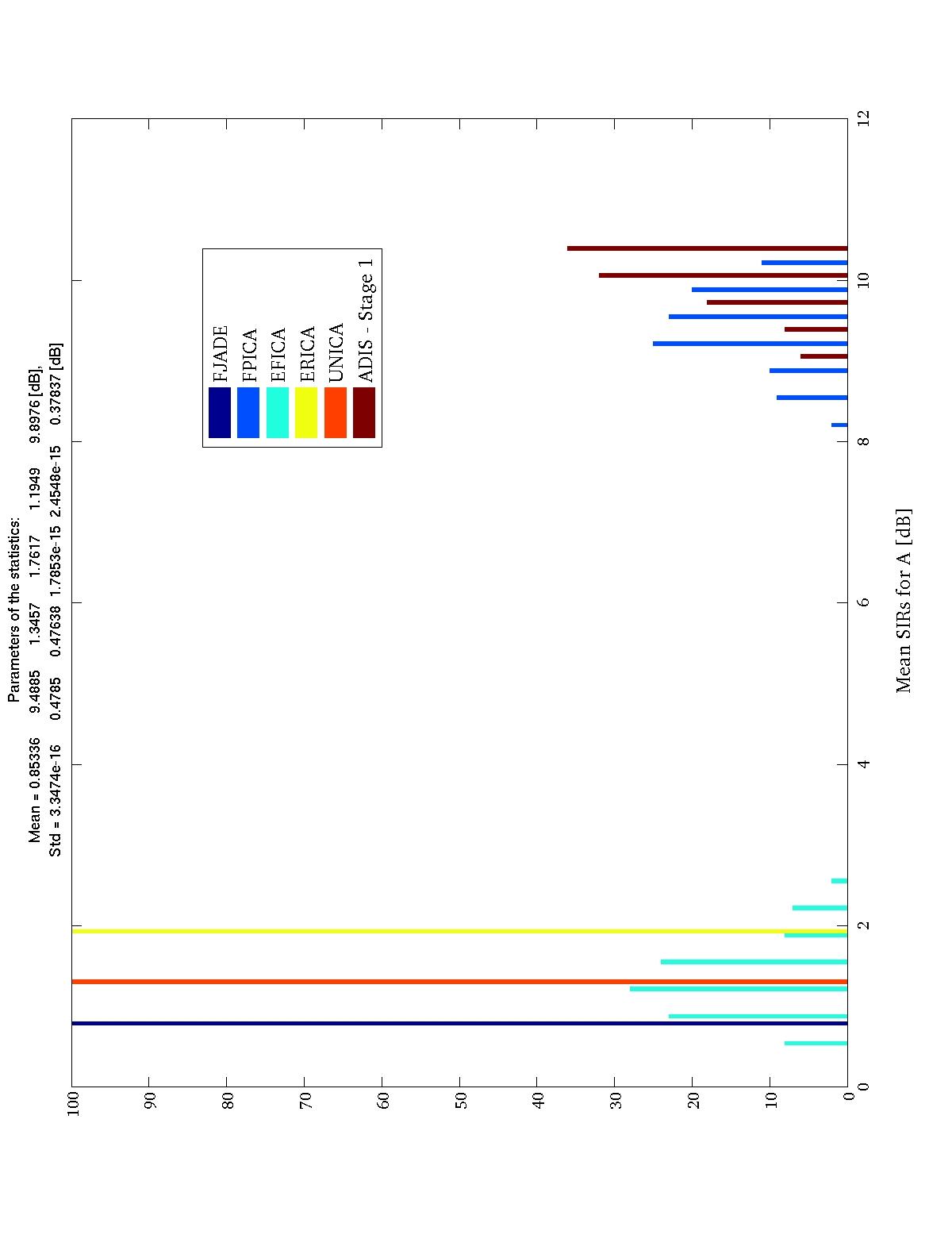}
}

& 

\subfigure[Vsparserand10]
{
\hspace{-0.5in}
\label{bench5d}
\includegraphics[width = 85mm, angle = -90]{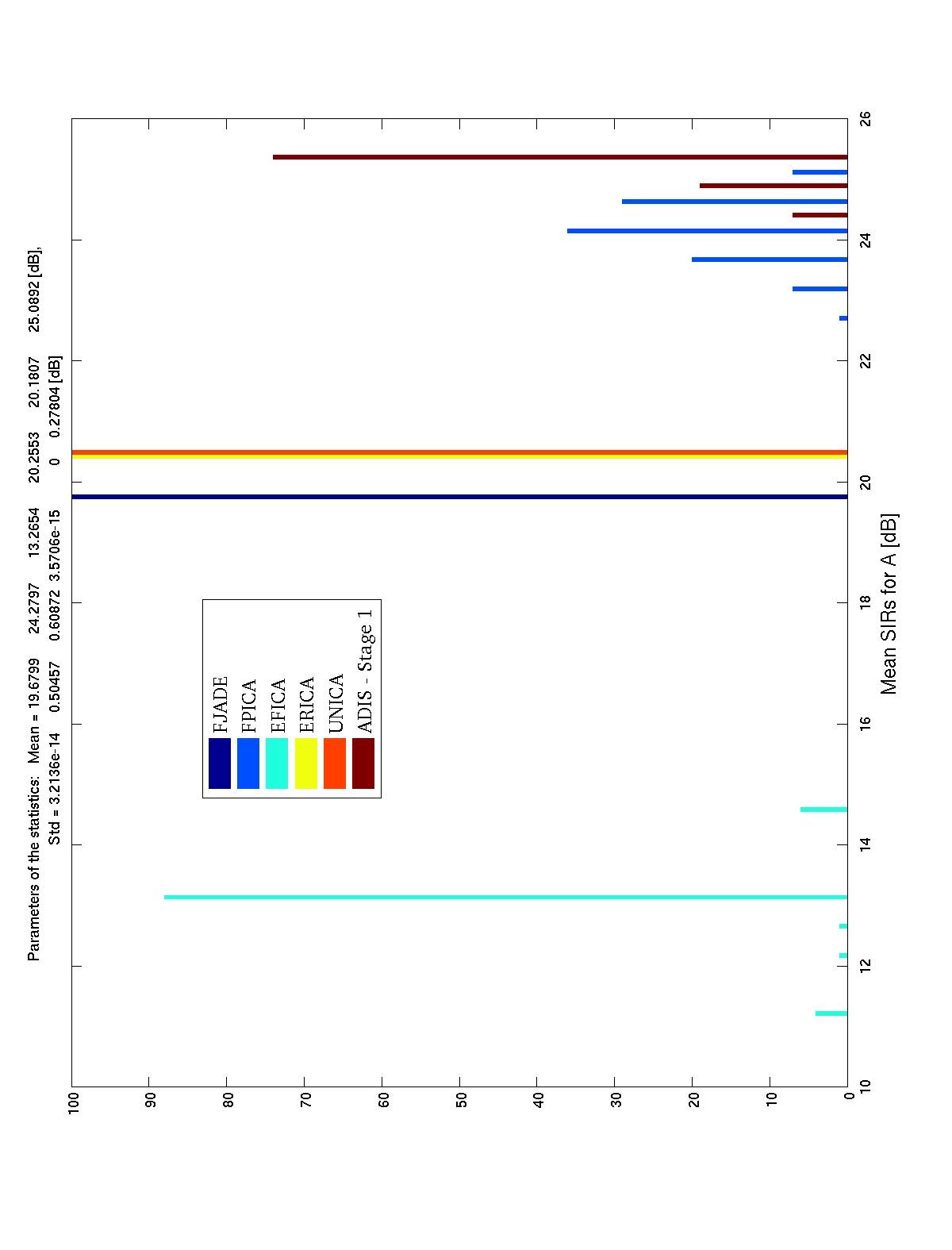}
}

\end{tabular}
\label{bench5}
\caption{Histograms of mean SIR for each algorithm over 100 Monte Carlo simulations using randomly generated mixing matrices for various benchmarking datasets. The properties of mixing matrices $A$ used for the simulations are as follows:
(a) Ill conditioned random (b) Hilbert (c) Toeplitz (d) Hankel}
\end{figure*}

\begin{figure*}[htbp]
\centering
\begin{tabular}{cc}

\subfigure[ACsincpos10]
{
\hspace{-1in}
\label{bench6a}
\includegraphics[width = 85mm, angle = -90]{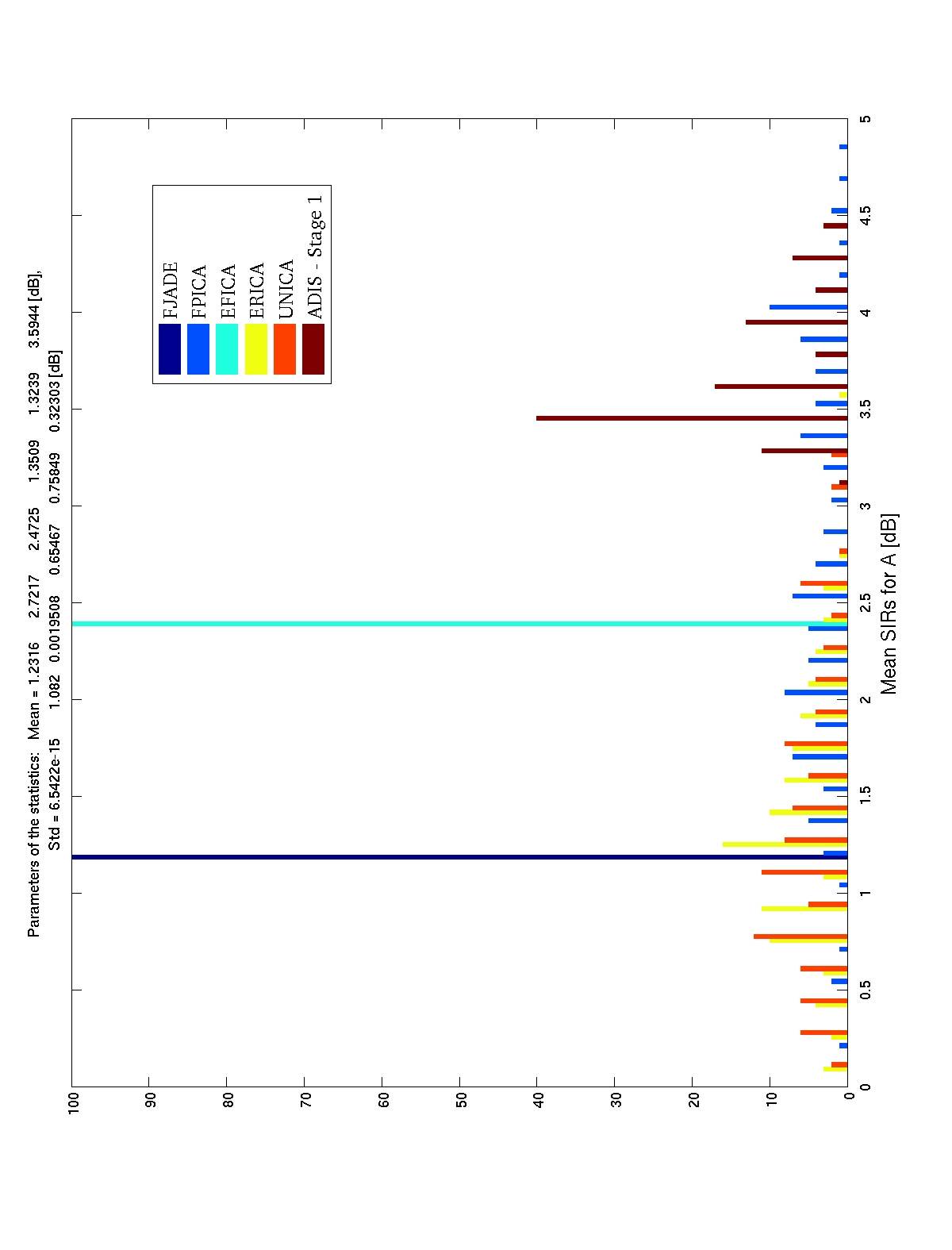}
}

& 

\subfigure[X5smooth]
{
\hspace{-0.5in}
\label{bench6b}
\includegraphics[width = 85mm, angle = -90]{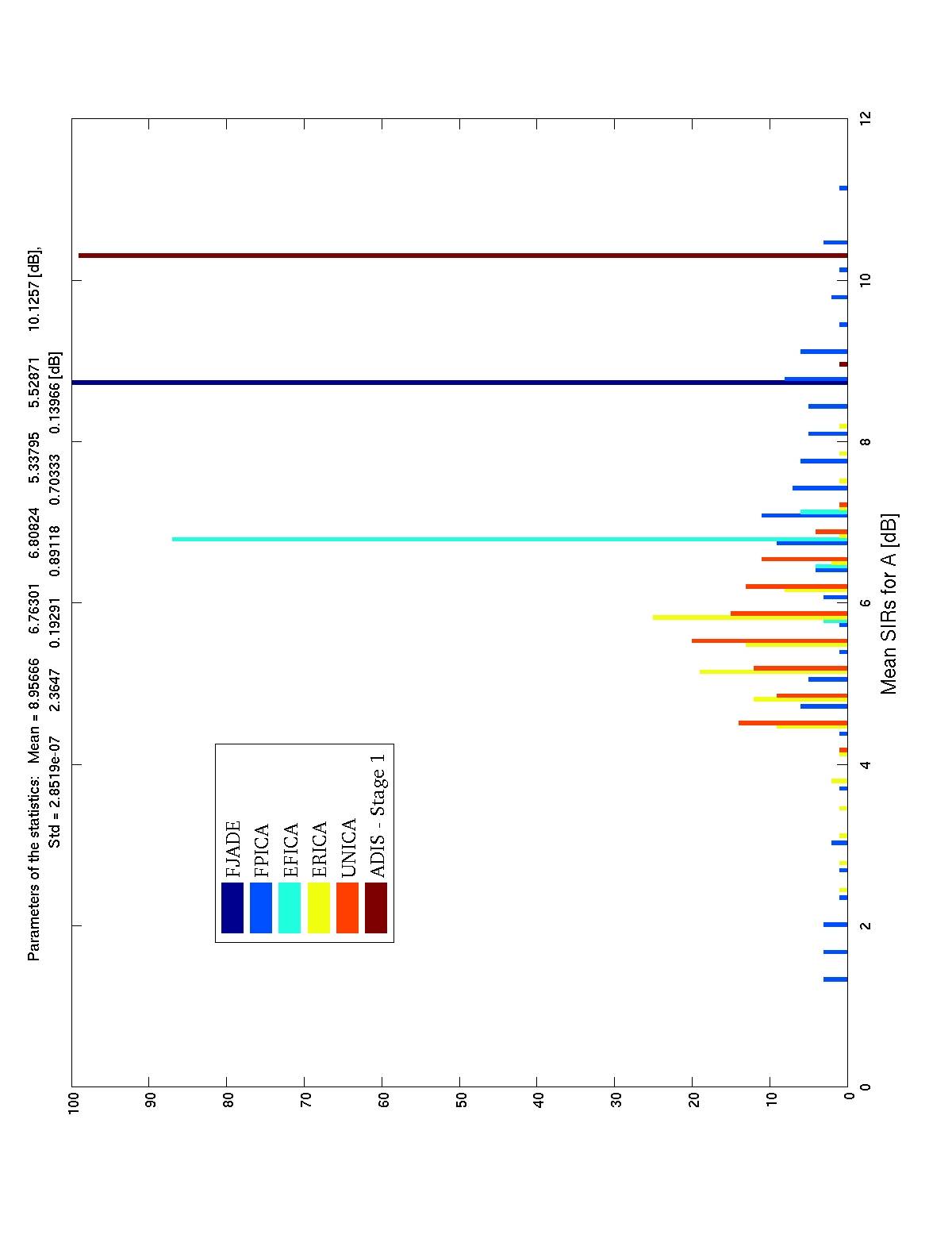}
}

\\

\subfigure[speech20]
{
\hspace{-1in}
\label{bench6c}
\includegraphics[width = 85mm, angle = -90]{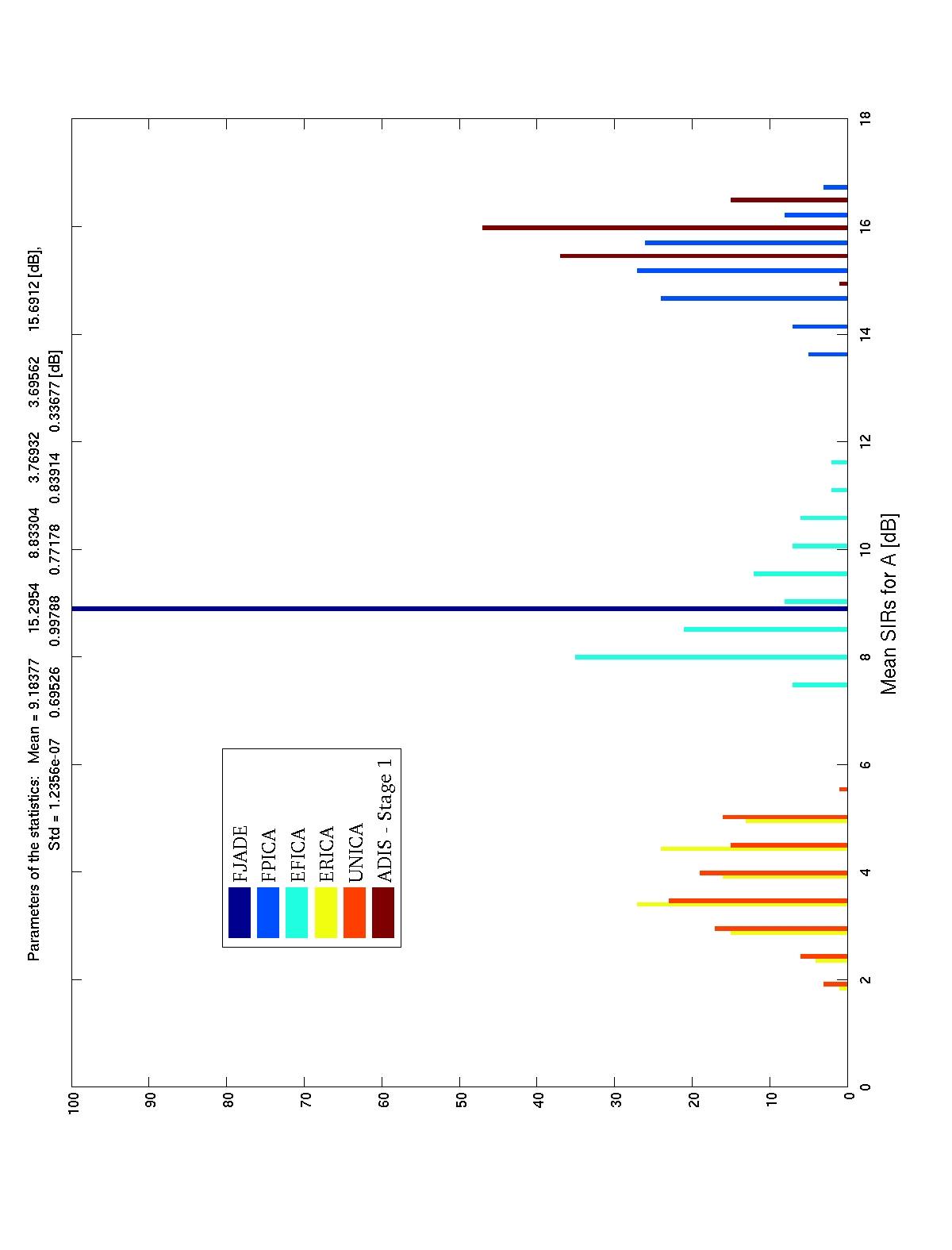}
}

& 

\subfigure[X10randsparse]
{
\hspace{-0.5in}
\label{bench6d}
\includegraphics[width = 85mm, angle = -90]{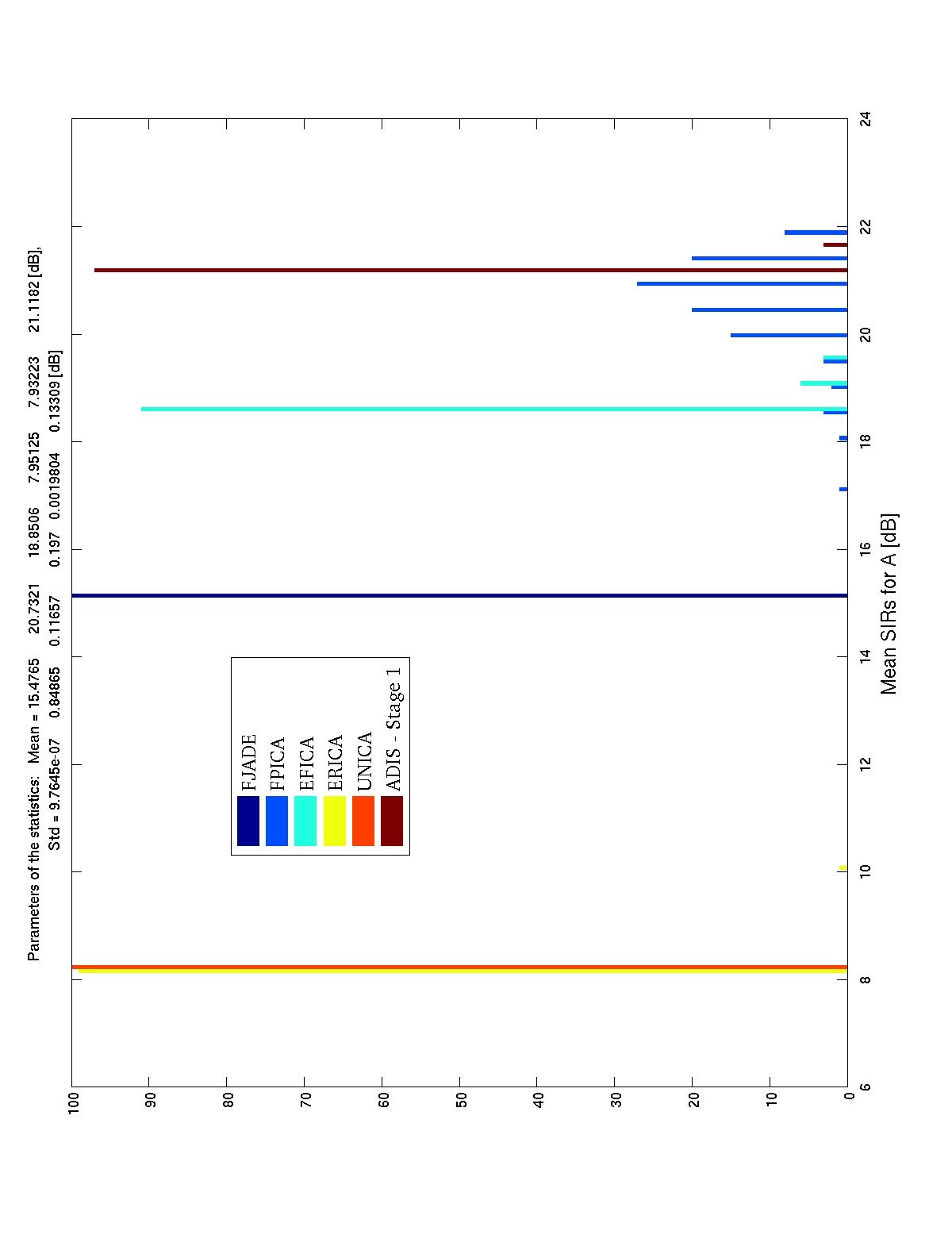}
}

\end{tabular}
\label{bench6}
\caption{Histograms of mean SIR for each algorithm over 100 Monte Carlo simulations using randomly generated mixing matrices for various benchmarking datasets. The properties of mixing matrices $A$ used for the simulations are as follows:
(a) Orthogonal (b) Nonnegative symmetric (c) Random bipolar (d) Skew symmetric}
\end{figure*}

The results are shown in figures \ref{bench1a} - \ref{bench3d} and table \ref{sirtable}. The key performance feature is the SIR index \cite{SIR:2006} (see appendix for definition), the higher the value the better. Another important feature is the variability of SIR over 100 mixtures each generated using a different mixing matrix but containing the same underlying sources. Ideally an algorithm should be robust enough to converge to the same solution regardless of variation in the mixing matrix. The standard deviation of SIR captures this variability, the lower the variability of SIR the better. Additional results showing simulation results with different types of mixing matrices $A$ are shown in \ref{bench4a} - \ref{bench6d}.

ADIS perfomed better than all other algorithms in almost all cases both in terms of the mean SIR index as well as the standard deviation of the mean SIR, which measures the robustness and convergence to the same solution. ADIS also produces extensive convergence diagnostics a sample of which is shown in figure \ref{samplediagnostic}. These diagnostics guarantee convergence and improve confidence in the estimated sources.

An illustration of performance improvement using ADIS (Stage 1 + 2) over Stage 1 is shown fir the ACsparse10 dataset in figure \ref{ADIS2stage}. The corresponding convergence diagnostics are shown in figure \ref{ADIS2stagediagnostic}. ADIS Stage 2 performs better than ADIS Stage 1 but we did not observe as dramatic an improvement as seen for ACsparse10. Other ICA algorithms were outperformed using only ADIS Stage 1.

All experiments were performed on a computer with an Intel Xeon (TM) processor (3.4 GHz) and 4GB of RAM. The runtimes of ADIS (Stage 1) were comparable to those of FPICA, ERICA and UNICA. We found EFICA and JADE to be faster than other algorithms in general.

%------more benchmarking stuff with various types of mixing mats--------

%---

%-----

\begin{figure*}[htbp]
\centering
\begin{tabular}{c}

\includegraphics[width = 100mm, angle = -90]{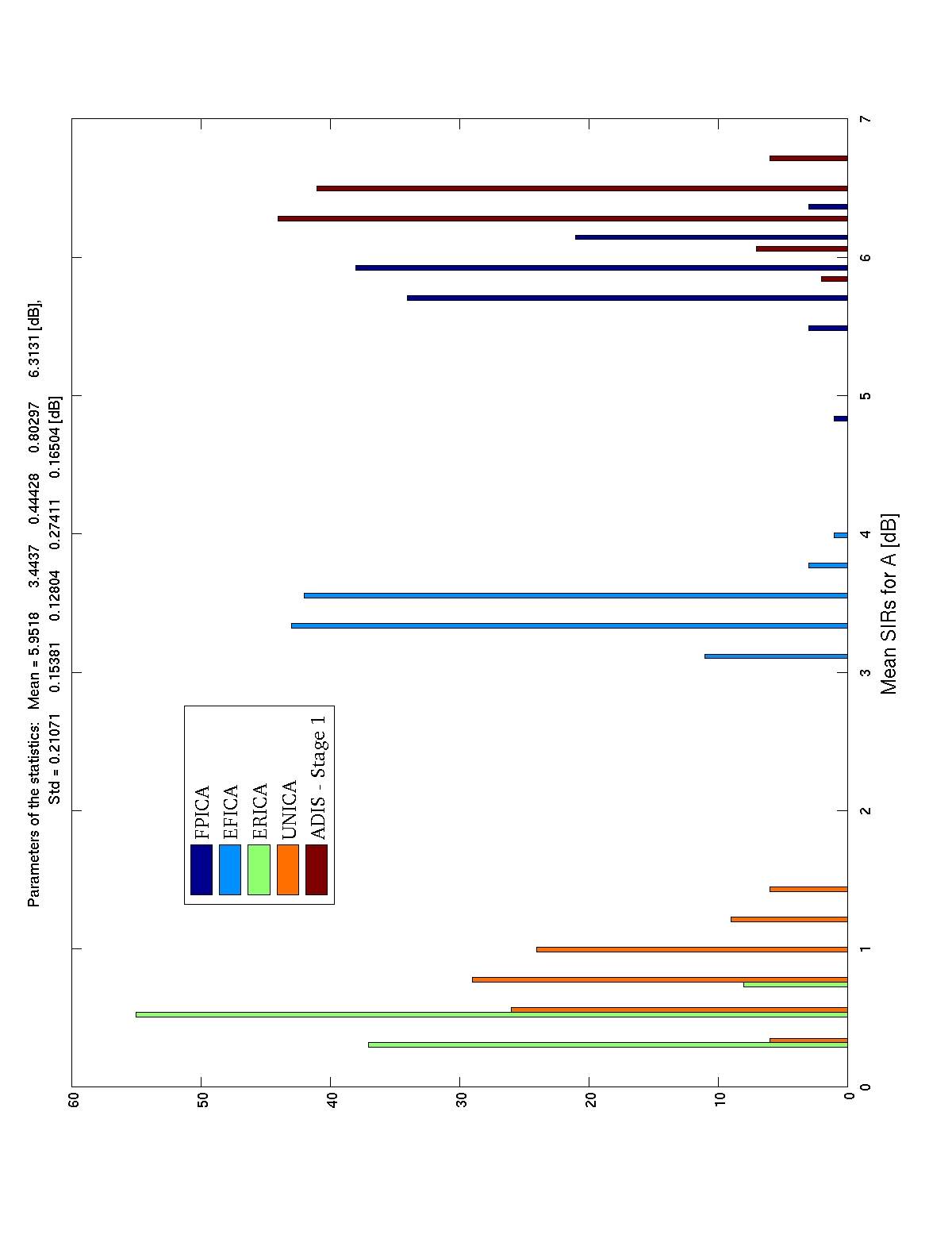}

\end{tabular}
\label{figspeech64_illcondrandom}
\caption{Histograms of mean SIR for each algorithm over 100 Monte Carlo simulations using randomly generated mixing matrices on the 64sounds benchmark dataset. This is a relatively large dataset with 64 sources. FJADE fails on this test case. $A$ was chosen to be an ill conditioned random mixing matrix for this simulation.}
\end{figure*}

%----------------------------------------------------------------------------------------------

\section{Case Study using real fMRI data}\label{realdata}

\subsection{fMRI case study}

To demonstrate how ADIS performs on a real dataset, we used the "FSL Evaluation and Example Data Suite" (FEEDS) from FMRIB Image Analysis Group, Oxford University. The URL for this data suite is: 
\url{http://www.fmrib.ox.ac.uk/fsl/feeds/doc/index.html}
One of the datasets in the example suite contains an audio visual experiment with two explanatory variables, the visual stimulus (30s off, 30s on) and an auditory stimulus (45s off, 45s on). Analysis was carried out using FEAT (FMRI Expert Analysis Tool) Version 5.4, part of FSL (FMRIB's Software Library). 
\begin{verbatim}www.fmrib.ox.ac.uk/fsl\end{verbatim}
The following pre-statistics processing was applied; motion correction using MCFLIRT [Jenkinson 2002]; non-brain removal using BET [Smith 2002]; spatial smoothing using a Gaussian kernel of FWHM 5mm; mean-based intensity normalisation of all volumes by the same factor; highpass temporal filtering (Gaussian-weighted LSF straight line fitting, with sigma=50.0s).

\begin{figure*}[htbp]
\begin{center}
\includegraphics[width = 120mm]{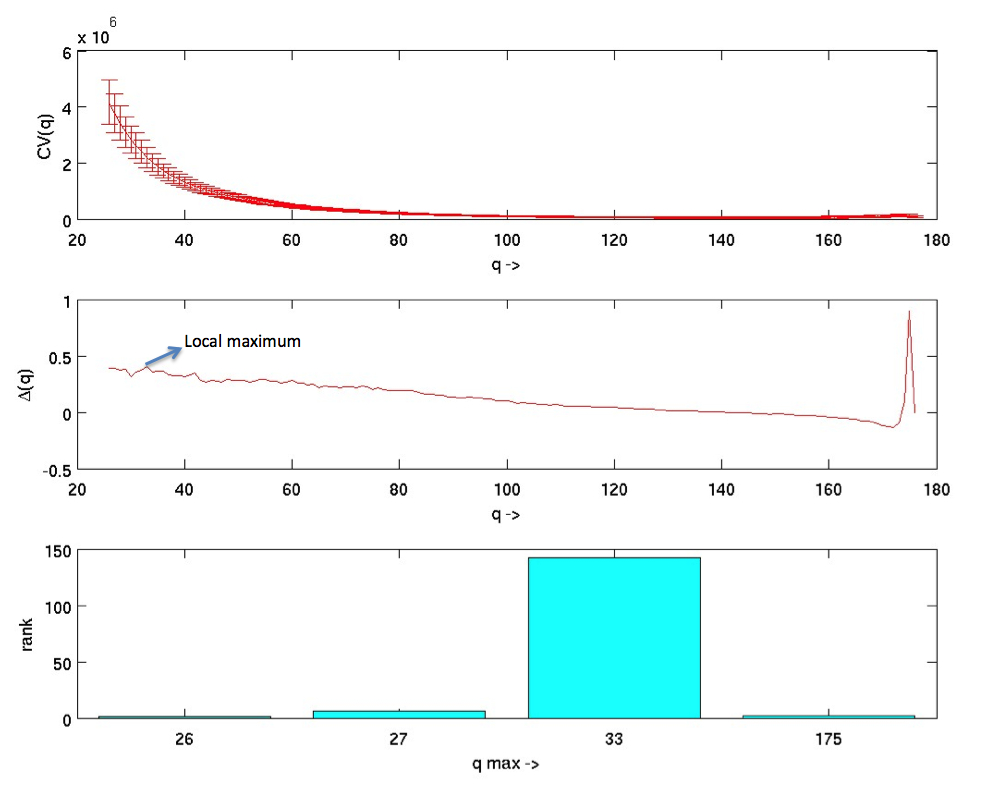}
\caption{Latent dimensionality estimation summary for fMRI data. The number of latent sources were estimated to be $q = 34$.}
\label{latent_dim_est_fmri}
\end{center}
\end{figure*}

\begin{figure*}[htbp]
\centering
\begin{tabular}{cc}
\subfigure[Estimated Auditory Cortex activating source $z$-statistic for the audio-visual fMRI data thresholded at $z > 5$]
{
\label{fmria}
\includegraphics[width = 65mm]{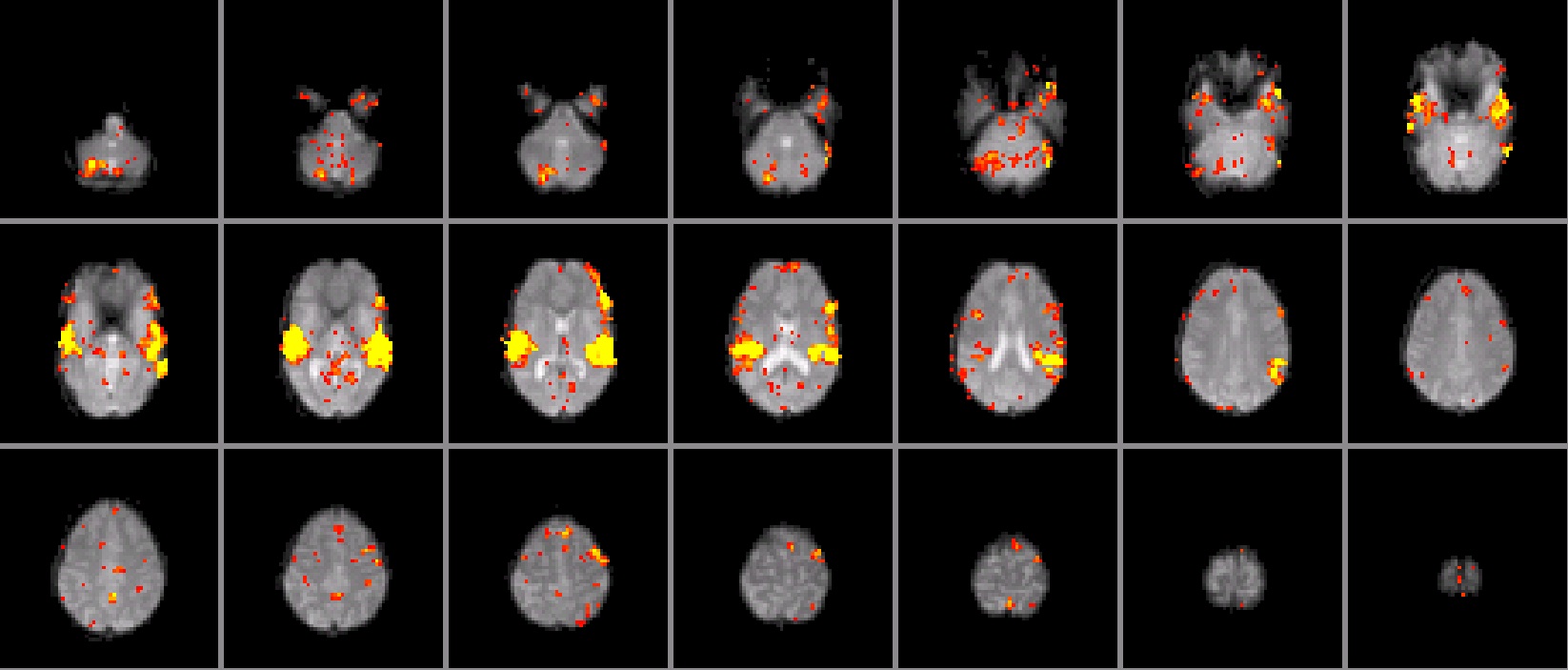}
}
&
\hspace{0mm}

\subfigure[Estimated Visual Cortex activating source $z$-statistic for the audio-visual fMRI data thresholded at $z > 5$]
{
\label{fmrib}
\includegraphics[width = 65mm]{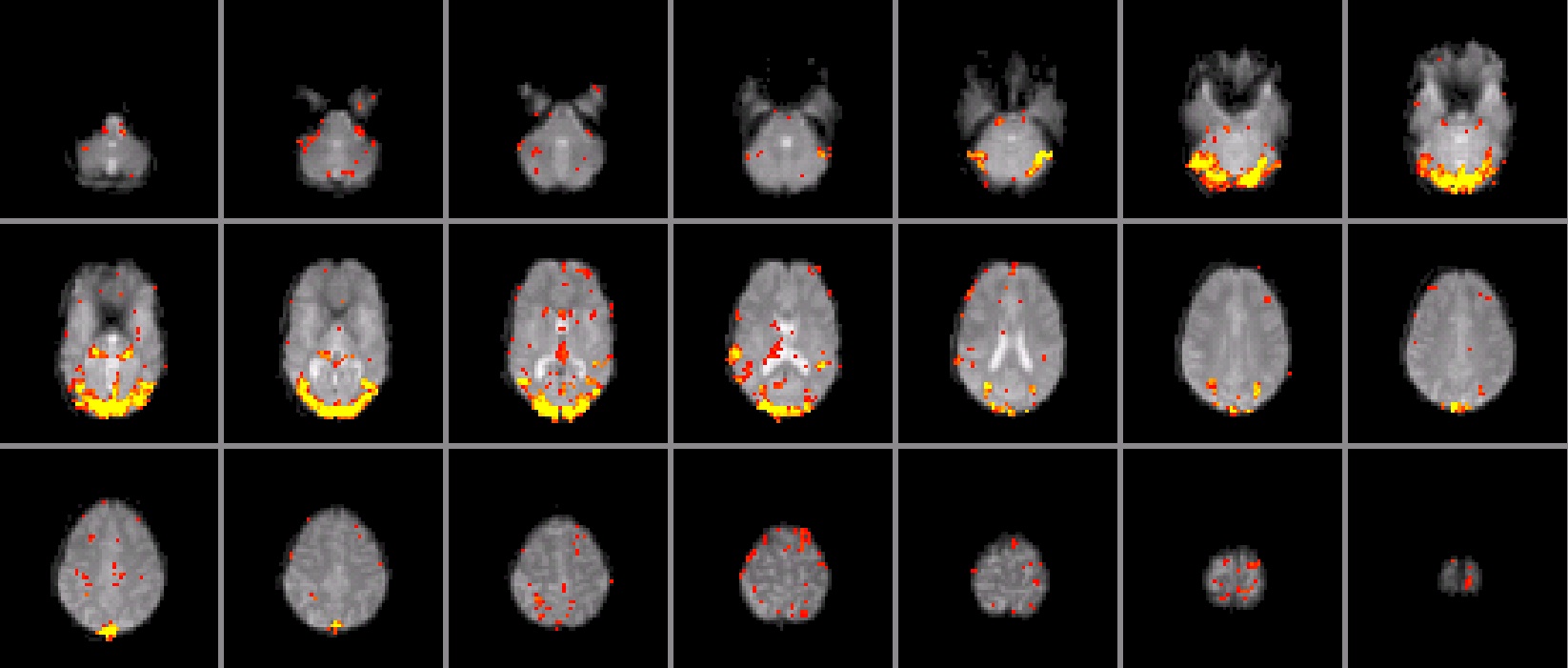}
} \\

\subfigure[Convergence Diagnostic plot for the Auditory source estimation]
{
\label{fmric}
\includegraphics[width = 65mm, angle = -90]{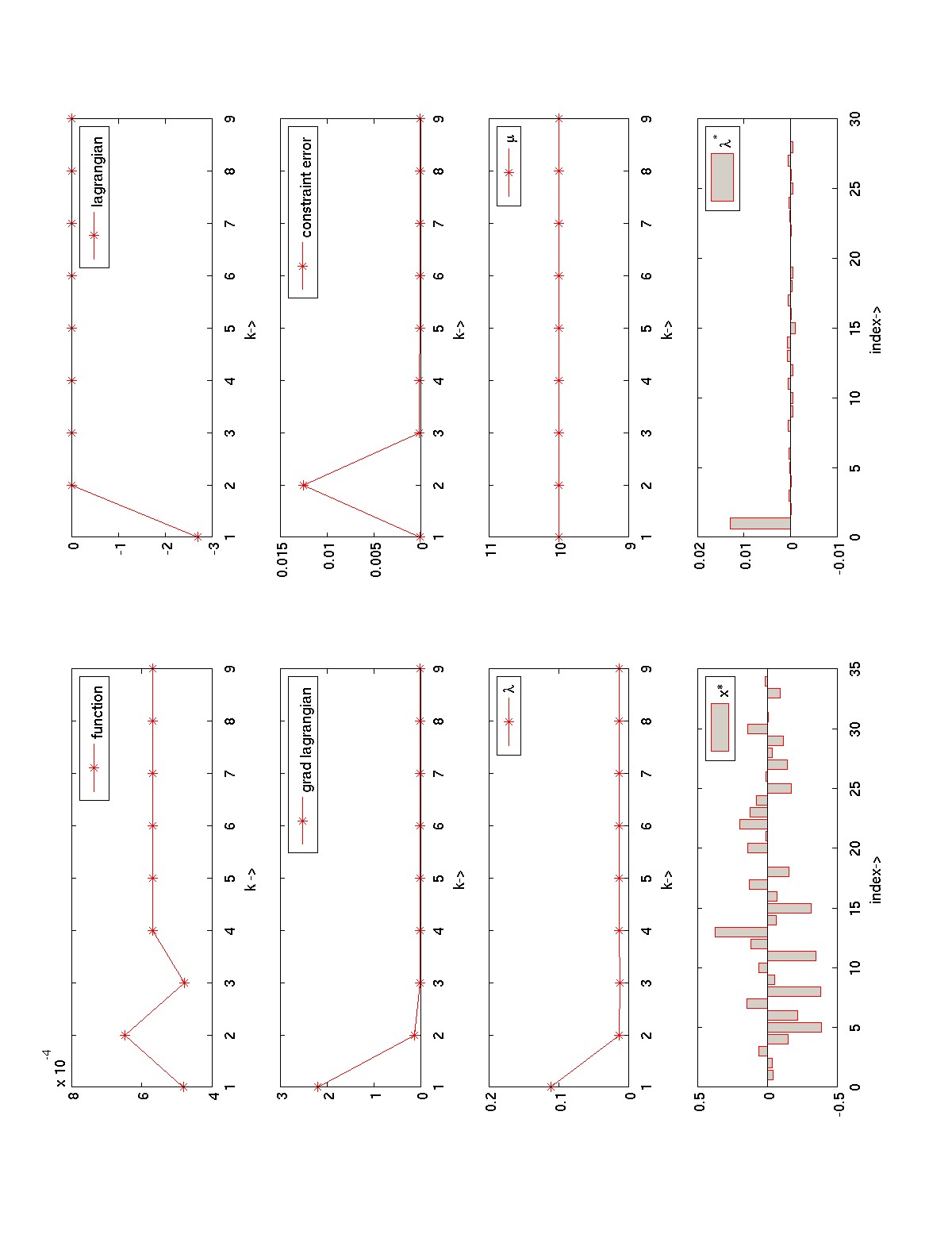}
} 

&

\subfigure[Convergence Diagnostic plot for the Visual source estimation]
{
\label{fmrie}
\includegraphics[width = 65mm, angle = -90]{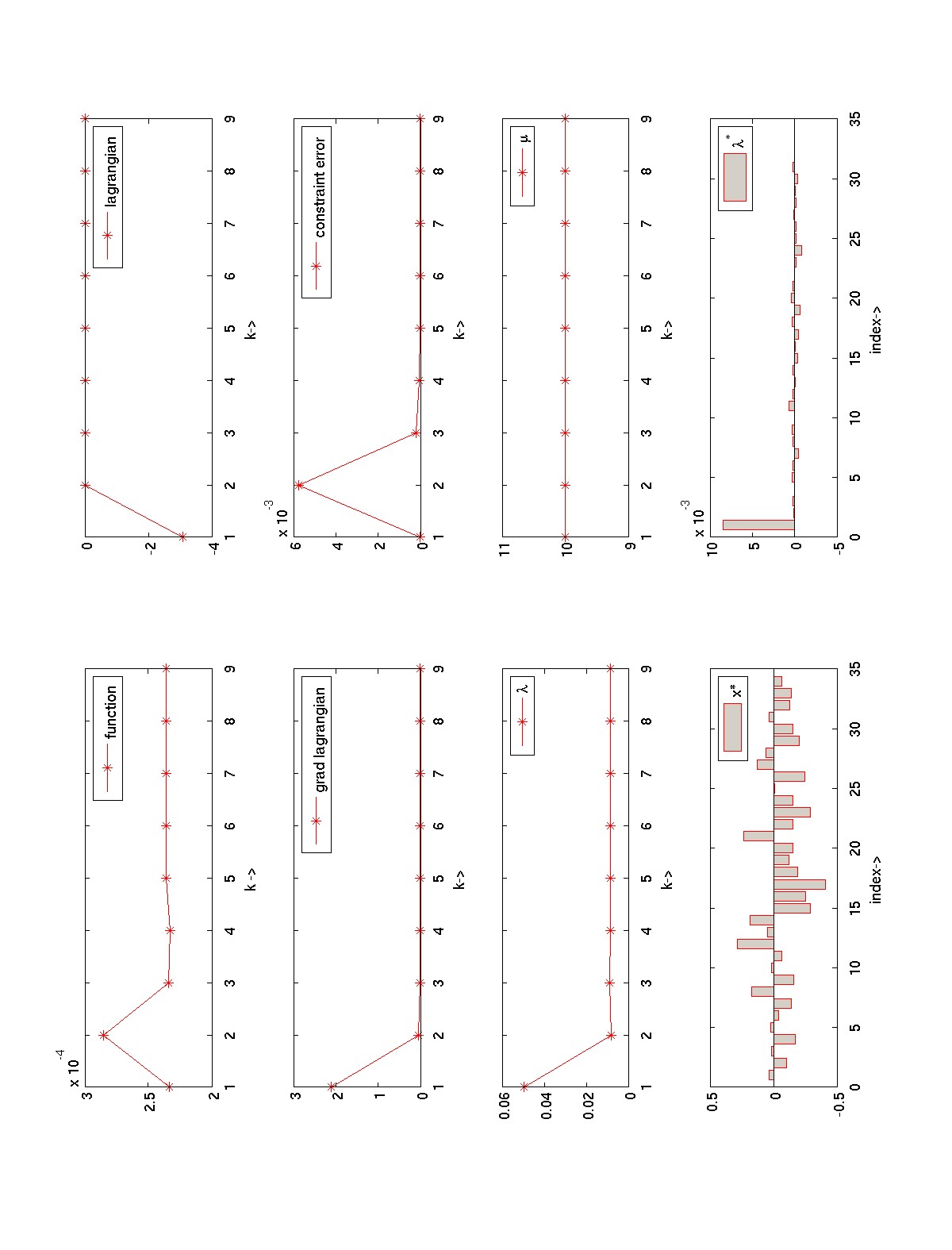}
} \\

\subfigure[Associated Timecourse (from mixing matrix) for the Auditory source]
{
\label{fmrif}
\includegraphics[width = 65mm]{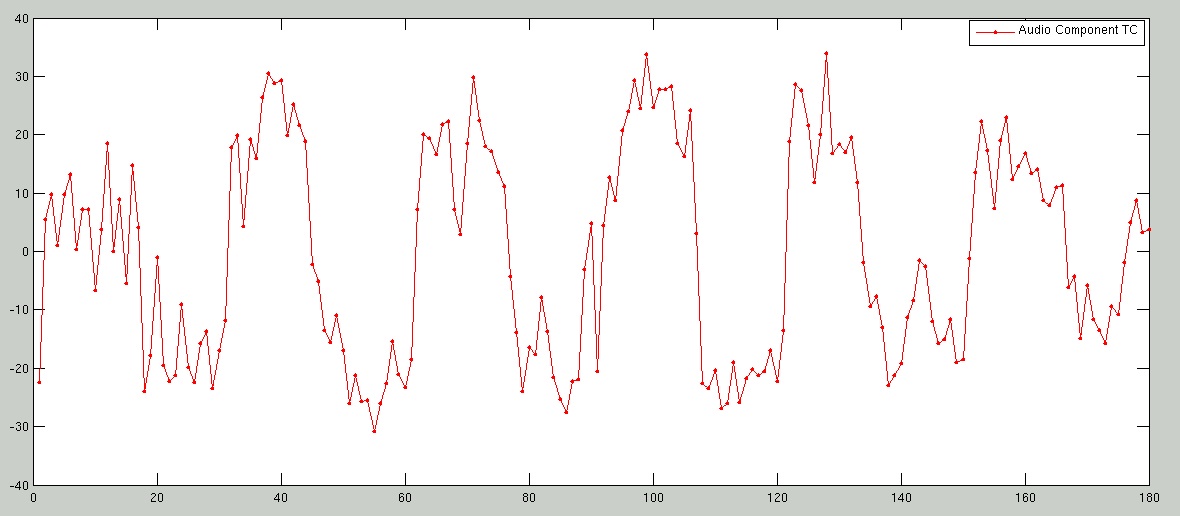}
} 

&
\hspace{0mm}

\subfigure[Associated Timecourse (from mixing matrix) for the Visual source]
{
\label{fmrig}
\includegraphics[width = 65mm]{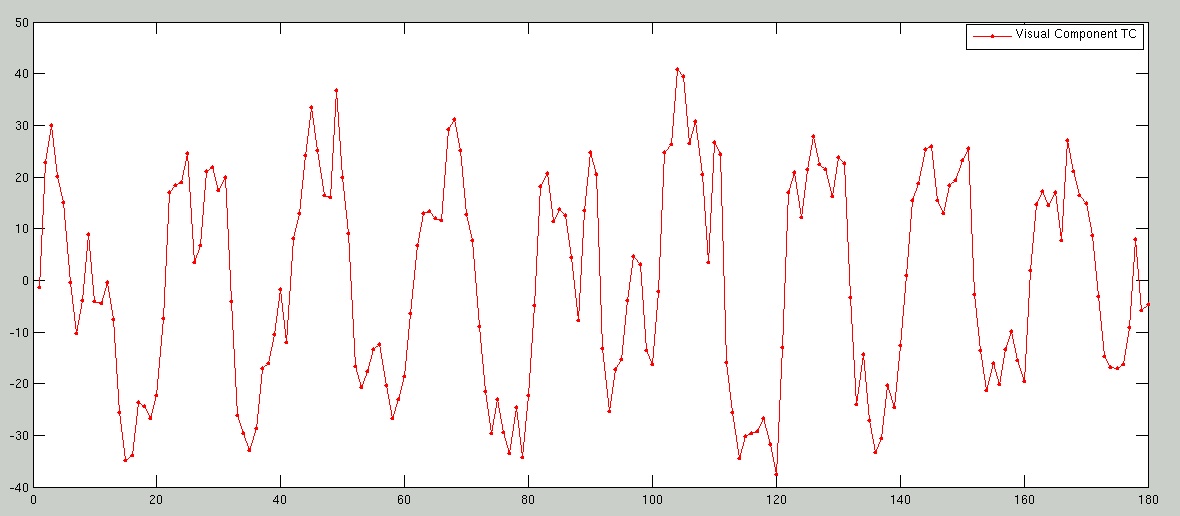}
} \\

\end{tabular}
\caption{Result of applying ADIS (Stage 1) to audio-visual fMRI data}
\label{fmri}
\end{figure*}

ADIS estimated a latent dimensionality of $q = 34$. The source components activating the auditory and visual cortex were identified by inspecting the estimated voxelwise variance explained map. The results are shown in figures \ref{latent_dim_est_fmri}-\ref{fmri}. 

%-----

% subfigure syntax

%----------------------------

\section{Conclusion}
We implemented ADIS, an algorithm for probabilistic, constrained and non-square projection pursuit. We validated all aspects of ADIS including the latent dimensionality estimation procedure and its optimization core. When compared to other algorithms using standard benchmarking datasets using ICALAB, we find our algorithm outperforms other standard algorithms such as FastICA, FPICA, JADE, ERICA and UNICA in terms of both robustness and separation quality. Our algorithm also guarantees "optimality" for each blind source via extensive convergence diagnostics and enables the user to use arbitrary contrast function and constraints for BSS. We hope it will be useful as a general BSS tool for the signal processing and fMRI community.

%\appendix
\section{Appendix}\label{supplemental}

\section{Negentropy Index}
Given a random variable $X$, the negative entropy a measure of non-Gaussianity. It is easy to show that imposition of independence on sources in BSS is equivalent to maximization of negentropy. %Several contrast functions that measure approximations to negative entropy are possible \cite{FPICA:2000}. 

Robust approximations to negative entropy were developed in \cite{negentropy:1998}. If $G$ is a non-quadratic, non-linear function and $v$ is a Gaussian random variable of the same variance as $X$ then the negentropy measure $J(X)$ is given as \cite{FPICA:2000}

\begin{equation}\label{negentropy}
J(X) \propto [E(G(X)) - E(G(v)]^2
\end{equation}

%We used the negentropy measure based on non-quadratic, non-linear functions proposed in \cite{FPICA:2000}.
In this paper, we used the following function \cite{negentropy:1998} for $G$:
\begin{equation}\label{functionG}
G(x) = \log{[\cosh{(x)}]}
\end{equation}
where $\cosh{(x)}$ is the hyperbolic cosine function
\begin{equation}
\cosh{(x)} = \frac{e^{x} + e^{-x}}{2}
\end{equation}

\section{Sources to Interferences Ratio (SIR) }
The SIR ratio is defined in \cite{SIR:2006}. We give here a brief summary of the key equations from that paper.
Let $ y = \{y_1,y_2,\ldots,y_k\}$ and let $P_y$ be the orthogonal projector onto the subspace spanned by $y_1,y_2,\ldots y_k$. If $s = [s_1,s_2,\ldots,s_k]$ are the true sources and if $\hat{s}_1, \hat{s}_2,\ldots, \hat{s}_k$ are the corresponding estimated values then define:
\begin{equation}\label{sir1}
s_{target} = P_{s_j} \hat{s}_j
\end{equation}
\begin{equation}\label{sir2}
e_{interf} = P_s \hat{s}_j - P_{s_j} \hat{s}_j
\end{equation}

The purity of source separation is measured using the SIR performance index defined as follows:

\begin{equation}\label{sir3}
SIR = 10 \log_{10} \frac{ ||s_{target}||^2} { ||e_{interf}||^2}
\end{equation}

\section{Benchmarking datasets}
The 13 benchmarking datasets and their short descriptions are as follows: \\ 
(\url{http://www.bsp.brain.riken.jp/ICALAB/ICALABSignalProc/}) :
\begin{itemize}
\item nband5 - contains 5 narrow band sources. This is a rather "easy" benchmark for second order separation algorithms but apprently presents challenges for higher order algorithms.
\item 10halo - contains 10 speech signals that are highly correlated (all 10 speakers say the same sentence).
\item GnBand - contains 5 fourth order colored sources with a distribution close to Gaussian. This is a rather "difficult" benchmark. 
\item acspeech16 - contains 16 typical speech signals which have a temporal structure but are not precisely independent
\item ABio5 - contains 5 typical biological sources
\item ACsparse10 - contains 10 sparse (smooth bell-shape) sources that are approximately independent. The SOS blind source separation algorithms fail to separate such sources.
\item 25SpeakersHALO - 25 highly correlated speech signals
\item Vsparserand10 - very sparse random signals
\item ACsincpos10 - positive sparse signals
\item X5smooth - smooth signals
\item Speech20 - 20 speech/music sources
\item X10randsparse - random sparse signals
%%\item ACvsparse10 - contains 10 very sparse (short regular pulses) signals. Second order BSS algorithms fail to separate these sources.
%%\item Sergio7 - this benchmark contains 7 sources (some of them are asymmetrically distributed)
\item 64soundsstd - a variety of sound sources (64 in total) 
\end{itemize}

\section{Details on Optimization Algorithm} \label{optimalgdetails}
Our optimization algorithm solves the general problem:
\begin{eqnarray}\label{a1}
\mbox{ min }_x f(x) & \\
\mbox{s.t. } c_i(x) = 0, & i = 1,2,\ldots, m \\
\mbox{s.t. } g_j(x) \ge 0, & j = 1,2, \ldots L
\end{eqnarray}

where $x \in R^n$.

We convert the inequality constraints into equality constraints via slack variables as follows:
\begin{eqnarray} \label{a2}
g_j(x) - s_j = 0 & \\ 
s_j \ge 0, & j = 1,2, \ldots L
\end{eqnarray}

Thus the optimization problem becomes:
\begin{eqnarray}\label{a3}
\mbox{ min } f(x) & \\
\mbox{s.t. } c_i(x) = 0, & i = 1,2,\ldots, m \\
\mbox{s.t. } g_j(x) - s_j = 0, &  j = 1,2, \ldots L \\
s_j \ge 0 & 
\end{eqnarray}

This problem is now an equality constrained problem where the inequalities have been replaced by the bound constraints on the slack variables. Thus it suffices to consider equality constrained problems with bounds on independent variables as follows:
\begin{eqnarray}\label{a4}
\mbox{ min } f(x) & \\
\mbox{s.t. } c_i(x) = 0, & i = 1,2,\ldots, m \\
\mbox{s.t. } l_i \le x_i \le u_i, & i = 1,2, \ldots n
\end{eqnarray}
where $x \in R^n$.

Our code uses a trust region based augmented lagrangian approach to solve these bound constrained problems following closely the LANCELOT software package \cite{LANCELOT:1992}, \cite{Conn:1991}. The augmented lagrangian function for the above problem is defined as:
\begin{equation}\label{a5}
\mathcal{L}(x, \lambda, \mu) = f(x) - \sum_{i = 1}^m \lambda_i c_i(x) + \frac{\mu}{2} \sum_{i = 1}^m c_i(x)^2
\end{equation}

At each outer iteration $k$, given current values of $\lambda^k$ and $\mu_k$ we solve the subproblem:

\begin{eqnarray}\label{a6}
\mbox{ min } \mathcal{L}(x, \lambda^k, \mu_k) \\
\mbox{ s.t. } l_i \le x_i \le u_i
\end{eqnarray}

If $P$ is the projection operator defined as
\begin{equation}\label{a7}
[P(z, l , u)]_i = \left \{
\begin{array}{ccc}
l_i & \mbox{ if } & z_i \le l_i \\
z_i & \mbox{ if }  & l_i \le z_i \le u_i \\
u_i & \mbox{ if } & z_i \ge u_i 
\end{array}
\right.
\end{equation}
then the Karush-Kuhn-Tucker (KKT) optimality condition for \ref{a6} is given as \cite{Conn:1991}:
\begin{equation}\label{a8}
x - P(x - \nabla_x \mathcal{L}(x, \lambda^k, \mu_k),  l , u) = 0
\end{equation}

The outer iteration code is given in Framework 1. Note that the penalty parameter $\mu_k$  is updated based on a feasibility monitoring strategy that allows for a decrease in $\mu_k$ if sufficient accuracy is not achieved in solving the subproblem \ref{a6}.

\begin{algorithm}
\begin{algorithmic}[1]
\REQUIRE Initial point $x_{init}$,  $\lambda^0$, $\mu_0$, $\theta^h \in (1, \infty)$, $\theta_l \in (0, 1)$
\STATE Choose tolerances $\eta^*_{con}$ and $\eta^*_{grad}$. The default in ADIS is $\eta^*_{con} = \eta^*_{grad} = 1e{-6}$. .
\STATE $\mu = \mu_0$, $\eta_{con} = 1/\mu_0^{0.1}$, $\eta_{grad} = 1/\mu_0$
\FOR{$k = 0,1,2,\ldots$}

\STATE $found = 0$
\WHILE{ $found \neq 1$}
\STATE Try to find $x_k$ such that \\ $||x_k - P(x_k - \nabla_x \mathcal{L}(x_k, \lambda^k, \mu_k),  l , u)||_\infty \le \eta_{grad}$ via F2 using starting point as $x_{k - 1}$.
	\IF{ above step is completed successfully }
	 	\STATE Set $found = 1$
	\ELSE 
		\STATE $\lambda^{k + 1} = \lambda_k$
		\STATE $\mu_{k + 1} = \theta_l \mu_k$
		\STATE $\eta_{con} = 1/\mu_k^{0.1}$
		\STATE $\eta_{grad} = 1/\mu_k$	
	\ENDIF
		
\ENDWHILE

		\IF{ $ || c(x_k) ||_\infty \le \eta_{con}$ }
			\IF{ $ || c(x_k) ||_\infty \le \eta^*_{con}$ and \\ $||x_k - P(x_k - \nabla_x \mathcal{L}(x_k, \lambda^k, 0),  l , u)||_\infty \le \eta^*_{grad}$ }
				\STATE Stop and return current solution $x_k$.
			\ENDIF
			\STATE $\lambda^{k + 1} = \lambda_k - \mu_k c(x_k)$
			\STATE $\mu_{k + 1} = \mu_k$
			\STATE $\eta_{con} = \eta_{con} / \mu_{k+1}^{0.9} $
			\STATE $\eta_{grad} = \eta_{grad} / \mu_{k + 1} $
		\ELSE
			\STATE $\lambda^{k + 1} = \lambda_k$
			\STATE $\mu_{k + 1} = \theta_h \mu_k$
			\STATE $\eta_{con} = 1/\mu_k^{0.1}$
			\STATE $\eta_{grad} = 1/\mu_k$		
		\ENDIF

\ENDFOR
\end{algorithmic}
\caption{F1: Outer Iteration}
\end{algorithm}

At each inner iteration we form a quadratic approximation to the augmented lagrangian and approximately solve the inequality constrained quadratic sub-problem:

\begin{eqnarray}\label{a9}
\mbox{ min }_p \,\,\, \frac{1}{2} p^T \nabla^2_{xx} \mathcal{L}(x, \lambda, \mu) p + \nabla_x \mathcal{L}(x, \lambda, \mu)^T p \\
\mbox{ s.t. } l_i \le x_i \le u_i \\
\mbox{ s.t. } || p ||_\infty \le \Delta 
\end{eqnarray}

The inner iteration code uses non-linear gradient projection \cite{byrd95limited} followed by Newton-CG-Steihaug conjugate gradient iterations \cite{Steihaug:1983}. Quasi-Newton updates are performed using either SR1 \cite{Conn:SR1} (recommended for non-convex functions) or BFGS \cite{Broyden:1970} (recommended for convex functions). For very large problems, we switch to the limited memory variants  \cite{Nocedal:LM} of these quasi-Newton approximations. The algorithm details are given in Framework 2.  The trust region update code is based on a standard progress monitoring strategy \cite{Nocedal:book} and is given in Framework 3.

\begin{algorithm}
\begin{algorithmic}[1]
\REQUIRE  $j_{max}$, $\eta_{grad}$, $\Delta$, $l$, $u$, $\lambda^k$, $\mu_k$, $\eta \in (0,1)$, $flag$
\STATE $found = 0$
\STATE $x = x_{k - 1}$, $j = 1$
\STATE Compute, $g = \nabla_x \mathcal{L}(x, \lambda^k, \mu_k)$
\STATE Estimate $B =  \nabla^2_{xx} \mathcal{L}(x, \lambda^k, \mu_k)$ using BFGS, SR1 or limited memory BFGS, limited memory SR1 quasi Newton Updates.
\WHILE{$found \neq 1$ and $j \le j_{max}$}
\STATE Calculate the Cauchy point $p_c$ for problem:
\begin{eqnarray}
\mbox{ min }_p \,\,\, \frac{1}{2} p^T B p + g^T p \\
\mbox{ s.t. } l - x\le p \le u - x \\
 \mbox{ s.t. } || p ||_\infty \le \Delta 
\end{eqnarray}
using non-linear gradient projection and calculate the current active set $\mathcal{A}$. Let $e_i$ be the unit vector with $1$ at position $i$ and zeros elsewhere. If $i_1, i_2, \ldots i_q \notin \mathcal{A}$ then
let $\tilde{Q} = [e_{i_1}, e_{i_2},\ldots, e_{i_q}]$.
\STATE $\tilde{g} = \tilde{Q}^T ( g + B \, p_c)$ and $\tilde{B} = \tilde{Q}^T B \tilde{Q}$
\STATE Compute the approximate solution $\hat{v}$ to the problem
\begin{eqnarray}
\mbox{ min }_v \,\,\, \frac{1}{2} v^T \tilde{B} v + \tilde{g}^T v \\
\mbox{ s.t. } l - x \le p_c + \tilde{Q} v \le u - x \\
 \mbox{ s.t. } || p_c + \tilde{Q} v ||_\infty \le \Delta 
\end{eqnarray}
using truncated conjugate gradient iteration (Newton-CG, Steihaug). If $flag = 1$ use preconditioned Newton-CG using the inexact-modified Cholesky factorization.
\STATE Compute $\hat{p} = p_c + \tilde{Q} \hat{v}$
\STATE Calculate $\delta_{\mathcal{L}} =  \mathcal{L}(x) - \mathcal{L}(x + \hat{p}) $, $\delta_m = 0.5 \hat{p}^T B \hat{p} + g^T \hat{p}$ and $\rho = \delta_{\mathcal{L}} / \delta_m$

\IF{ $\rho > \eta$ }
\STATE	$x = x + \hat{p}$
\ENDIF

\STATE Compute new trust region radius $\Delta$ using Framework F3.

\STATE Compute, $g = \nabla_x \mathcal{L}(x, \lambda^k, \mu_k)$ if $\rho > \eta$ holds otherwise use the previous value.
\STATE Estimate $B =  \nabla^2_{xx} \mathcal{L}(x, \lambda^k, \mu_k)$ using BFGS, SR1 or limited memory BFGS, limited memory SR1 quasi Newton Updates. Do the update even if $\rho < \eta$.

\IF{$||x - P(x - \nabla_x \mathcal{L}(x, \lambda^k, \mu_k),  l , u)||_\infty \le \eta_{grad}$}
\STATE $found = 1$
\ENDIF

\STATE $j = j + 1$

\ENDWHILE
\end{algorithmic}
\caption{F2: Inner Iteration}
\end{algorithm}

\begin{algorithm}
\begin{algorithmic}[1]
\REQUIRE $\rho$, $\hat{p}$, $\Delta$
\IF{$\rho > 0.75$}
	\IF{$ ||\hat{p}||_\infty \le 0.8 \Delta $}
		\STATE $\Delta = \Delta$
	\ELSE
		\STATE $\Delta = 2 \Delta$
	\ENDIF
\ENDIF
\IF{$0.1 \le \rho \le 0.75$}
	\STATE $\Delta = \Delta$
\ELSE
	\STATE $\Delta = 0.5 \Delta$
\ENDIF
\RETURN $\Delta$
\end{algorithmic}
\caption{F3:Trust Region Update}
\end{algorithm}

\section{Optimization Benchmarks}\label{supplemental}

The optimization core of ADIS has been tested on many benchmark problems from the GAMS library at \url{http://www.gamsworld.org/performance} as well as benchmarks from MINOS \cite{MINOS:1993}.

This section will describe some numerical experiments on interesting and difficult optimization benchmarks used to test the optimization core of ADIS. For these benchmarks, the gradient information was generated using automatic differentiation from the software package INTLAB \cite{rump95intlab}. A limited memory variant of symmetric rank 1 (SR1) updating was used. The CG iterations were not preconditioned. The convergence tolerances $\eta_{con}$ and $\eta_{tol}$ were set to their default value of $1e{-6}$.

\subsubsection*{Electron 50}
Given $n_p$ electrons, find the equilibrium state distribution (of minimal Columb potential) of the electrons positioned on a conducting sphere. This problem is from COPS3 \cite{COPS3}  benchmark dataset.

The problem is to find a configuration of low energy for a given set of point charges on a conducting sphere. It originated with Thomson's plum pudding model of the atomic nucleus and is representative of an important class of problems in physics and chemistry that determine a structure with respect to atomic positions. Mathematically, the problem is:

\begin{eqnarray}\label{p1}
\mbox{ min }_{x,y,z} \,\,\,\, f(x,y,z) =  \\ \sum_{i = 1}^{n_p - 1} \sum_{j = i + 1}^{n_p} [ ( x_i - x_j)^2 + (y_i - y_j)^2 + (z_i - z_j)^2 ]^{-0.5} \nonumber
\end{eqnarray}
subject to 
\begin{equation}\label{p2}
x_i^2 + y_i^2 + z_i^2 = 1, i = 1, \ldots, n_p
\end{equation}

\begin{figure*}[htbp]
\begin{center}
\includegraphics[width = 120mm, angle = -90]{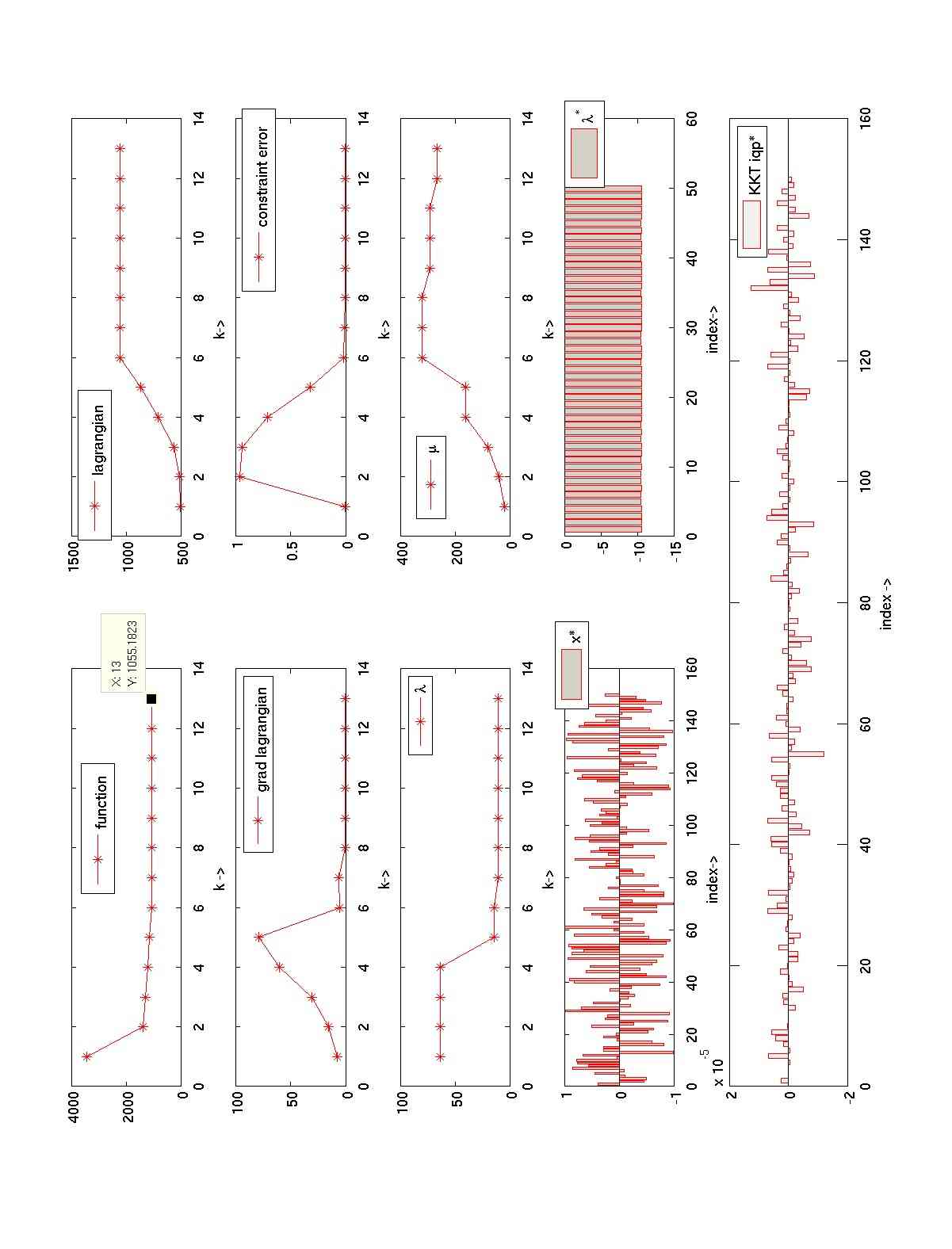}
\caption{Electron 50 convergence diagnostics. The best objective of $f^* = 1055.1823$ was attained in 13 outer iterations. Figure shows the evolution of objective function, the Lagrangian, norm of the gradient of the Lagrangian, norm of the constraint satisfaction error, norm of the Lagrange multipliers and penalty parameter over algorithm iterations along with verification of KKT optimality conditions.}
\label{figelec50}
\end{center}
\end{figure*}

The 150 variable problem for $n_p = 50$ was taken from GAMS performance library (PrincetonLib (NLP)) at \url{http://www.gamsworld.org/performance/princetonlib/htm/fekete/ fekete2.htm}. The best known objective for this problem for $n_p = 50$ is $f^* = 1055.1823$. Our code attains this best objective in 13 outer iterations. See figure \ref{figelec50} for convergence diagnostics.

\subsubsection*{Non-negative Least Square (NNLS)}

In NNLS we solve the problem:
\begin{equation}\label{p3}
\mbox{ min }_{x} \,\,\,\, f = || A x - b ||_2 
\end{equation}
subject to:
\begin{equation}
C x - d \ge 0
\end{equation}

We solve the 300 variable optimization problem taken from GAMS performance library (PrincetonLib (NLP)) at \url{http://www.gamsworld.org/performance/princetonlib/htm/nnls/nnls.htm}. 

\begin{figure*}[htbp]
\begin{center}
\includegraphics[width = 120mm, angle = -90]{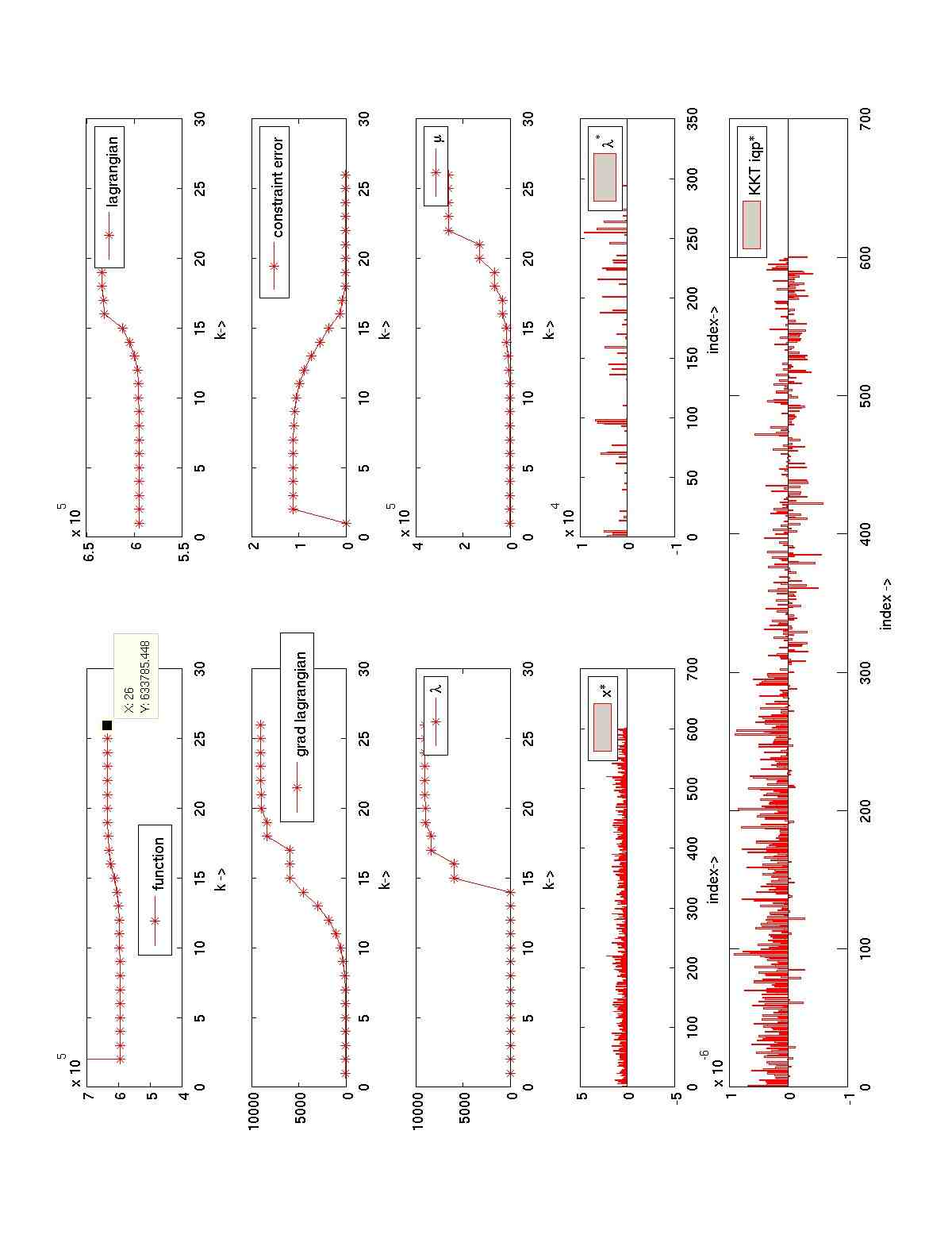}
\caption{Non-negative least squares convergence diagnostics. The best objective of $f^* = 633785.44$ was attained in 26 outer iterations. Figure shows the evolution of objective function, the Lagrangian, norm of the gradient of the Lagrangian, norm of the constraint satisfaction error, norm of the Lagrange multipliers and penalty parameter over algorithm iterations along with verification of KKT optimality conditions.}
\label{fignnls}
\end{center}
\end{figure*}

The best known object for this problem is $f^* =  633785.4462$. Our code attains this best objective in 26 outer iterations. See figure \ref{fignnls} for convergence diagnostics.

%\begin{figure*}
%\centering
%\begin{tabular}{cc}

%\hspace{-2.5in}
%\subfigure[Electron 50]{
%\includegraphics[height = 120mm, angle = 0]{figures/elec50_diagnostic2.jpg}
%\label{figelec50}}

%\hfill
%&

%\hspace{-2.8in}
%\subfigure[Non-negative least squares]{
%\includegraphics[height = 120mm, angle = 0]{figures/nnls2_diagnostic2.jpg}
%\label{fignnls}
%}

%\\
%\hfill
%\hspace{1.0in}
%\subfigure[Largest small polygon]{
%\includegraphics[height = 120mm, angle = -90]{figures/pgon_diagnostics2_May7.jpg}
%\label{figpolygon}
%}

%& 

%{}

%\end{tabular}
%\caption{ (a) Electron 50 convergence diagnostics. The best objective of $f^* = 1055.1823$ was attained in 13 outer iterations. (b) Non-negative least squares convergence diagnostics. The best objective of $f^* = 633785.44$ was attained in 26 outer iterations. (c) Largest Small Polygon convergence diagnostics for $n_v = 6$. Note that we first convert the problem into a minimization problem by multiplying the objective function with -1. The optimal objective value of $ -0.675$ for the minimization problem was attained in 11 outer iterations.}
%\label{benchmarkopt}
%\end{figure*}

\begin{figure*}[htbp]
\begin{center}
\includegraphics[width = 120mm, angle = -90]{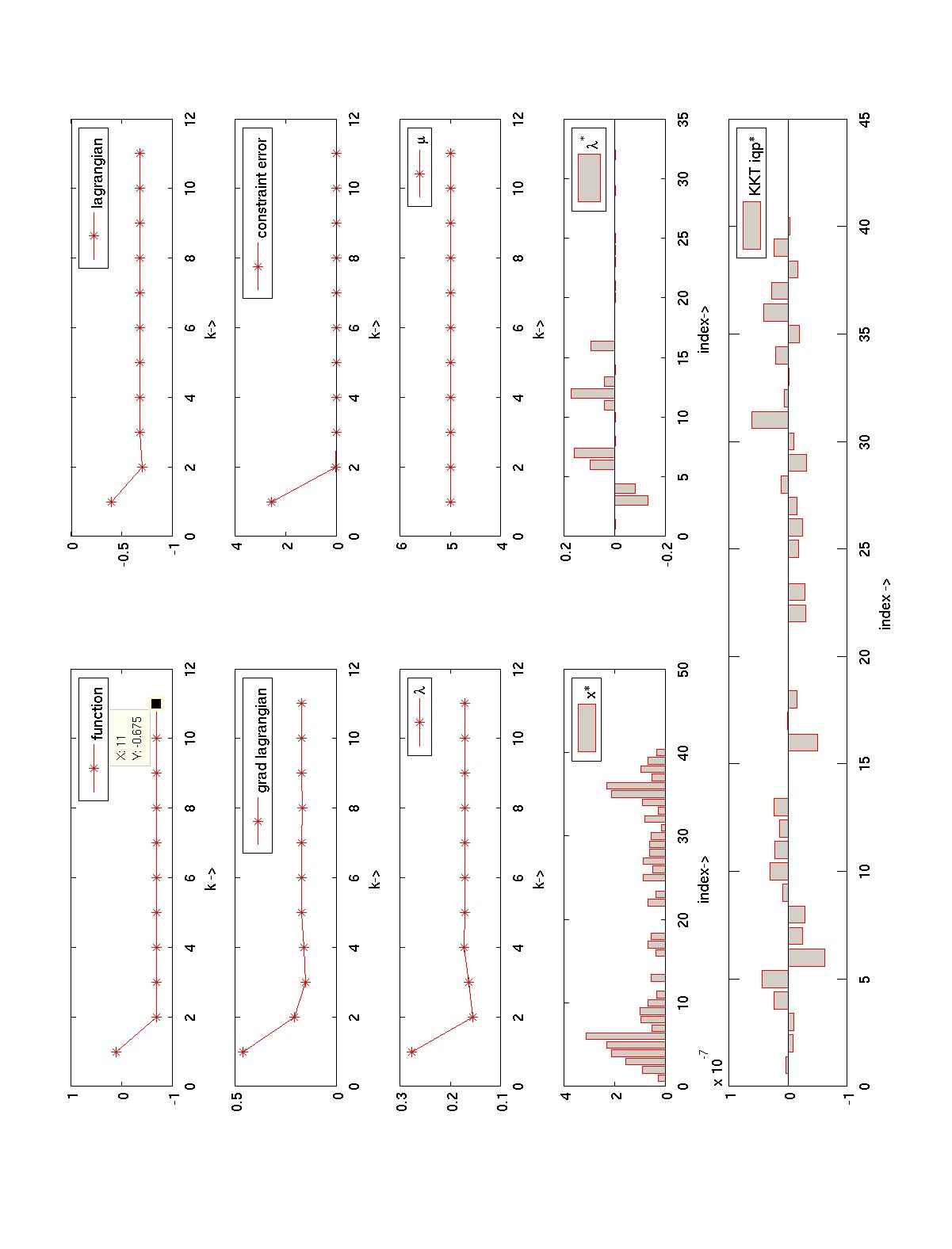}
\caption{Largest Small Polygon convergence diagnostics for $n_v = 6$. Note that we first convert the problem into a minimization problem by multiplying the objective function with -1. The optimal objective value of $ -0.675$ for the minimization problem was attained in 11 outer iterations. Figure shows the evolution of objective function, the Lagrangian, norm of the gradient of the Lagrangian, norm of the constraint satisfaction error, norm of the Lagrange multipliers and penalty parameter over algorithm iterations along with verification of KKT optimality conditions.}
\label{figpolygon}
\end{center}
\end{figure*}

\subsubsection*{Largest Small PolyGon}

This is a classic problem also from COPS3 \cite{COPS3}  benchmark dataset. Given coordinates $(r_i,\theta_i)$ of the $n_v$ vertices of a polygon, we wish to solve the problem:

\begin{equation}\label{p4}
\mbox{ max }_{r,\theta}  \,\,\, f(r,\theta) = 0.5 \sum_{i = 1}^{n_v - 1} r_{i+1} r_i \sin(\theta_{i + 1} - \theta_i)
\end{equation}
subject to:
\begin{eqnarray}
r_i^2 + r_j^2 - 2 \, r_i r_j \cos(\theta_i - \theta_j) \le 1, \, 1 \le i < n_v, \, i < j \le n_v \\
\theta_i \le \theta_{i + 1}, \, 1 \le i < n_v \\
\theta_i \in [0, \pi], r_i \ge 0, \, 1 \le i \le n_v
\end{eqnarray}

Some of the interesting features of this problem that make it difficult include the presence of the order of $n_v^2$ nonlinear nonconvex inequality constraints and the presence of $O(n_v!)$ local minima. See \cite{COPS3} for more details. We solve the $n_v=6$ problem taken from GAMS performance library (PrincetonLib (NLP)) at \url{http://www.gamsworld.org/performance/princetonlib/htm/polygon/pgon.htm}. 

The best known objective in GAMS library for this problem is $f^* =  0.5$. 

We first convert the maximization problem \ref{p4}  into a minimization problem by multiplying the objective function by -1. We solve this problem and then evaluate the original objective \ref{p4} at the solution.
We find that our algorithm attains an objective of $\hat{f} = 0.675$ in 11 outer iterations. 

This is better than that reported by GAMS solvers. We were a little surprised by this observation. On researching this problem further we found that Graham \cite{Graham:1975} had solved this problem in 1975 and the best solution is indeed 0.675 (see \url{http://mathworld.wolfram.com/GrahamsBiggestLittleHexagon.html}). On plotting the optimal hexagon estimated by our code alongside the solution from \cite{Graham:1975}, we find that they are identical. See figure \ref{figpolygon} for convergence diagnostics.

%---

%\end{comment}

\nocite{JADE:1996}
\nocite{FPICA:2000}
\nocite{FPICA:1999}
\nocite{EFICA:2006}
\nocite{EFICA2:2006}
\nocite{ERICA:2002}
\nocite{UNICA:2000}
\nocite{ICALAB_BOOK:2003}
\nocite{ICALAB}
\nocite{SIR:2006}
\nocite{LANCELOT:1992}
\nocite{Conn:1991}

\nocite{Nocedal:book}

\newpage
\bibliographystyle{plain}
\bibliography{bibliography_proj_pursuit_arxiv.bib}

\end{document}